\documentstyle[12pt,epsfig]{article}
\makeatletter
\renewcommand{\theequation}{\thesection.\arabic{equation}}
\@addtoreset{equation}{section}
\makeatother
\def\appendix#1{
  \addtocounter{section}{1}
  \setcounter{equation}{0}
  \renewcommand{\thesection}{\Alph{section}}
  \renewcommand{\theequation}{\Alph{section}.\arabic{equation}}
  \section*{Appendix \thesection\protect\indent #1}
  \addcontentsline{toc}{section}{Appendix \thesection #1}
  }
\newcommand{\beq}{\begin{equation}}
\newcommand{\eeq}{\end{equation}}
\newcommand{\beqy}{\begin{eqnarray}}
\newcommand{\eeqy}{\end{eqnarray}}

\def\tpi{\tilde{\pi}}
\def\tPi{\tilde{\Pi}}
\def\tB{\tilde{B}}
\def\te{\tilde{e}}
\def \ut#1{\rlap{\lower1ex\hbox{$\sim$}}#1{}}
\def \UT#1{\rlap{\lower1ex\hbox{\scriptsize$\sim$}}#1{}}
\newcommand{\UI}[1]{^{\mbox{ \ } #1}}
\newcommand{\LI}[1]{_{\mbox{ \ } #1}}
\def\tN{\ut{N}}
\def\tM{\ut{M}}
\def\ep{\epsilon}
\def\otep{\tilde{\epsilon}}
\def\utep{\UT{\epsilon}}
\def\BC{{\bf C}}
\def\CA{{\cal A}}
\def\CB{{\cal B}}
\def\CF{{\cal F}}
\def\CD{{\cal D}}
\def\CH{{\cal H}}
\def\CG{{\cal G}}
\def\CN{{\cal N}}

\def\CL{{\cal L}}
\def\CP{{\cal P}}
\def\CS{{\cal S}}
\def\CV{{\cal V}}
\def\M3{M^{(3)}}
\def\TM{\widetilde{M}}
\def\tx{\tilde{x}}
\def\Tr{{\rm Tr}}
\def\Str{{\rm STr}}
\newsavebox{\LBRA}
\savebox{\LBRA}(5,30){%
\begin{picture}(5,30)
\put(2,2){\thicklines\line(0,1){26}}
\put(3,2){\thicklines\oval(2,2)[bl]}
\put(3,28){\thicklines\oval(2,2)[tl]}
\end{picture}}
\newsavebox{\RBRA}
\savebox{\RBRA}(5,30){%
\begin{picture}(5,30)
\put(3,2){\thicklines\line(0,1){26}}
\put(2,2){\thicklines\oval(2,2)[br]}
\put(2,28){\thicklines\oval(2,2)[tr]}
\end{picture}}
\begin{document}
\begin{flushright}
gr-qc/9601050\\January 1996
\end{flushright}
\renewcommand{\thefootnote}{\fnsymbol{footnote}}
\vspace{0.5in}
\begin{center}\Large{\bf Nonperturbative solutions for\\
canonical quantum gravity: an overview}\footnote[1]{
A thesis for partial fulfillment of the rquirements
for the degree of Doctor of Science}\\
\vspace{1cm}
\normalsize\ Kiyoshi Ezawa\footnote[2]
{e-mail address: ezawa@funpth.phys.sci.osaka-u.ac.jp}
\vspace{0.5in}

        Department of Physics \\
        Osaka University, Toyonaka, Osaka 560, Japan\\
\vspace{0.1in}
\end{center}
\renewcommand{\thefootnote}{\arabic{footnote}}
\setcounter{footnote}{0}
\vspace{1.2in}
\baselineskip 17pt
\begin{abstract}
In this paper we will make a survey of solutions to the
Wheeler-De Witt equation which have been found up to now in
Ashtekar's formulation for canonical quantum gravity.
Roughly speaking they are classified into two categories,
namely, Wilson-loop solutions and topological solutions.
While the program of finding solutions which are composed of
Wilson loops is still in its infancy, it is expected
to be developed in the near future.
Topological solutions are the only solutions at present which we
can give their interpretation in terms of spacetime geometry.
While the analysis made here is formal in the sense that
we do not deal with rigorously regularized constraint
equations, these topological solutions are expected to exist
even in the fully regularized theory and they are considered to
yield vacuum states of quantum gravity.
We also make an attempt to review the spin network states
as intuitively as possible. In particular, the explicit formulae
for two kinds of measures on the space of
spin network states are given.

\end{abstract}

\vspace{0.2in}

\newpage

\tableofcontents

\baselineskip 16pt

\section{Introduction}

\subsection{Motivations for quantum gravity}

By a large amount of empirical evidence in astrophysics,
it turns out that general relativity describes almost (more than
99 percent) correctly classical behavior of the gravitational
interaction. According to the singularity theorem demonstrated
by Penrose and Hawking\cite{hawk}, however,
general relativity necessarily leads to spacetime singularity
after the stellar collapse or at the beginning of the
cosmological evolution as long as the energy momentum tensor
satisfies certain positivity conditions.
This intrinsic pathology suggests that general relativity
is in fact the effective theory of a certain more fundamental
theory of gravity, which is called the theory of
quantum gravity. It is natural to think that, in the very early
period of the universe in which the characteristic scale
of the system is the Planck length $L_{P}\equiv(G\hbar/c^{3})^{1/2}
\sim 10^{-33}cm$, or the Planck energy scale $\Lambda_{P}
\equiv\hbar c/L_{P}\sim10^{19}GeV$, nonperturbative effects of
quantum gravity spoil the validity of general relativity and
therefore the initial singularity is circumvented.
This is the main motivation for quantum gravity from the
perspective of general relativitists.

Particle physicists have other motivations.
Most quantum field theories including the
Standard model are plagued with the problem of
ultraviolet divergences. Among these UV divergences
some relatively tamable ones are handled by the renormalization
prescription. Here a natural hope arises that the UV divergences
may be removed by taking account
of quantum gravity, because it is conceivable in quantum gravity
that the physics at the Planck scale is quite different from
that at our scale.

More ambitious particle physicists attempt to
construct the unified theory in which all the four interactions---
gravitational, weak, electromagnetic and strong
interactions--- are described by a single entity. To do so
gravitational interaction needs to be quantized because the other
three interactions have already been quantized.

Because quantum gravity is essentially the physics
at the Planck scale which is far beyond our experience,
it is no wonder that we have not yet acquired a consensus on
what the theory of quantum gravity would look like.
In consequence there are many approaches to quantum gravity
\cite{isham}. Among them, the following three approaches
are vigorously investigated:

i) superstring theory. In the mid seventies
perturbative quantization of general relativity turned out to be
non-renormalizable. This result lead particle physicists to consider
that general relativity is the low energy effective theory
of a more fundamental theory which is presumably renormalizable
(or finite).
One such candidate is the theory of
superstrings, in which the gravity
is incorporated as a massless excitation mode of the closed string.
Recently it has been clarified that the strong coupling phase of
a superstring theory is described by the weak coupling phase
of another superstring theory\cite{HT}\cite{wittS}.
A series of these recent discoveries is expected to accelerate
developments in nonperturbative aspects of superstring theory;

ii) discretized quantum gravity. Another natural reaction to the
non-renormalizability of general relativity is to consider
the nonperturbative quantization of general relativity.
There are two main approaches to implement this, discretized
quantum gravity and canonical quantum gravity. In discretized
quantum gravity, we first approximate the spacetime manifold
and the Einstein-Hilbert action by a
simplicial complex (a connected set of four-simplices)
and by the Regge action respectively\cite{regge}. Quantization
is performed by numerical path integral methods, namely by
taking a summation of various numerical configurations.
There are several choices of independent variables, the two
most typical ones among which are seen in
Regge calculus and in dynamical triangulations\cite{ambj}.
In Regge calculus we fix a triangulation and use the edge
lengths of four-simplices as dynamical variables, whereas in
dynamical triangulations we set the edge lengths to unity
and sum up all the possible ways of triangulations.
While we have not yet extracted physically meaningful
information on four dimensional gravity,
this approach is expected to yield some intuitive
picture on quantum gravity in the near future;

iii) canonical quantization of general relativity.
Some general relativitists consider that splitting the metric
into background and quantum fluctuation parts is the origin
of the failure in the perturbative quantization of
general relativity. They therefore anticipate that it may be
possible to quantize general relativity if we deal with the
metric as a whole. Inspired by this anticipation canonical
quantization for general relativity has been investigated for
about thirty years. While the metric formulation seems to be
too complicated to solve, since Ashtekar discovered new variables
which simplifies the canonical formulation\cite{ashte},
Ashtekar's formalism has been actively investigated by many people
\cite{schil}. This Ashtekar's formulation for general relativity
is now considered to be one of the promising approaches
to quantum gravity.

In the following we will explain more precisely
this canonical approach.

\subsection{Issues on canonical quantum gravity}

Canonical quantization in the metric formulation starts from
the (3+1)-decomposition of the spacetime metric:
\beq
ds^{2}=g_{\mu\nu}dx^{\mu}dx^{\nu}=-N^{2}dt^{2}
+q_{ab}(dx^{a}+N^{a}dt)(dx^{b}+N^{b}dt).
\eeq
By plugging this into the Einstein-Hilbert action $I_{EH}=
\int_{M}d^{4}x\sqrt{-g}R$, we obtain the ADM action\cite{ADM}
\beq
I_{ADM}=\int dt\int_{\M3}d^{3}x[\tpi^{ab}\dot{q}_{ab}-
N\CH-N_{a}\CH^{a}],
\eeq
where $\CH^{a}$ and $\CH$ are respectively the momentum constraint
and the Hamiltonian constraint. If we perform Dirac's quantization,
these two first class constraints yield the constraint
equations imposed on the wavefunction $\Psi[q_{ab}]$.
The momentum constraint $\hat{\CH}^{a}$ requires
$\Psi[q_{ab}]$ to be invariant under spatial diffeomorphisms.
The Hamiltonian constraint $\hat{\CH}$ yields the notorious
Wheeler-De Witt (WD) equation\cite{De Witt}
\beq
\left\{\sqrt{q}^{-1}(q_{ac}q_{bd}-\frac{1}{2}q_{ab}q_{cd})
\frac{\delta}{\delta q_{ab}}\frac{\delta}{\delta q_{cd}}
+\sqrt{q}^{(3)}R(q)\right\}\Psi[q_{ab}]=0,
\label{eq:WDEQ}
\eeq
where $^{(3)}R(q)$ is the 3-dimensional scalar curvature.
This WD equation is considered as the dynamical equation
in canonical quantum gravity.

Canonical quantum gravity is known to possess intrinsically
the following conceptual problems:\\
i) Because the WD equation is the Klein-Gordon type equation
with signature
$$
(-,+,+,+,+,+)\times\infty^{3},
$$
we cannot
construct any positive semi-definite inner-product which
is conserved by the WD equation. Thus we do not know how to
interpret the wave function $\Psi[q_{ab}]$.\\
ii) The \lq\lq issue of time" in quantum gravity\cite{kuch}. The
Hamiltonian of canonical quantum gravity is given by a linear
combination of the first class constraints $(\CH,\CH^{a})$.
As a result the evolution of the physical wavefunction
$\Psi[q_{ab}]$ (or of physical observables) w.r.t. the
coordinate time $t$ is trivial. This naturally follows from
general covariance of the theory. Several remedies for this problem
have been considered. For example, to identify a gravitational
degree of freedom or a matter degree of freedom with a physical
time. The former is called the \lq\lq intrinsic time"
and the latter is called the \lq\lq extrinsic time".

These conceptual problems originally stems from
lack of our knowledge on the quantization of diffeomorphism
invariant field theories. In order to resolve these problems,
we therefore have to investigate more extensively
these diffeomorphism invariant quantum field theories such as
quantum gravity.
When we try to carry out the program of
canonical quantum gravity, however,
we always run into the serious technical problem.
Namely, the Wheeler-De Witt equation (\ref{eq:WDEQ}) is so
complicated that we have not found even one solution
to it.\footnote{Quite recently a class of solutions
to the WD equation (\ref{eq:WDEQ}) have been found\cite{kowa}.
Their physical significance has not been clarified as yet.
} This severe fact had confronted the researchers
of canonical quantum gravity until 1986 when
the new canonical variables were discovered by Ashtekar
\cite{ashte}.

\subsection{Ashtekar's formalism and solutions to the
Wheeler-De Witt equation}

As we will see in \S 2, the Wheeler-De Witt equation takes
a simple form if written in terms of Ashtekar's new variables.
We therefore expect that the WD equation is solved
in Ashtekar's formalism. In fact several types of solutions
have been found. Roughly speaking they are classified into
two classes.

One is the class of \lq\lq Wilson loop solutions"
\cite{jacob}\cite{HBP}\cite{ezawa3}\cite{ezawa4} whose fundamental
ingredient is the parallel propagator of the Ashtekar connection
$A^{i}_{a}$ along a curve $\alpha:[0,1]\rightarrow\M3$:
$$
h_{\alpha}[0,1]=\CP\exp\{\int_{0}^{1}ds\dot{\alpha}^{a}(s)A^{i}_{a}
(\alpha(s))J_{i}\}.
$$
Recently there have been remarkable developments in the application
of spin network states
(the extended objects of Wilson loops, see \S 3)
to Ashtekar's formalism \cite{asle} \cite{ALMMT} \cite{ALMMT2}
\cite{baez3} \cite{baez2} \cite{RS2} \cite{RS3}.
Inspired by these developments, progress in the area of
constructing solutions from parallel propagators
is expected to be made in the near future.

While finding Wilson loop solutions is a promising program,
as yet we have not reached the point where we can
extract physically interesting results.
The other class of solutions, however, are considered to
have definite significance in quantum gravity.
They are the topological solutions\cite{kodama}\cite{bren}
which are regarded as \lq\lq vacuum states" in quantum gravity
\cite{smol2}\cite{mats}.
These solutions therefore guarantees at least that the efforts
made in canonical quantum gravity will not completely be
in vain.

\subsection{An outline of this paper}

In this paper we will make a survey of developments and issues
on the solutions for canonical quantum gravity.

\S 2 and \S 3 are devoted to an extensive review of the
bases necessary to read this paper.

In \S 2 we explain classical and quantum theory
of Ashtekar's formalism\cite{ashte2}. We followed the derivation
by way of the chiral Lagrangian\cite{capo}.
In \S 3 the results on the spin network states are reviewed.
The notion of spin network states is a kind of generalization
of link-functionals in lattice gauge theory\cite{KS}.
It was introduced into quantum general relativity
in order to construct a mathematically rigorous formulation
of the loop representation\cite{RS}, which is expected to
describe neatly the nonperturbative nature of canonical
quantum gravity.
While recent progress on the spin network states is
striking\cite{asle}\cite{ALMMT}\cite{ALMMT2}
\cite{baez3}\cite{baez2}\cite{RS2}\cite{RS3}, their underlying
ideas have been obscured because of the necessity for
mathematical rigor. This makes this area
of \lq\lq applied" spin network states inaccessible
to the usual physicists. So in that section we will make
a maximal attempt to provide the results in the forms
which are familiar to the physicists\footnote{
To a reader who is interested in the mathematically rigorous
formulation, we recommend to read \cite{ashte3}
and references therein.}. In particular we
give several explicit formulae for the measures defined on
spin network states. They are useful in proving the consistency
of the measures and, in particular, in showing completely that the
set of spin network states forms an orthonormal basis
w.r.t. these measures (the induced Haar measure and the induced
heat-kernel measure).

Most of the subject in \S4$\sim$\S 6 is based on the author's
recent works\cite{ezawa3}\cite{ezawa4}\cite{ezawa5}\cite{ezawa6}.

In \S 4 we present Wilson loop solutions. We first compute
the action of \lq\lq renormalized" scalar constraint on
the spin network states.
Then we provide simple solutions which we call
\lq\lq combinatorial solutions" to the renormalized WD equation.
After that we
explain why we cannot extract physically interesting results
from these solutions. As an attempt to construct physically
interesting solutions, we will introduce in \S 5 the lattice
approach to Ashtekar's formalism. There a set of nontrivial
solutions---\lq\lq multi-plaquette solutions"---
to a discretized WD equation is given.
Some properties of these solutions indicate a direction of future 
developments in this area.
Topological solutions are investigated in \S 6.
We will focus on the semiclassical interpretation of
these solutions and show that they semiclassically represent
the spacetimes which are vacuum solutions to the Einstein equation.
In particular in the case of a vanishing cosmological constant,
each topological solution
corresponds to a family of (3+1)-dimensional
Lorentzian structures with a fixed projection
to the Lorentz group. In that section we will also show
that Ashtekar's formulation for $N=1$ supergravity also
possesses topological solutions by unraveling the relation
between $GSU(2)$ BF theory and $N=1$ Ashtekar's formalism.
In \S 7 a brief comment is made on solutions\cite{csordas},
including the Jones polynomial
\cite{brug}\cite{BGP}\cite{GP}\cite{GP2} in
the loop representation\cite{RS},
which are not properly contained in the above two categories.

Here we will explain the notation used in this paper:
\begin{enumerate}
\item $\alpha,\beta,\cdots$ denote $SO(3,1)$ local Lorentz indices;
\item $\mu,\nu,\cdots$ stand for spacetime indices;
\item $a,b,\cdots$ are used for spatial indices;
\item $A,B,\cdots$ represent
left-handed $SL(2,{\bf C})$ spinor indices;
\item $i,j,\cdots$
denote indices for the  adjoint
representation of (the left-handed part of)
$SL(2,{\bf C})$. $J_{i}$ stands for the $SL(2,{\bf C})$ generator
subject to the commutation relation $[J_{i},J_{j}]=\ep_{ijk}J_{k}$;
\item $\otep^{abc}$( $\utep_{abc}$) is the Levi-Civita
alternating tensor density of weight $+1$ ($-1$) with
$\otep^{123}=\utep_{123}=1$;
\item $\ep^{ijk}$
is the antisymmetric (pseudo-)tensor with $\ep^{123}=1$;
\item $\ep^{AB}$ ($\ep_{AB}$) is the antisymmetric spinor with
$\ep^{12}=\ep_{12}=1$;\footnote{These antisymmetric spinors are used
to raise and lower the spinor index:
$\varphi^{A}=\ep^{AB}\varphi_{B},
\quad \varphi_{A}=\varphi^{B}\ep_{BA}$.}
\item relation between a
symmetric rank-2 spinor $\phi^{AB}$ and its equivalent vector
$\phi^{i}$ in the adjoint representation is given by
$\phi^{AB}=\phi^{i}(J_{i})^{AB}$, where
$(J_{i})^{A}\LI{B}$ is the $SL(2,\BC)$ generator
in the spinor representation subject to
$(J_{i})^{A}\LI{C}(J_{j})^{C}\LI{B}
=-\frac{1}{4}\delta^{ij}\delta^{A}_{B}+\frac{1}{2}
\ep^{ijk}(J_{k})^{A}\LI{B}$ ;
\item $D=dx^{\mu}D_{\mu}$ denotes the covariant exterior derivative
with respect to the $SL(2,\BC)$ connection $A=A^{i}J_{i}$ and ;
\item indices located between $($ and $)$ ($[$ and $]$) are regarded
as symmetrized (antisymmetrized).
\end{enumerate}

For analytical simplicity, we will consider in this paper
that the spacetime $M$
has the topology ${\bf R}\times\M3$ with $\M3$ being
a compact, oriented, 3 dimensional manifold without boundary.


\section{Ashtekar's formalism}


\subsection{Classical theory}

In this section we provide a review of Ashtekar's formulation
for canonical general relativity. Ashtekar's formalism were
originally derived from the first order ADM formalism
by exploiting the complex canonical transformation
\cite{ashte}\cite{ashte2}. In order to derive it
from the outset, however, it is more convenient to use
the chiral Lagrangian formalism\cite{capo}.
So we will take the latter route.

First we consider the usual first-order formalism of general
relativity, whose action is called the Einstein-Palatini action
with a cosmological constant $\Lambda$:
\beq
I_{EP}=-\frac{1}{2}
\int_{M}\ep_{\alpha\beta\gamma\delta}e^{\alpha}\wedge
e^{\beta}\wedge(R^{\gamma\delta}-\frac{\Lambda}{6}e^{\gamma}\wedge
e^{\delta}),
\eeq
where $e^{\alpha}$ is the vierbein and $R^{\alpha\beta}=
d\omega^{\alpha\beta}+\omega^{\alpha}\LI{\gamma}\wedge
\omega^{\gamma\beta}$ denotes the curvature of the spin-connection
$\omega^{\alpha\beta}$. Equations of motion are derived from the
variational principle w.r.t. the vierbein $e^{\alpha}$
and the spin-connection $\omega^{\alpha\beta}$.
The results are respectively given by
\beqy
0&=&\ep_{\alpha\beta\gamma\delta}e^{\beta}\wedge
(R^{\gamma\delta}-\frac{\Lambda}{3}e^{\gamma}\wedge e^{\delta})
\nonumber \\
0&=&\ep_{\alpha\beta\gamma\delta}e^{\gamma}\wedge
(de^{\delta}+\omega^{\delta}\LI{\ep}\wedge e^{\ep}).
\label{eq:EPeq}
\eeqy
If we assume that the vierbein $e^{\alpha}$ is
non-degenerate, the second equation is equivalent to the
torsion-free condition
\beq
de^{\alpha}+\omega^{\alpha}\LI{\beta}\wedge
e^{\beta}=0.
\eeq
The result of plugging its solution into $I_{EP}$ yields the
Einstein-Hilbert action
\beq
I_{EH}=\int_{M}d^{4}x\sqrt{g}(R-2\Lambda),
\eeq
where $R$ denotes the scalar curvature. This implies that
the two equations in (\ref{eq:EPeq}) are equivalent to
the Einstein equation provided that the vierbein is non-degenerate.

The (3+1)-decomposition of $I_{EP}$ yields the first-order
ADM formalism which is essentially equivalent to the usual
ADM formalism\cite{ADM}. The resulting Wheeler-De Witt (WD) equation
is as complicated as that in the ADM formalism.
This Einstein-Palatini action is therefore not suitable for
canonical quantization of general relativity.

In deriving Ashtekar's formalism from the outset, we need
another action which is classically equivalent to the
Einstein-Hilbert action. This is the complex chiral action
(or the Plebanski action \cite{pleb}):\footnote{
The definitions and properties of
$P^{(-)\alpha\beta}_{\quad\gamma\delta}$ and
$P^{(-)i}_{\alpha\beta}$ are listed in Appendix A.}
\beqy
I_{CC}&=&-i\int_{M}P^{(-)\alpha\beta}_{\quad\gamma\delta}
e_{\alpha}\wedge e_{\beta}\wedge(R^{\gamma\delta}-\frac{\Lambda}{6}
e^{\gamma}\wedge e^{\delta})\nonumber \\
&=&\frac{1}{2}I_{EP}-\frac{i}{2}\int_{M}e_{\alpha}\wedge e_{\beta}
\wedge R^{\alpha\beta}.
\eeqy
We can easily see that this action is equivalent to
$\frac{1}{2}I_{EH}$ under the torsion-free condition $de^{\alpha}+
\omega^{\alpha}\LI{\beta}\wedge e^{\beta}=0$, which is the equation
of motion derived from the real part of $I_{CC}$. As long as we
regard $(e^{\alpha},\omega^{\beta\gamma})$ to be real-valued,
we can deal with the real and imaginary parts of $I_{CC}$ separately
in deriving equations of motion. The complex chiral action
$I_{CC}$ is therefore classically equivalent to the Einstein-Hilbert
action.\footnote{The difference of the overall factor by $2$ is not
important because we can always change the overall factor by taking
different unit of length.}

Using eq.(\ref{eq:projection}) this complex chiral action
can be cast into the BF action:
\beq
I_{CC}=i\int_{M}(\Sigma^{i}\wedge F^{i}+\frac{\Lambda}{6}
\Sigma^{i}\wedge\Sigma^{i}),\label{eq:CCac}
\eeq
where $F^{i}=dA^{i}+\frac{1}{2}\ep^{ijk}A^{j}\wedge A^{k}$
is the curvature of the $SL(2,{\bf C})$ connection $A^{i}$
and $\Sigma^{i}$ is an $SL(2,{\bf C})$ Lie algebra-valued two-form.
The new variables $(A^{i},\Sigma^{i})$ are related with the
old ones $(e^{\alpha},\omega^{\alpha\beta})$ by:
\beqy
A^{i}&\equiv&-iP^{(-)i}_{\alpha\beta}\omega^{\alpha\beta}=
-\frac{1}{2}\ep^{ijk}\omega^{jk}-i\omega^{0i}\label{eq:creal1} \\
\Sigma^{i}&\equiv&iP^{(-)i}_{\alpha\beta}e^{\alpha}\wedge e^{\beta}
=\frac{1}{2}\ep^{ijk}e^{j}\wedge e^{k}+ie^{0}\wedge e^{k}.
\label{eq:alcon1}
\eeqy
Namely, $A^{i}$ is the anti-self-dual part of the spin-connection
and $\Sigma^{i}$ is the anti-self-dual two-form constructed from
the vierbein.

In order to rewrite the action in the canonical form, we have to
perform a (3+1)-decomposition. We consider the spacetime $M$ to be
homeomorphic to ${\bf R}\times \M3$ and use $t$ and $(x^{a})$ as
coordinates for ${\bf R}$ and the spatial hypersurface $\M3$
respectively. The result is
\beq
I_{CC}=i\int dt\int_{\M3}(\tpi^{ai}\dot{A^{i}_{a}}+A^{i}_{t}G^{i}
+\Sigma^{i}_{ta}\Phi^{ai}),\label{eq:bfac2}
\eeq
where $\tpi^{ai}\equiv\frac{1}{2}\otep^{abc}
\Sigma^{i}_{bc}$ plays the role
of the momentum conjugate to $A^{i}_{a}$, and $G^{i}$ and $\Phi^{ai}$
are the first class constraints {\em in BF theory}:
\beqy
G^{i}&=&D_{a}\tpi^{ai}\equiv\partial_{a}\tpi^{ai}+\ep^{ijk}A^{j}_{a}
\tpi^{ak} \label{eq:gauss} \\
\Phi^{ai}&=&\frac{1}{2}\otep^{abc}F^{i}_{bc}+\frac{\Lambda}{3}
\tpi^{ai} \label{eq:krcon}.
\eeqy

In order to obtain the action for Ashtekar's formalism we further
have to express $\Sigma^{i}_{ta}$ in terms of $\tpi^{ai}$ by solving $\Sigma^{i}=iP^{(-)i}_{\alpha\beta}e^{\alpha}\wedge e^{\beta}$.
Let us first fix the Lorentz
boost part of $SL(2,{\bf C})$ gauge transformation by choosing
the spatial gauge:
\beq
e^{0}=-Ndt,\quad e^{i}=e^{i}_{a}(dx^{a}+N^{a}dt),
\label{eq:spatialgauge}
\eeq
where $N$ and $N^{a}$ are the lapse function and the shift vector,
and $e^{i}_{a}dx^{a}$ yields the induced dreibein on $\M3$.
Plugging this expression into eq.(\ref{eq:alcon1}) and performing
a straightforward calculation, we find
\beqy
\tpi^{ai}&=&{\rm det}(e^{j}_{b})e^{a}_{i}\equiv\te^{ai}
\label{eq:spatial} \\
\Sigma^{i}_{ta}&=&N^{b}\utep_{bac}\tpi^{ci}+\frac{i}{2}\tN\ep^{ijk}
\utep_{abc}\tpi^{bj}\tpi^{ck},\label{eq:lagmul}
\eeqy
where $e^{a}_{i}$ denotes the co-dreibein $e^{a}_{i}e^{i}_{b}=
\delta^{a}_{b}$ and $\tN\equiv N\{{\rm det}
(\tpi^{ai})\}^{-\frac{1}{2}}$
is an $SO(3,{\bf C})$ invariant scalar density of weight $-1$
which plays the role of a new Lagrange multiplier.

Here we should note that eq.(\ref{eq:lagmul}) is
covariant under $SO(3,{\bf C})$ gauge transformations.
Eq.(\ref{eq:lagmul}) therefore holds under
arbitrary gauge, as long as the vierbein $e^{\alpha}$ can be written
in the form of eq.(\ref{eq:spatialgauge}) in a particular gauge.
This is always possible if the dreibein $(e^{i}_{a})$ is
non-degenerate. A more detailed investigation\cite{mats2} shows that
this is possible if there is a spacelike hypersurface at each
point in the spacetime.

Now the desired action is obtained by substituting
eq.(\ref{eq:lagmul}) into the BF action (\ref{eq:bfac2}):
\beq
I_{CC}=i\int dt\int_{\M3}d^{3}x(\tpi^{ai}\dot{A^{i}_{a}}+A_{t}^{i}
G^{i}+N^{a}{\cal V}_{a}+\frac{i}{2}\tN\CS).
\eeq
From this action we see that there are three kinds of first class
constraints in Ashtekar's formalism, namely, Gauss' law constraint
(\ref{eq:gauss}), the vector constraint ${\cal V}_{a}$, and the
scalar constraint $\CS$.\footnote{
We should note that $G^{i}$, ${\cal V}_{a}$ and $\CS$ are
respectively of density weight $+1$, $+1$ and $+2$, while
they are not explicitly shown by the tilde.}
The latter two constraints are of the
following form
\beqy
{\cal V}_{a}&\equiv&\utep_{abc}\tpi^{ci}\Phi^{bi}
=-\tpi^{bi}F^{i}_{ab}\label{eq:vector} \\
\CS&\equiv&\ep^{ijk}\utep_{abc}\tpi^{ai}\tpi^{bj}\Phi^{ck}
=\ep^{ijk}\tpi^{ai}\tpi^{bj}(F^{k}_{ab}+\frac{\Lambda}{3}
\utep_{abc}\tpi^{ck}) \label{eq:scalar}
\eeqy

Next we look into the constraint algebra. For this aim it is
convenient to use smeared constraints:
\beqy
G(\theta)&=&i\int_{\M3}d^{3}x\theta^{i}G^{i}
=-i\int_{\M3}d^{3}xD_{a}\theta^{i}\tpi^{ai} \nonumber \\
\CD(\vec{N})&=&-i\int_{\M3}d^{3}xN^{a}({\cal V}_{a}+A_{a}^{i}G^{i})
=i\int_{\M3}d^{3}x\tpi^{ai}\CL_{\vec{N}}A_{a}^{i} \nonumber \\
\CS(\tN)&=&\frac{1}{2}\int_{\M3}d^{3}x\tN\CS, \label{eq:smecon}
\eeqy
where $\theta^{i}$ is an $SO(3,{\bf C})$ Lie algebra-valued scalar,
$\vec{N}=(N^{a})$ is a vector on $\M3$ and $\tN$ is a scalar density
of weight $-1$.
We will refer to $\CD(\vec{N})$ as the \lq diffeomorphism constraint'.

Under the Poisson bracket
\beq
\{A^{i}_{a}(x),\tpi^{bj}(y)\}_{PB}=-i\delta^{ij}\delta^{b}_{a}
\delta^{3}(x,y),
\eeq
these smeared constraints generate gauge transformations
in a broad sense. Gauss' law constraint and the diffeomorphism
constraint respectively generate small $SL(2,{\bf C})$ gauge
transformations and small spatial diffeomorphisms:
\beqy
\{(A^{i}_{a},\tpi^{ai}),G(\theta)\}_{PB}&=&(-D_{a}\theta^{i},
(\theta\times\tpi^{a})^{i})\nonumber \\
\{(A^{i}_{a},\tpi^{ai}),\CD(\vec{N})\}_{PB}&=&
(\CL_{\vec{N}}A^{i}_{a},\CL_{\vec{N}}\tpi^{ai}),
\eeqy
where we have used the notation $(\theta\times\tpi^{a})^{i}\equiv
\ep^{ijk}\theta^{j}\tpi^{ak}$. The scalar constraint generates
(infinitesimal) many-fingered time evolutions
\beqy
\{A^{i}_{a},\CS(\tN)\}_{PB}&=&-i\tN\ep^{ijk}\utep_{abc}\tpi^{bj}
\Phi^{ck}+\frac{\Lambda}{3}\varphi^{i}_{a} \nonumber \\
\{\tpi^{ai},S(\tN)\}_{PB}&=&-\otep^{abc}D_{b}\varphi^{i}_{c},
\eeqy
where $\Phi^{ai}$ is the constraint (\ref{eq:krcon}) in BF theory
and $\varphi^{i}_{a}\equiv-\frac{i}{2}\tN\ep^{ijk}\utep_{abc}\tpi^{bj}
\tpi^{ck}$ is an $SO(3,{\bf C})$ Lie algebra-valued one-form.

Using these results we can now easily compute the Poisson brackets
between smeared constraints. We find
\beqy
\{G(\theta^{\prime}),G(\theta)\}_{PB}&=&
-G(\theta\times\theta^{\prime})\nonumber \\
\{G(\theta),\CD(\vec{N})\}_{PB}&=&-G(\CL_{\vec{N}}\theta)\nonumber \\
\{\CD(\vec{M}),\CD(\vec{N})\}_{PB}&=&-\CD([\vec{N},\vec{M}]_{LB})
\nonumber \\
\{S(\tN),G(\theta)\}_{PB}&=&0 \nonumber \\
\{S(\tM),\CD(\vec{N})\}_{PB}&=&-S(\CL_{\vec{N}}\tM)\nonumber \\
\{S(\tN),S(\tM)\}_{PB}&=&-i\int_{\M3}d^{3}xK^{a}{\cal V}_{a}
=\CD(\vec{K})+G(K^{a}A_{a}),\label{eq:conal}
\eeqy
where $[\vec{N},\vec{M}]_{LB}=(N^{b}\partial_{b}M^{a}-
M^{b}\partial_{b}N^{a})$ stands for the Lie bracket and
$K^{a}\equiv(\tN\partial_{b}\tM-\tM\partial_{b}\tN)
\tpi^{bj}\tpi^{aj}$ is a vector on $\M3$.
The first three equations are the manifestation of the fact that
$SL(2,{\bf C})$ gauge transformations and spatial diffeomorphisms
form a semi-direct product group. The fourth and the fifth equations
mean that the scalar constraint $\CS$ transforms
trivially under $SL(2,{\bf C})$ gauge transformations and
as a scalar density of weight $+2$ under spatial diffeomorphisms.
It is only the last equation which is somewhat difficult to derive.
It involves non-trivial structure
functionals and is expected to cause
one of the most formidable obstacles to quantum Ashtekar's formalism.

Next we will explain the reality conditions\cite{ashte}\cite{ashte2}.
In demonstrating the classical equivalence of the complex chiral
action with the Einstein-Hilbert action, it has been indispensable
that the vierbein $e^{\alpha}$ and the spin-connection
$\omega^{\alpha\beta}$ are real-valued. This tells us that,
in order to extract
full information on general relativity from Ashtekar's formalism,
we have to impose some particular conditions called
\lq\lq reality conditions" on the canonical variables $(A^{i}_{a},
\tpi^{ai})$. According to the treatment of the $SL(2,{\bf C})$ gauge
degrees of freedom, there are two alternative ways of imposing reality
conditions:
\begin{enumerate}
\item Fix the Lorentz boost degrees of freedom by choosing the 
spatial gauge (\ref{eq:spatialgauge}). As a consequence the gauge
group reduces to $SU(2)$. The classical reality conditions in this
case take the following form
(the bar denotes the complex conjugation)
\beq
\overline{\tpi^{ai}}=\tpi^{ai},\quad A^{i}_{a}+\overline{A^{i}_{a}}
=-\ep^{ijk}\omega^{jk}_{a}(e),\label{eq:creal2}
\eeq
where $^{(3)}\omega^{ij}\equiv\omega^{ij}_{a}(e)dx^{a}$ is the
spin-connection on $\M3$ which satisfies the torsion-free
condition w.r.t. the dreibein $^{(3)}e^{i}\equiv e^{i}_{a}dx^{a}$
$$
^{(3)}d^{(3)}e^{i}+^{(3)}\omega^{ij}\wedge^{(3)}e^{j}=0.
$$
\item Do not fix the gauge and keep the full $SL(2,{\bf C})$ gauge
degrees of freedom. The reality conditions in this case are merely
pullbacks of eqs.(\ref{eq:creal1})(\ref{eq:alcon1}) to $\M3$, namely
\beqy
A^{i}_{a}&=&-\frac{1}{2}\ep^{ijk}\omega^{jk}_{a}-i\omega^{0i}_{a}
\nonumber \\
\tpi^{ai}&=&\frac{1}{2}\otep^{abc}(\ep^{ijk}e^{j}_{b}e^{k}_{c}
+2ie^{0}_{b}e^{i}_{c})\label{eq:creal3}
\eeqy
must hold for real $e^{\alpha}_{a}$ and real
$\omega^{\alpha\beta}_{a}$.
\end{enumerate}

We cannot say which is absolutely better than the other.
It is clever to make a relevant choice according to each problem
to work with.

Before concluding this subsection we will comment on the Euclidean
case. The Euclidean counterpart of the Lorentz group
$SO(3,1)^{\uparrow}\cong SL(2,{\bf C})/\{\pm1\}$ is the
four-dimensional rotation group $SO(4)\cong(SU(2)\times SU(2))/
\{\pm1\}$. As a result the anti-self-dual part of a rank two
tensor takes a real value, for example,
\beqy
A_{E}^{i}&=&-\frac{1}{2}\ep^{ijk}\omega^{jk}+\omega^{0i},
\nonumber \\
\Sigma_{E}^{i}&=&-\frac{1}{2}\ep^{ijk}e^{j}\wedge e^{k}+
e^{0}\wedge e^{i}.
\eeqy
Using this the Euclidean counterpart of the complex chiral action
are given by
\beqy
(I_{CC})^{E}&=&\int_{M}(\Sigma_{E}^{i}\wedge F^{i}_{E}-
\frac{\Lambda}{6}\Sigma^{i}_{E}\wedge\Sigma^{i}_{E})\nonumber \\
&=&\frac{1}{2}(I_{EP})^{E}-\int_{M}e^{\alpha}\wedge e^{\beta}\wedge
R^{\alpha\beta},
\eeqy
where $(I_{EP})^{E}$ is the Euclidean version of the
Einstein-Palatini action.

If we deal with this action in the first order formalism, namely if
we regard $A^{i}_{E}$ and $e^{\alpha}$ as independent variables,
$(I_{CC})^{E}$ does not coincide with the Euclidean Einstein-Hilbert
action $(I_{EH})^{E}$ {\em even on-shell}.
This is because information
on the self-dual part of the torsion-free condition is lost.
However, if we impose in advance the torsion-free condition
$de^{\alpha}+\omega^{\alpha\beta}\wedge e^{\beta}=0$,
$(I_{CC})^{E}$ coincides with $(I_{EH})^{E}$ owing to
the first Bianchi identity. In other words, in order to reproduce
the result of Euclidean general relativity from that of Euclidean
Ashtekar's formalism, we have to impose the torsion-free condition
by hand. In the canonical treatment, this amounts to the real
canonical transformation from $\omega^{0i}_{a}$ to $(A_{E}^{i})_{a}$
which is generated by $\frac{1}{2}\int_{\M3}\ep^{ijk}\omega^{ij}_{a}
\tpi^{ak}$.


\subsection{Quantization}

Next we consider the quantization of Ashtekar's formalism.
There are two well-known quantization methods for systems involving
first order constraints such as Ashtekar's formalism, namely the
reduced phase space quantization and the Dirac quantization.

In the reduced phase space quantization\footnote{
For more detailed explanations of the reduced phase space
quantization, we refer the reader to \cite{wood} or to Appendix D of
\cite{ashte2}.}, we first construct
the reduced phase space which consists only of the physical degrees
of freedom and then perform the canonical quantization
on this reduced phase space. The reduced phase space is
constructed by first solving the first class constraints completely
and then removing all the gauge degrees of freedom which are
generated by the first class constraints.
In general relativity this amounts to finding all the diffeomorphism
equivalence classes of the solutions to
the Einstein equation. It is in practice
impossible to carry out this. So we usually adopt Dirac's
quantization procedure\cite{dirac} when we canonically quantize
gravity.

In Dirac's quantization, we first promote canonical variables
in the unconstrained phase space to quantum operators
by replacing $i$ times the Poisson brackets with the corresponding
quantum commutation relations. The first class constraints
become the operator equations imposed on physical
wave functions. When the first class constraints are
at most linear in momenta, this prescription is known to yield
the same result as that in the reduced phase space method,
up to a minor subtlety (see Appendix D of \cite{ashte2}).
Because the scalar constraint is at least quadratic in momenta,
we expect that these two methods lead to different results
for quantum Ashtekar's formalism.

Let us now perform Dirac's quantization. The canonical variables
$(A^{i}_{a},\tpi^{ai})$ are promoted to the fundamental quantum
operators $(\hat{A}^{i}_{a},\hat{\tpi}^{ai})$ subject to
the commutation relations
\beq
[\hat{A}^{i}_{a}(x),\hat{\tpi}^{bj}(y)]=\delta^{ij}\delta^{b}_{a}
\delta^{3}(x,y).\label{eq:commut}
\eeq
In this paper we will take the connection representation (or
holomorphic representation) in which the $SL(2,\BC)$
connection $\hat{A}^{i}_{a}$ is diagonalized. Wave
functions are thus given by holomorphic functionals
$\Psi[A]=<A|\Psi>$ of the $SL(2,{\bf C})$ connection. The
action of $\hat{A}^{i}_{a}$ and $\hat{\tpi}^{ai}$ on these
wavefunctions are respectively represented by multiplication
by $A^{i}_{a}$ and by functional derivative w.r.t. $A^{i}_{a}$:
\beqy
\hat{A}^{i}_{a}(x)\Psi[A]&=&A^{i}_{a}\cdot\Psi[A]\nonumber \\
\hat{\tpi}^{ai}(x)\Psi[A]&=&-\frac{\delta}{\delta A^{i}_{a}(x)}
\Psi[A].
\eeqy

Next we impose the constraint equations:
\beqy
\hat{G}^{i}\Psi[A]&=&0 \label{eq:qgauss} \\
\hat{\CD}_{a}\Psi[A]&=&0 \label{eq:qdiffeo} \\
\hat{\CS}\Psi[A]&=&0. \label{eq:nwd}
\eeqy
Gauss' law constraint (\ref{eq:qgauss}) and
the diffeomorphism constraint
(\ref{eq:qdiffeo}) respectively require the physical wavefunctions
to be invariant under small $SL(2,{\bf C})$ gauge transformations
and small spatial diffeomorphisms\footnote{
$A^{g}_{a}=(A^{g})^{i}_{a}J_{i}=gA^{i}_{a}J_{i}g^{-1}+
g\partial_{a}g^{-1}$ and $\phi^{\ast}A^{i}_{a}(x)=\partial_{a}
\phi^{b}(x)A^{i}_{b}(\phi(x))$ respectively denote
the image of the connection $A^{i}_{a}(x)$ under the small
$SL(2,{\bf C})$ gauge transformation $g(x)$ and the pullback
of $A^{i}_{a}(x)$ by the spatial diffeomorphism
$\phi:\M3\rightarrow\M3$.}
\beqy
\Psi[A^{g}]&=&\Psi[A] \\
\Psi[\phi^{\ast}A]&=&\Psi[A].\label{eq:intqdiff}
\eeqy
The scalar constraint (\ref{eq:nwd}) yields Ashtekar's version of
the Wheeler-De Witt equation. Because this equation involves
at least second order functional derivative, its rigorous
treatment requires some regularization. In association with this
there is an issue on the operator ordering in the constraints.
At the formal level there are two plausible candidates:
\begin{enumerate}
\item Putting $\hat{\tpi}^{ai}$ to the right
\cite{jacob}\cite{bori}. A virtue of this
ordering is that $\hat{G}^{i}$ and $\hat{\CD}_{a}$ correctly
generate $SL(2,{\bf C})$ gauge transformations and spatial
diffeomorphisms. In this ordering, however, commutator
$[\hat{\CS},\hat{\CS}]$ fails to vanish weakly because the
structure functionals appear to the right of the constraints.
\item Putting $\hat{\tpi}^{ai}$ to the left
\cite{ashte}\cite{kodama}. This ordering
has a merit that the commutator algebra of the constraints
$\hat{G}^{i}$, $\hat{{\cal V}}_{a}$ (not $\hat{\CD}_{a}$)
and $\hat{\CS}$ formally closes. A demerit of this ordering is
that $\hat{\CD}_{a}$ (or $\hat{{\cal V}}_{a}$) does not generate
diffeomorphisms correctly.\footnote{
In the loop representation, however, $\hat{\cal V}_{a}$ generates
diffeomorphism correctly by virtue of the relation
(\ref{eq:connloop}).
For a detailed analysis, see e.g. ref.\cite{brug2}.}
\end{enumerate}

We should note that the above discussion is formal in the sense
that we deal with non-regularized constraints.

In order to extract physical information from the physical wave
functions, we also need to construct physical observables and
to specify a physical inner product. Physical observables are
self-adjoint operators which commutes with the constraints
at least weakly.
Finding the physical inner product has been a longstanding
problem in quantum Ashtekar's formalism because it is intimately
related with the reality conditions.
As we have seen before, Ashtekar's canonical variables have to
satisfy the nontrivial reality conditions (\ref{eq:creal2}) or
(\ref{eq:creal3}), which have to be implemented in the quantum
theory as nontrivial adjointness conditions.
Because the adjointness conditions can be attributed to
the problem of inner product, the issue of reality conditions
has not been taken so seriously as yet. Quite recently, however,
a promising candidate for the physical inner product
has been proposed\cite{thie}. If it turns out that
this inner product is genuinely physical and useful, we can say that
the program of quantizing Ashtekar's formalism has made
great progress.


\section{Spin network states}

Recently there have been a large number of remarkable developments
in the program of applying spin network states
\cite{penrose} to quantum gravity
in terms of Ashtekar's new variables \cite{baez2} \cite{asle}
\cite{ALMMT} \cite{ALMMT2} \cite{RS2} \cite{RS3}.
As one of the promising approaches to canonical quantum gravity,
this program is expected to make further progress in the near
future. Unfortunately, however, most of the results are given
in the form of mathematical theorems and propositions which are
unfamiliar to most of the physicists. This makes it obscure
that these notions have originary been brought into
quantum general relativity in order to concretize
the simple physical ideas. In this section
we make a maximal attempt to explain some of these results on
the spin network states in the language of physics and to
clarify the underlying ideas.
In particular, we provide explicit formulae for two types of measures
defined on spin network states\cite{asle}\cite{ALMMT}.
Because we are only interested in its application to quantum gravity,
we will restrict our attention to the case where the gauge group is
$SU(2)$ or $SL(2,{\bf C})$. Our discussion is, however, applicable
to any compact gauge group $G$ and its complexification $G^{{\bf C}}$
if some necessary modifications are made.
For simplicity we will only deal with the unitary representations
of $SU(2)$.


\subsection{Backgrounds and the definition}

As is seen in \S 2, we can regard Ashtekar's formalism as a kind of
$SL(2,{\bf C})$ gauge theory. A common feature of gauge
theories is that, by virtue of Gauss' law constraint
(\ref{eq:qgauss}), the wavefunction in the connection representation
is given by a functional of the connection which is invariant under
small gauge transformations. We will henceforth restrict our
attention to the wavefunctions which are invariant under all
the gauge transformations including large gauge transformations.

In order to construct gauge invariant functionals of the connection
$A^{i}$, it is convenient to use the parallel propagator
$h_{\alpha}[0,1]$ of $A^{i}$
evaluated along the curve $\alpha:[0,1]\rightarrow\M3$
\beq
h_{\alpha}[0,1]=\CP\exp\{\int_{0}^{1}ds\dot{\alpha}^{a}(s)
A^{i}_{a}(\alpha(s))J_{i}\},
\eeq
where $\CP$ stands for the path ordering with smaller $s$ to
the left.
In spite of the
fact that the gauge transformation of the connection is
inhomogenous, the parallel propagator transforms homogeneously
under the gauge transformation $g(x)$:
\beqy
A_{a}(x)&\rightarrow&g(x)A_{a}g^{-1}(x)-\partial_{a}g(x)g^{-1}(x)
\nonumber \\
h_{\alpha}[0,1]&\rightarrow&g(\alpha(0))h_{\alpha}[0,1]
g^{-1}(\alpha(1)), \label{eq:paragauge}
\eeqy
where we have set $A_{a}\equiv A^{i}_{a}J_{i}$.

Thus it is considerably straightforward to construct gauge invariant
functionals from these parallel propagators $h_{\alpha}[0,1]$.
The simplest example is the Wilson loop along a loop $\gamma:[0,1]
\rightarrow\M3$ $(\gamma(0)=\gamma(1))$
\beq
W(\gamma,\pi)\equiv\Tr\pi(h_{\gamma}[0,1]).
\eeq
Its gauge invariance follows immediately from
eq.(\ref{eq:paragauge}).
Considerably many early works on Ashtekar's formalism
were based on the use of these Wilson loops. In fact
the \lq\lq reconstruction theorem" proposed by Giles\cite{gile}
guarantees that, in the case where the gauge group is compact,
Wilson loops suffice to extract all the
gauge invariant information on the connection.\footnote{
In the $SL(2,{\bf C})$ case it was shown that the Wilson loop
is sufficient to separate all the separable points of the space
$\CA/\CG$ of gauge equivalence classes of connections
\cite{asle2}.In other words the Wilson loop misses some information
on the null rotation part of the holonomies (parallel propagators
along the loops).}

However, the framework based on the Wilson loop has several
drawbacks in order to provide efficient tools. Among them,
particularly serious ones are the following:

i) (Overcompleteness)
it is known that the set of Wilson loops form an overcomplete
basis in $\CA/\CG$. In order to extract necessary information
on the connection, we have to impose algebraic constraints
such as Mandelstam identities\cite{migdal} on
Wilson loops.

ii) (Non-locality)
The action of local operators on Wilson loops frequently induces
the change of global properties of the loops on which the Wilson
loops are defined. For example, it often happens that orientation
of some loops has to be inverted
in order to make the resulting composite loop to be consistent.

These drawbacks become the sources of extra intricacy when we carry
out calculations by means of Wilson loops. In this sense the Wilson
loops are not so suitable for practical calculations.
Naturally a question arises as to whether or not there exist
more convenient tools composed of parallel propagators
which give rise to complete
(but not overcomplete) basis in $\CA/\CG$
and on which the action of local operators is described by
purely local manipulation. The answer is yes.
Spin network states satisfy these requirements.

A spin network state is defined on a \lq graph'.
A graph $\Gamma=(\{e\},\{v\})$ consists of a set $\{v\}$ of
\lq vertices' and a set $\{e\}$ of \lq edges'.
Each vertex $v\in\M3$ is a point on $\M3$ and each edge
$e:[0,1]\rightarrow\M3$ is a curve
on $\M3$ which connects two vertices (or a loop on $\M3$ which
is based at a vertex). An example of a graph is shown in figure1.

\begin{figure}[t]
\begin{center}
\epsfig{file=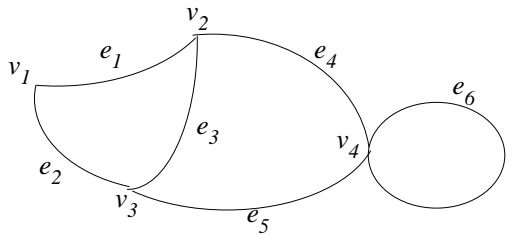,height=6cm}
\end{center}
\caption{An example of a graph.}
\end{figure}

In order to construct a unique spin network state, we have only to
determine a \lq colored graph' $(\Gamma,\{\pi_{e}\},\{i_{v}\})$ as
follows: i) evaluate along each edge $e$ a parallel propagator of
the connection $A_{a}dx^{a}$ in a representation $\pi_{e}$ of
$SL(2,{\bf C})$; and ii) equip each vertex $v$ with an intertwining
operator\footnote{We can regard
each intertwining operator as an invariant tensor
in the corresponding tensor product representation.}
\beq
i_{v}:\left(\bigotimes_{e(1)=v}\pi^{\ast}_{e}\right)\otimes
\left(\bigotimes_{e(0)=v}\pi_{e}\right)\longrightarrow
1 \mbox{ (trivial)},\label{eq:Intertwiner}
\eeq
where $\pi^{\ast}$ denotes the conjugate representation of $\pi$
defined by $\pi^{\ast}(g)\equiv
^{T}\!\!\pi(g^{-1})$ ($^{T}$ represents
taking the transpose). The corresponding spin network state is
defined as
\beq
<A|\Gamma,\{\pi_{e}\},\{i_{v}\}>\equiv\prod_{v\in\{v\}}i_{v}\cdot
\left(\bigotimes_{e\in\{e\}}\pi_{e}(h_{e}[0,1])\right).
\label{eq:SPN}
\eeq
Gauge invariance of the spin network states thus defined is
manifest from the transformation law(\ref{eq:paragauge})
of the parallel propagators and from the defining property
(\ref{eq:Intertwiner}) of the intertwining operators.

It is obvious that the spin network states form a complete basis
because the Wilson loops are regarded as particular spin network
states. Moreover, it turns out that they are not overcomplete,
namely, we do not have to impose Mandelstam identities.
We will see in the next subsections that they actually yield
an orthonormal basis w.r.t. some inner products.
Another merit of spin network states is that the action of
local operators is described in terms of purely local
operations on the graph. Thus we can concentrate only on
the local structures in the practical calculation.
While we do not show this generically, an example is
demonstrated in the next section.

Here we should explain the consistency of the spin network states.
In practical calculation it is sometimes convenient
to consider the graph $\Gamma$ as a subgraph
$\Gamma<\Gamma^{\prime}$ of a larger graph $\Gamma^{\prime}$,
where $\Gamma<\Gamma^{\prime}$ means that all the edges
and the vertices in $\Gamma$ is contained in $\Gamma^{\prime}$,
namely, $\{e\},\{v\}\in\Gamma^{\prime}$. The spin network state
$<A|\Gamma,\{\pi_{e}\},\{i_{v}\}>$ is then regarded as
the spin network state $<A|\Gamma^{\prime},
\{\pi^{\prime}_{e^{\prime}}\},\{i^{\prime}_{v^{\prime}}\}>$
which is determined as follows.

A graph $\Gamma^{\prime}$ which is larger than $\Gamma$ is
obtained from $\Gamma$ by a sequence of the following four
moves\cite{baez2}:\\
i) adding a vertex $v^{\prime}\not\in\Gamma$;\\
ii) adding an edge $e^{\prime}\not\in\Gamma$;\\
iii) subdividing an
edge $\Gamma\in e=e_{1}\cdot e_{2}$ with $e_{1}(1)=e_{2}(0)=
v^{\prime}\not\in\Gamma$; and\\
iv) reverting the orientation of an
edge $\Gamma\ni e\rightarrow e^{\prime}=e^{-1}\in\Gamma^{\prime}$.

In order to obtain the spin network state $<A|\Gamma^{\prime},
\{\pi^{\prime}_{e^{\prime}}\},\{i^{\prime}_{v^{\prime}}\}>$
which is equal to $<A|\Gamma,\{\pi_{e}\},\{i_{v}\}>$,
we have to associate with each move the following operation:\\
for i) we equip the vertex $v^{\prime}$ with the unity
$i^{\prime}_{v^{\prime}}=1$;\\
for ii) we provide the edge
$e^{\prime}$ with the trivial representation
$\pi^{\prime}_{e^{\prime}}(h_{e^{\prime}}[0,1])=1$;\\
for iii) we fix the representation and the intertwining operator
as $\pi^{\prime}_{e_{1}}=\pi^{\prime}_{e_{2}}=\pi_{e}$ and
as $i^{\prime}_{v^{\prime}}=\delta_{I}^{J}$ respectively, where
$\delta_{I}^{J}$ denotes the Cronecker delta in the representation
$\pi_{e}$; and\\
for iv) the corresponding representation for $e^{\prime}$
is given by the conjugate representation
$\pi^{\prime}_{e^{\prime}}=(\pi_{e})^{\ast}$.

Action of local operators on spin network states
is necessarily independent of the choice of the graph, because
the operators do not perceive the difference of the graphs
on which the identical spin network state is defined.
However this consistency gives rise to a criterion for
defining well-defined measures on $\overline{\CA/\CG}$,\footnote{
The space $\overline{\CA/\CG}$ of equivalence classes of
generalized connections modulo generalized gauge transformations
is a completion of the space $\CA/\CG$.
Roughly speaking it is obtained by allowing the parallel propagators
$h_{e}[0,1]$ which cannot be obtained by integrating the
well-defined connection $A_{a}$ on $\M3$. The cylindrical functions
on $\overline{\CA/\CG}$ are obtained from the superposition
of spin network states on a graph $\Gamma$ in the limit of
making $\Gamma$ larger and larger\cite{asle}.} namely, a
well-defined measure should be invariant under the above four moves.

Because we only deal with the case where the gauge group is
$SL(2,{\bf C})$ or $SU(2)$, some simplifications are possible.
In particular, because the conjugate representation $\pi^{\ast}$
in $SL(2,{\bf C})$ is unitary equivalent to the original
representation $\pi$,\footnote{
In fact if we use the convention in which $J_{2}$ in each
representation is a real skew-symmetric matrix, then the
representation $\pi$ and its conjugate $\pi^{\ast}$ are
related by the unitary matrix $\pi(\exp(\pi J_{2}))$.}
we do not have to worry about the orientation of
the edges. We can therefore neglect the consistency under
the fourth move.


\subsection{The induced Haar measure on $SU(2)$ gauge theories}

As is seen in \S 2, Ashtekar's formalism for Euclidean gravity
is embedded in the $SU(2)$ gauge theory.
Besides, an Ashtekar-like formulation exists also
in Lorentzian gravity which deals with the $SU(2)$ connection
\cite{barb}. For these formulations the induced Haar measure
is used for a natural measure of the theory.

The induced Haar measure on $\overline{\CA/\CG}$
is defined as follows. First we define
a \lq generalized connection' $A_{\Gamma}$ on the graph
$\Gamma\subset\M3$ as a map
$$
A_{\Gamma}:\{e\}=(e_{1},\ldots,e_{n_{\Gamma}})\rightarrow
(SU(2))^{n_{\Gamma}}.
$$
The space $\overline{\CA}_{\Gamma}$ is the set of all
the generalized connections on $\Gamma$.
Of course the usual connection
$A_{a}$ gives rise to a generalized connection by the relation
$A_{\Gamma}(e)=h_{e}[0,1]$. The space $\overline{(\CA/\CG)}
_{\Gamma}$ is the quotient space of $\overline{\CA}_{\Gamma}$
modulo generalized gauge transformations:
\beqy
&&g:\{v\}=(v_{1},\ldots,v_{N_{\Gamma}})\rightarrow
(SU(2))^{N_{\Gamma}},\nonumber \\
&&A_{\Gamma}(e)\rightarrow g(e(0))A_{\Gamma}(e)g^{-1}(e(1)).
\nonumber
\eeqy

Next we consider a function
$f_{\Gamma}(A_{\Gamma})$
on $\overline{(\CA/\CG)}_{\Gamma}$. We will assume that
the pull-back $(p_{\Gamma})^{\ast}f_{\Gamma}$ of
$f_{\Gamma}$ becomes a cylindrical function
on $\overline{\CA/\CG}$, where $p_{\Gamma}$ denotes the projection
map
$$
p_{\Gamma}:\overline{\CA}\rightarrow \overline{\CA}_{\Gamma}.
$$
It is obvious that $(p_{\Gamma})^{\ast}f_{\Gamma}$
can be expressed by a linear
combination of spin network states defined on $\Gamma$.
For $(p_{\Gamma})^{\ast}f_{\Gamma}$
the induced Haar measure $d\mu_{H}$ on
$\overline{\CA/\CG}$ is defined as
\beq
\int d\mu_{H}(A)(p_{\Gamma})^{\ast}
f_{\Gamma}(A)\equiv\int d\mu_{\Gamma}
(A_{\Gamma})f_{\Gamma}(A_{\Gamma})\equiv
\int_{\overline{\CA}_{\Gamma}}\left(\prod_{e\in\{e\}}
d\mu(A_{\Gamma}(e))\right)f_{\Gamma}(A_{\Gamma}),\label{eq:HAAR}
\eeq
where $d\mu$ denotes the Haar measure on $SU(2)$.
The consistency condition of this measure is
provided by the following equation
\beq
\int d\mu_{\Gamma}(A_{\Gamma})f_{\Gamma}(A_{\Gamma})=
\int d\mu_{\Gamma^{\prime}}(A_{\Gamma^{\prime}})
(p_{\Gamma\Gamma^{\prime}})^{\ast}f_{\Gamma}(A_{\Gamma^{\prime}}),
\label{eq:consistent}
\eeq
where $\Gamma^{\prime}>\Gamma$, and
$$
p_{\Gamma\Gamma^{\prime}}:\overline{\CA}_{\Gamma^{\prime}}
\rightarrow\overline{\CA}_{\Gamma}
$$
is the projection from $\overline{\CA}_{\Gamma^{\prime}}$
onto $\overline{\CA}_{\Gamma}$.
It was shown that this consistency condition indeed holds
\cite{asle}. We will presently see this by a concrete calculation.
Before doing so
we provide the explicit formula for the induced Haar
measure on the spin network states.

Let us first calculate the inner product between two spin network
states\\
$<A|\Gamma,\{\pi_{p_{e}}\},\{i_{v}\}>$ and
$<A|\Gamma,\{\pi_{q_{e}}\},\{i^{\prime}_{v}\}>$
which are defined on the same graph $\Gamma$:\footnote{As in Appendix
B $\pi_{p}$ denotes the spin-$\frac{p}{2}$ representation of
$SL(2,{\bf C})$.}
\beqy
&&<\Gamma,\{\pi_{p_{e}}\},\{i_{v}\}|\Gamma,\{\pi_{q_{e}}\},
\{i^{\prime}_{v}\}>_{|d\mu_{H}}\nonumber \\
&&\!\!\!\!\equiv\int d\mu_{H}(A)
\overline{<A|\Gamma,\{\pi_{p_{e}}\},\{i_{v}\}>}
<A|\Gamma,\{\pi_{q_{e}}\},\{i^{\prime}_{v}\}> \nonumber \\
&&\!\!\!\!
=\overline{(\prod_{v}i_{v})}\cdot(\prod_{v}i^{\prime}_{v})\cdot
\prod_{e}\left(\int d\mu(A_{\Gamma}(e))
\overline{\pi_{p_{e}}(A_{\Gamma}(e))_{I_{e}}\UI{J_{e}}}\pi_{q_{e}}
(A_{\Gamma}(e))_{I^{\prime}_{e}}\UI{J^{\prime}_{e}}\right).
\eeqy
In obtaining the last expression we have used eq.(\ref{eq:SPN}).
By plugging eq.(\ref{eq:haar}) into this equation we find
\beqy
<\Gamma,\{\pi_{p_{e}}\},\{i_{v}\}|\Gamma,\{\pi_{q_{e}}\},
\{i^{\prime}_{v}\}>_{|d\mu_{H}}&=&
\overline{(\prod_{v}i_{v})}\cdot(\prod_{v}i^{\prime}_{v})\cdot
\left(\prod_{e}\frac{\delta_{p_{e},q_{e}}}{p_{e}+1}
\delta^{I_{e}}_{I^{\prime}_{e}}\delta^{J^{\prime}_{e}}_{J_{e}}
\right)\nonumber \\
&=&\prod_{e}\frac{1}{p_{e}+1}\delta_{p_{e},q_{e}}
\prod_{v}<i_{v}|i^{\prime}_{v}>,\label{eq:HAAR1}
\eeqy
where we have used the symbol $<i_{v}|i^{\prime}_{v}>$ to mean
the complete contraction of two intertwining operators
\beq
<i_{v}|i^{\prime}_{v}>\equiv\overline{i_{v}}\cdot i_{v}^{\prime}
\cdot\left(\prod_{e:e(0)=v}\delta^{I_{e}}_{I^{\prime}_{e}}
\prod_{e:e(1)=v}\delta^{J^{\prime}_{e}}_{J_{e}}\right).
\label{eq:contract}
\eeq

We are now in a position to prove the consistency
(\ref{eq:consistent}). Consistency under the moves i) and ii)
are trivial and thus we have only to show consistency
under the move iii). This immediately follows from
the first equality in eq.(\ref{eq:HAAR1}) if we
substitute expressions like
$\pi_{p_{e}}(A_{\Gamma}(e))_{I}\UI{J}=\pi_{p_{e}}
(A_{\Gamma}(e_{1}))_{I}\UI{K}\pi_{p_{e}}(A_{\Gamma}
(e_{2}))_{K}\UI{J}$ into the induced Haar measure
defined on $\Gamma^{\prime}$. An only possibility of
the change stems from the subdivided
edge $e=e_{1}\cdot e_{2}$ which yields the following contribution
to the measure
$$
\frac{\delta_{p_{e},q_{e}}}{p_{e}+1}\delta^{I}_{I^{\prime}}
\delta_{K}^{K^{\prime}}
\times\frac{\delta_{p_{e},q_{e}}}{p_{e}+1}\delta^{K}_{K^{\prime}}
\delta^{J^{\prime}}_{J}.
$$
By taking care not to contract two
$(\delta_{p_{e},q_{e}})$'s, we get the same result as that from
the measure defined on $\Gamma$:
$$
\frac{\delta_{p_{e},q_{e}}}{p_{e}+1}\delta_{I^{\prime}}^{I}
\delta_{J}^{J^{\prime}}.
$$
Thus we have demonstrated concretely that the consistency holds
for the induced Haar measure on $SU(2)$ gauge theories.

Eq.(\ref{eq:HAAR1}) also tells us that two spin network states
are orthogonal with each other if the representations
$\pi_{p_{e}}$ and $\pi_{q_{e}}$ do not coincide on one or more
edges $e\in\Gamma$. As a corollary of this result and the
consistency it follows that the inner product of
two spin network states
$<A|\Gamma,\{\pi_{e}\},\{i_{v}\}>$ and
$<A|\Gamma^{\prime},\{\pi^{\prime}_{e^{\prime}}\},\{i^{\prime}
_{v^{\prime}}\}>$ vanishes if the former is not equivalent to
a spin network state on $\Gamma^{\prime}$ and also if
the latter is not equivalent to that on $\Gamma$.

In order to show completely
that the spin network states form an orthonormal
basis, the contraction $<i_{v}|i^{\prime}_{v}>$ of intertwining
operators remains to be calculated. For this goal we first have
to provide a complete set of intertwining operators. A convenient
choice is given as follows.

We know that the intertwining operator for a trivalent
vertex is, up to an overall constant,
uniquely fixed by the Clebsch-Gordan coefficient
$$
<p,J;q,K|r,L>:\pi_{p}\otimes\pi_{q}\otimes\pi^{\ast}_{r}
\rightarrow1\mbox{ (trivial)}.
$$
Thus a complete set of intertwining operators
for an $n$-valent vertex $v$ is given
by decomposing the vertex into $n-2$
trivalent vertices $(v_{1},v_{2},\ldots,v_{n-2})$
and by assigning to the resulting $n-3$ virtual edges
$(e_{V}^{1},\ldots,e_{V}^{n-3})$ $n-3$ irreducible representations
$(\pi_{r_{1}},\ldots,\pi_{r_{n-3}})$ which satisfy triangular
inequations \cite{KS}\cite{RS3}.
In this way each basic intertwining operator for an $n$-valent
vertex is specified by an non-negative integer-valued
$n-3$ dimensional vector $(r_{1},\ldots,r_{n-3})$. The way of
decomposing an $n$-valent vertex into $n-2$ trivalent vertices
is not unique and thus we have to choose one.
We can obtain the complete set starting from an arbitrary choice.
An example for the $n=6$ case is shown in figure2.

\begin{figure}[t]
\begin{center}
\epsfig{file=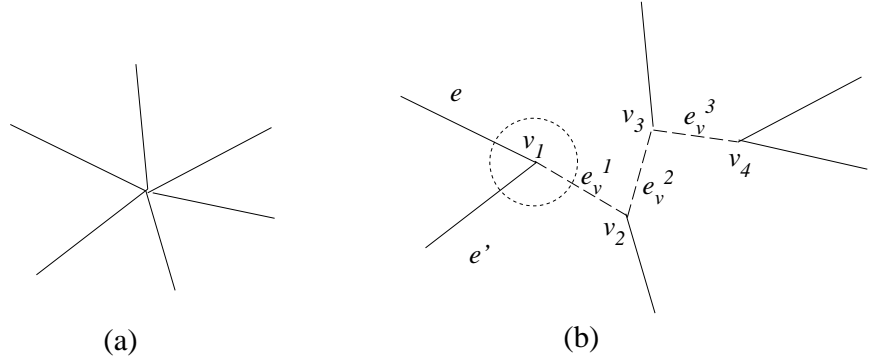,height=5cm}
\end{center}
\caption{An example of decomposing a six-valent vertex
(a) into four trivalent vertices (b). Dashed lines
denote the virtual edges.}
\end{figure}

Let us now show that any such choice of the basis for intertwining
operators yields an orthonormal basis for $<i_{v}|i^{\prime}_{v}>$
(eq.(\ref{eq:contract})).
Because there exist at least one trivalent vertex connecting
two true edges, like the one which is enclosed by the circle in
figure2, we will start from one such edge, say $v_{1}$.
We assume that
the two true edges $e$ and $e^{\prime}$ emanating from $v_{1}$
are respectively equipped with the spin-$\frac{p}{2}$ and the
spin-$\frac{q}{2}$ representations. The two non-negative
integer-valued vectors which specify $i_{v}$ and
$i^{\prime}_{v}$ are respectively supposed to have
components $r_{1}$ and $r^{\prime}_{1}$ which correspond to
the virtual edge $e_{V}^{1}$ with an end point $v_{1}$.
By paying attention only to the vertex $v_{1}$, we find
\begin{eqnarray}
<i_{v}|i^{\prime}_{v}>&\propto&\sum_{J,J^{\prime}=1}^{p+1}
\sum_{K,K^{\prime}=1}^{q+1}\overline{<p,J;q,K|r_{1},M>}
<p,J^{\prime};q,K^{\prime}|r_{1}^{\prime},M^{\prime}>
\delta_{J}^{J^{\prime}}\delta_{K}^{K^{\prime}} \nonumber \\
&=&\delta_{r_{1},r_{1}^{\prime}}\delta^{M}_{M^{\prime}}.
\label{eq:trivalent}
\end{eqnarray}
In deriving the last equation we have used the fact that
the Clebsch-Gordan Coefficients ${<p,J;q,K|r,M>}$
yield the unitary transformation:
$$
\pi_{p}\otimes\pi_{q}\rightarrow\pi_{|p-q|}\oplus
\pi_{|p-q|+2}\oplus\cdots\oplus\pi_{p+q}.
$$
Eq.(\ref{eq:trivalent}) shows that
the trivalent vertex $v_{1}$ turns the Cronecker delta in the
tensor product representation $\pi_{p}\otimes\pi_{q}$
into that in the irreducible representation $\pi_{r_{1}}$
($=\pi_{r_{1}^{\prime}}$). This holds also at the other
trivalent vertices.
Thus we can successively perform the calculation like
eq.(\ref{eq:trivalent}). The final result is
\beq
<i_{v}=(r_{1},\ldots,r_{n-3})|i^{\prime}_{v}=(r^{\prime}_{1},\ldots,
r^{\prime}_{n-3})>=C(v)\prod_{k=1}^{n-3}
\delta_{r_{k},r^{\prime}_{k}}, \label{eq:HAAR2}
\eeq
where $C(v)$ is an overall constant factor which depends on
the way of decomposing the $n$-valent vertex $v$.

Eq.(\ref{eq:HAAR1}) together with eq.(\ref{eq:HAAR2}) shows that
the spin network states form an orthonormal basis w.r.t.
the induced Haar measure (\ref{eq:HAAR}) if they are multiplied
by appropriate overall constant factors.

A remarkable fact is that this measure is invariant
under spatial diffeomorphisms. This is obvious because
it does not depend on any background structure.
The induced Haar measure is therefore expected to be
a \lq prototype' whose appropriate modification
yields the physical measure in canonical quantum gravity.


\subsection{The coherent state transform and the induced heat-kernel
measure}

In the last subsection we have derived the explicit formula
for the inner product of the spin network states w.r.t.
the induced Haar measure on $SU(2)$ gauge theories.
However, because the gauge group of Ashtekar's formulation
for Lorentzian gravity is $SL(2,{\bf C})$ which is noncompact,
we cannot use the induced Haar measure as it is.
Nevertheless we can define some types of measures
on $SL(2,{\bf C})$ gauge theories, one of which is the
induced heat-kernel measure\cite{ALMMT}.
This measure is closely related to the \lq coherent state
transform' which maps a functional of the $SU(2)$ connection
to a holomorphic functional of the $SL(2,{\bf C})$ connection.
We expect that, using this kind of transform,
many of the machineries obtained in the framework of $SU(2)$ spin
network states can be brought into that of $SL(2,{\bf C})$
ones.
In this subsection we distinguish the objects in $SL(2,{\bf C})$
gauge theories from those in $SU(2)$ by attaching the superscript
$^{{\bf C}}$ to the former.

In order to define the coherent state transform and
the induced heat-kernel measure we first furnish each
edge $e\in\Gamma$ with a length function $l(e)>0$.
The coherent state transform
$$
C_{t}^{l}:L^{2}(\overline{\CA/\CG},d\mu_{H})
\rightarrow L^{2}(\overline{\CA^{{\bf C}}/\CG^{{\bf C}}},
d\nu_{t}^{l})\cap\CH(\overline{\CA^{{\bf C}}/\CG^{{\bf C}}}),
$$
from the space of square integrable functionals w.r.t. $d\mu_{H}$ on
$\overline{\CA/\CG}$ to the space of holomorphic functionals
which are square integrable w.r.t. the induced heat-kernel measure
$d\nu_{t}^{l}$ on $\overline{\CA^{{\bf C}}/\CG^{{\bf C}}}$
is given by a consistent family $\{C_{t,\Gamma}^{l}\}_{|\Gamma}$
of transforms
$$
C_{t,\Gamma}^{l}:L^{2}(\overline{(\CA/\CG)}_{\Gamma},
d\mu_{\Gamma})\rightarrow L^{2}(\overline{(\CA^{{\bf C}}/
\CG^{{\bf C}})}_{\Gamma},d\nu_{t,\Gamma}^{l})\cap
\CH(\overline{(\CA^{{\bf C}}/\CG^{{\bf C}})}_{\Gamma}).
$$
For a function $f_{\Gamma}(A_{\Gamma})$ on
$\overline{(\CA/\CG)}_{\Gamma}$ the transform
$C_{t,\Gamma}^{l}$ is defined as
\beq
C_{t,\Gamma}^{l}[f_{\Gamma}](A_{\Gamma}^{{\bf C}})\equiv
\int_{\overline{\CA_{\Gamma}}}\left(\prod_{e}
d\mu(A_{\Gamma}(e))\rho_{l(e)t}(A_{\Gamma}(e)^{-1}A_{\Gamma}
^{{\bf C}}(e))\right)f_{\Gamma}(A_{\Gamma}(e)),
\eeq
where, as in Appendix B,
$\rho_{t}(g)$ is the heat-kernel for the Casimir operator
$\Delta=J_{i}J_{i}$ of $SU(2)$.
For $(p_{\Gamma})^{\ast}f_{\Gamma}(A)\in L^{2}(
\overline{\CA/\CG},d\mu_{H})$
the coherent state transform is given by
\beq
C_{t}^{l}[(p_{\Gamma})^{\ast}f_{\Gamma}](A^{{\bf C}})=
C_{t,\Gamma}^{l}[f_{\Gamma}](p_{\Gamma}(A^{{\bf C}})).
\eeq
In particular, by using eq.(\ref{eq:coherent}), we find
\beq
C_{t}^{l}[<A|\Gamma,\{\pi_{p_{e}}\},\{i_{v}\}>](A^{{\bf C}})
=\exp(-\frac{t}{2}\sum_{e}l(e)\frac{p_{e}(p_{e}+2)}{4})
<A^{{\bf C}}|\Gamma,\{\pi_{p_{e}}\},\{i_{v}\}>.
\label{eq:COHERE}
\eeq
From this equation we immediately see that $C_{t}^{l}$ satisfies
the consistency condition iff $l(e\cdot e^{\prime})=
l(e)+l(e^{\prime})$ and $l(e^{-1})=l(e)$ hold.

The induced heat-kernel measure $d\nu_{t}^{l}$ for the functional
$(p_{\Gamma})^{\ast}f_{\Gamma}$ on
$\overline{(\CA^{\BC}/\CG^{\BC})}$ is defined as:
\beqy
\int_{\overline{\CA^{\BC}}}d\nu_{t}^{l}(A^{\BC})
(p_{\Gamma})^{\ast}f_{\Gamma}(A^{\BC})&\equiv&
\int_{\overline{\CA^{\BC}}_{\Gamma}}d\nu_{t,\Gamma}^{l}
(A_{\Gamma}^{\BC})f_{\Gamma}(A_{\Gamma}^{\BC})\nonumber \\
&\equiv&\int_{\overline{\CA^{\BC}}_{\Gamma}}\left(\prod_{e}
d\nu_{l(e)t}(A^{\BC}_{\Gamma}(e))\right)
f_{\Gamma}(A^{\BC}_{\Gamma}),
\eeqy
where $d\nu_{t}(g^{\BC})$ is Hall's averaged heat-kernel measure
\cite{hall}. 
 
Now, from Hall's theorem which is explained in Appendix B,
it follows that the coherent state transform
$C_{t}^{l}$ is an isometric isomorphism of $L^{2}(\overline{
\CA/\CG},d\mu_{H})$ onto $L^{2}(\overline{\CA^{\BC}/\CG^{\BC}},
d\nu_{t}^{l})\cap\CH(\overline{\CA^{\BC}/\CG^{\BC}})$
\cite{ALMMT}. Thus we see that the spin network
states form an orthonormal basis w.r.t. the induced heat-kernel
measure:
\beqy
&&\!\!\!\!<\Gamma,\{\pi_{p_{e}}\},\{i_{v}\}|\Gamma,\{\pi_{q_{e}}\},
\{i^{\prime}_{v}\}>_{|d\nu_{t}^{l}}\nonumber \\
&&\!\!\!\!\equiv\int_{\overline{\CA^{\BC}}}d\nu_{t}^{l}(A^{\BC})
\overline{<A^{\BC}|\Gamma,\{\pi_{p_{e}}\},\{i_{v}\}>}
<A^{\BC}|\Gamma,\{\pi_{q_{e}}\},\{i^{\prime}_{v}\}>\nonumber \\
&&\!\!\!\!=\exp(t\sum_{e}l(e)\frac{p_{e}(p_{e}+2)}{4})<\Gamma,
\{\pi_{p_{e}}\},\{i_{v}\}|\Gamma,\{\pi_{q_{e}}\},
\{i^{\prime}_{v}\}>_{|d\mu_{H}} \nonumber \\
&&\!\!\!\!=\exp(t\sum_{e}l(e)\frac{p_{e}(p_{e}+2)}{4})\prod_{e}
\frac{\delta_{p_{e},q_{e}}}{p_{e}+1}\prod_{v}
<i_{v}|i^{\prime}_{v}>. \label{eq:HKmeasure}
\eeqy
Similarly to the coherent state transform, the consistency
of this measure is satisfied iff $l(e\cdot e^{\prime})=
l(e)+l(e^{\prime})$ and $l(e^{-1})=l(e)$ hold.

Here we make a few remarks. The coherent state transform
$C_{t}^{l}$ and the induced heat-kernel measure $d\nu_{t}^{l}$
are not invariant under diffeomorphisms, because their definitions
require to introduce a length function $l(e)$ which is
not diffeomorphism invariant.
Thus we can use this measure only in the context which do not
respect the diffeomorphism invariance.
In the lattice formulation
explained in \S 5, however, we do not have to worry so much about
the diffeomorphism invariance because it is manifestly violated by
the introduction of a lattice. So it is possible
that $d\nu_{t}^{l}$ play an important role
in the lattice formulation. 

The invariance of the induced heat-kernel measure
$d\nu_{t}^{l}(A^{\BC})$ under generalized
$SU(2)$ gauge transformations follows from the
bi-$SU(2)$ invariance of the averaged heat-kernel measure
$d\nu_{t}(g^{\BC})$. However, because $d\nu_{t}(g^{\BC})$
is not bi-$SL(2,{\BC})$ invariant, $d\nu_{t}^{l}(A^{\BC})$
is not invariant under full $SL(2,\BC)$ gauge
transformations. This measure is therefore useful when
we fix the Lorentz boost gauge degrees of freedom.

We should note that
the induced heat-kernel
measure does not yield a physical measure for quantum
gravity because it does not implement the correct reality
conditions. This measure, however,
may be useful for examining some property
of the wavefunctions.


\subsection{The Wick rotation transform}

Finding the physical inner product for quantum Ashtekar's formalism
in the Lorentzian signature is one of the most formidable
problem because the physical inner product must implement
non-trivial reality conditions. To the author's knowledge
no works on the inner product which have been made so far
have reached a decisive conclusion.
Quite recently, however, a promising candidate has appeared for
the construction scheme of the physical inner product.
Leaving the detail to the original literature\cite{thie},
here we will briefly explain the underlying idea.

As is mentioned before, there is a formulation for
Lorentzian gravity\cite{barb} whose configuration variable
is a real $SU(2)$ connection 
\beq
(A^{\prime})_{a}^{i}=-\frac{1}{2}\ep^{ijk}\omega_{a}^{jk}
+\omega^{0i}_{a}. \label{eq:realsu2}
\eeq
While the Wheeler-De Witt equation in this formulation is
as complicated as that in the ADM formalism,
this formulation has a merit that the reality conditions are given
by trivial self-adjointness conditions. A physical measure
in this formalism is therefore given by the induced Haar measure
explained in \S\S 3.2. The main idea in ref.\cite{thie} is
to obtain Ashtekar's $SL(2,\BC)$ connection
$$
A_{a}^{i}=-\frac{1}{2}\ep^{ijk}\omega_{a}^{jk}-i\omega_{a}^{0i}
$$
from the real $SU(2)$ connection (\ref{eq:realsu2}), by way of
the \lq\lq Wick rotation transform":
\beq
(A^{\prime})_{a}^{i}\rightarrow A_{a}^{i}=e^{\hat{C}}(A^{\prime})
_{a}^{i}e^{-\hat{C}},\label{eq:wick}
\eeq
where $\hat{C}$ is the operator version of the \lq\lq Wick rotator"
$$
C\equiv\frac{\pi}{2}\int_{\M3}d^{3}x\omega_{a}^{0i}\te^{ai}.
$$
We expect that the \lq\lq Wick rotated measure", the measure
which is induced from the induced Haar measure
by this Wick rotation, implements the
correct reality conditions. However, to find a fully regularized
operator version of this transform seems to be considerably
difficult, because $C$ involves an expression which is non-polynomial
in the canonical variables $(A^{i}_{a},\te^{ai})$ (or
$((A^{\prime})^{i}_{a},\te^{ai})$).
If we find a well-defined Wick rotation transform, we can assert
that we have made a large step forward toward completing
the program of canonical quantum gravity in terms of
Ashtekar's new variables.


\subsection{Diffeomorphism invariant states}

As we have seen in the previous subsections spin network states
provide us with a complete basis in the space $\overline{\CA/\CG}$
of equivalence classes of (generalized) connections up to
gauge transformations. Thus, as long as we work with the spin
network states, Gauss' law constraint (\ref{eq:qgauss}) is
automatically solved. As it is, however, spin network states
are not invariant under spatial diffeomorphisms. A further device
is therefore required in order to
construct diffeomorphism invariant states, namely to solve the
integrated diffeomorphism constraint (\ref{eq:intqdiff}).

We know that an integrated version $\hat{U}(\phi)$ of the
diffeomorphism constraint operator $\hat{\CD}(\vec{N})=-i\int_{\M3}
d^{3}x\CL_{\vec{N}}A^{i}_{a}\delta/\delta A^{i}_{a}$
generates a small spatial diffeomorphism
$$
\hat{U}(\phi)A_{a}^{i}(x)\hat{U}(\phi^{-1})=
(\phi^{-1})^{\ast}A_{a}^{i}(x)
\equiv\partial_{a}(\phi^{-1}(x))^{b}A_{b}^{i}(\phi^{-1}(x)).
$$
This affects the transformation of the parallel propagator
$h_{\alpha}[0,1]$ as a distortion of the curve $\alpha$
\beq
\hat{U}(\phi)h_{\alpha}[0,1]\hat{U}(\phi^{-1})=
h_{\phi^{-1}\circ\alpha}[0,1]. \label{eq:intpdiff}
\eeq
Thus we see that the integrated diffeomorphism operator
$\hat{U}(\phi)$ induces a diffeomorphism of the graph $\Gamma$
on which the spin network states are defined:
\beqy
\hat{U}(\phi)<A|\Gamma,\{\pi_{e}\},\{i_{v}\}>&=&
<A|\phi^{-1}\circ\Gamma,\{\pi^{\prime}_{
\phi^{-1}\circ e}=\pi_{e}\},
\{i^{\prime}_{\phi^{-1}(v)}=i_{v}\}>\nonumber \\
&\equiv&<A|\phi^{-1}\circ\Gamma,\{\pi_{e}\},\{i_{v}\}>,
\label{eq:spidiff}
\eeqy
where $\phi\circ\Gamma\equiv(\{\phi\circ e\},\{\phi(v)\})$.

Naively considering,
taking the following average yields the basis of
diffeomorphism invariant states:
\beq
\{Vol({\rm Diff}_{0}(\M3))\}^{-1}
\int_{{\rm Diff}_{0}(\M3)}[\CD\phi]
<A|\phi\circ\Gamma,\{\pi_{e}\},\{i_{v}\}>,\label{eq:formal}
\eeq
where $[\CD\phi]$ is an \lq invariant measure' on the space
${\rm Diff}_{0}(\M3)$ of small diffeomorphisms on $\M3$.
This average is formal in the sense that
it is indefinite because it involves
the ratio of two divergent expressions,
namely, $Vol({\rm Diff}_{0}(\M3))$ and
$\int_{{\rm Diff}_{0}(\M3)}[{\cal D}\phi]$.

A prescription to make this
formal average be mathematically rigorous
was given in ref.\cite{ALMMT2}.
A rough outline of the strategy used there is the following.
We first consider the space $\Phi\equiv Cyl^{\infty}
(\overline{\CA/\CG})$ of smooth cylindrical functions
each of which is given by a linear combination of
the spin network states defined on a sufficiently large
graph.
We can thus perform an orthogonal decomposition
of $\varphi\in\Phi$ as:
$$
\varphi[A]=\sum_{\Gamma^{\prime}}\sum_{\{\pi_{e^{\prime}}\}}
\varphi_{\Gamma^{\prime},\{\pi_{e^{\prime}}\}}[A],
$$
where $\varphi_{\Gamma^{\prime},\{\pi_{e^{\prime}}\}}$
denotes the projection of $\varphi$ onto the space of spin network
states defined on a graph $\Gamma^{\prime}$ with fixed
representations $\{\pi_{e^{\prime}}\}$ assigned to the edges
$\{e^{\prime}\}$. Now we can define
the average of $<A|\Gamma,\{\pi_{e}\},\{i_{v}\}>$ over the
group ${\rm Diff}_{0}(\M3)$ of small diffeomorphisms:
\beq
\Psi^{DI}_{([\Gamma],\{\pi_{e}\},\{i_{v}\})}[\varphi]\equiv
\sum_{\Gamma_{2}\in[\Gamma]}
<\Gamma_{2},\{\pi_{e}\},\{i_{v}\}|\varphi>,\label{eq:diffinv}
\eeq
where $[\Gamma]$ stands for the orbit of the graph
$\Gamma$ under ${\rm Diff}_{0}(\M3)$,
and $<|>$ represents some inner product on $\overline{\CA/\CG}$
which is invariant under diffeomorphisms.
Eq.(\ref{eq:diffinv}) is well-defined as a distributional
wavefunction which belongs to
the topological dual $\Phi^{\prime}$ of $\Phi$.
This is because, if we use for example the induced Haar measure as
the inner product $<|>$,
nonvanishing contributions to
eq.(\ref{eq:diffinv}) are only from the inner
products in which the spin network states appearing
in the bra $<|$ and in the ket $|>$ correspond to the identical
graph $\Gamma$ with coincident colored edges $\{(e,\pi_{e})\}$.
Hence, from the spin network states $<A|\Gamma,\{\pi_{e}\},
\{i_{v}\}>$, we can construct a basis $\{\Psi^{DI}_{([\Gamma],\{\pi_{e}\},\{i_{v}\})}\}$ of diffeomorphism
(and gauge) invariant states.\footnote{
The notion of a discretized measure on loops which leads to
diffeo invariant measures was first introduced into quantum
gravity by ref.\cite{rayner}. This kind of diffeo invariant
measures were first constructd by Gel'fand and collaborators
using various point sets.}

Here a remark should be made. In the defining equation
(\ref{eq:diffinv}) we have assumed the existence of a diffeomorphism
invariant inner product $<|>$. Because the induced Haar measure
enjoys this property, the prescription sketched here
applies to Ashtekar's formulation for Euclidean gravity.
In the Lorentzian signature, however, we do not yet know
whether we can define diffeomorphism invariant states through
eq.(\ref{eq:diffinv}). This is because the induced heat-kernel
measure is not invariant under diffeomorphisms.
While we can construct a diffeomorphism invariant measure
\cite{ALMMT} by way of the Baez measure\cite{baez3},
it does not seem to be suitable for the present purpose
because it is not faithful. However, it is noteworthy
that, if appropriately regularized, the Wick rotated
measure explained in the
last subsection may provide a desired inner product
owing to the diffeomorphism invariance of the Wick rotator $C$.
A hard task remaining there is to find a convenient orthonormal
basis w.r.t. this Wick rotated measure such as the spin
network state basis w.r.t. the induced Haar measure.


\subsection{Area and volume operators}

In order to extract physical information from the wavefunctions
we need physical observables each of which is a self-adjoint
operator which commutes with the
constraint operators at least weakly.
Necessarily physical observables have to be invariant under
$SL(2,\BC)$ gauge transformations and spatial diffeomorphisms.

While we do not know any physical observables in pure gravity,
we can construct a large set of gauge and diffeomorphism invariant
operators, namely, area operators and volume operators
\cite{RS2}. In matter-coupled theories these operators are expected
to become physical observables.

First we explain the area operators. Classically the area $A(D)$
of a two-dimensional region $D\in\M3$ is given by
\beq
A(D)=\int_{D}d^{2}\sigma\left(\UT{n}_{a}\UT{n}_{b}
\tpi^{ai}\tpi^{bi}\right)^{\frac{1}{2}},
\eeq
where $\UT{n}_{a}\equiv\utep_{abc}\partial_{\sigma^{1}}x^{b}
\partial_{\sigma^{2}}x^{c}$ is the \lq\lq normal vector" to $D$.
Because this expression contains second order functional derivative,
some regularization is necessary in passing to the quantum theory.
A clever regularization was provided in \cite{RS2}. We first
divide the region $D$ into subregions $D=\sum_{I}D_{I}$
with the coordinate area of each subregion $D_{I}$ assumed to be
of order $\delta^{2}$.
Then the operator version of $A(D)$ is defined as
\beqy
\hat{A}(D)&=&\lim_{\delta\rightarrow0}\sum_{I}\hat{A}(D_{I})
\nonumber \\
\{\hat{A}(D_{I})\}^{2}&\equiv&
\int_{D_{I}}d^{2}\sigma\UT{n}_{a}(x(\sigma))
\int_{D_{I}}d^{2}\sigma^{\prime}\UT{n}_{b}(x(\sigma^{\prime}))
\hat{\tpi}^{ai}(x(\sigma))\hat{\tpi}^{bi}(x(\sigma^{\prime})).
\label{eq:defarea}
\eeqy

We can easily compute
the action of this area operator on the parallel
propagator $\pi(h_{\alpha}[0,1])$ in the representation $\pi$
evaluated along a smooth curve $\alpha$ which does not
intersect with itself. Because the action
of $\hat{\tpi}^{ai}=-\delta/\delta A_{a}^{i}$ on the parallel
propagator is given by
$$
\hat{\tpi}^{ai}(x)h_{\alpha}[0,1]=-\int_{0}^{1}ds\delta^{3}
(x,\alpha(s))\dot{\alpha}^{a}(s)h_{\alpha}[0,s]J_{i}h_{\alpha}[s,1],
$$
the action of $\{\hat{A}(D_{I})\}^{2}$ on the parallel propagator
is computed as
\beq
\{\hat{A}(D_{I})\}^{2}\cdot\pi(h_{\alpha}[0,1])=\left\{
\begin{array}{lll}
\pi(\Delta)\pi(h_{\alpha}[0,1])&\mbox{if}&
\alpha\cap D_{I}=x_{0}\in\M3 \\
0 &\mbox{if}& \alpha\cap D_{I}=\emptyset,
\end{array}\right.
\eeq
where $\Delta\equiv J_{i}J_{i}$ is the Casimir operator of
$SL(2,\BC)$. We assumed that $D_{I}$ is so small and
$\alpha$ is so well-behaved that they intersect with each other
at most once.

Here we should note that, if $\pi$ is the spin-$\frac{p}{2}$
representation, the Casimir $\pi(\Delta)=-\frac{p(p+2)}{4}$
is negative. Thus $\pi(h_{\alpha}[0,1])$ is an eigenfunction
of the squared-area operator $\{\hat{A}(D_{I})\}^{2}$
{\em with a negative eigenvalue}.
If we take the minus sign seriously, it seems to
follow that the spin network state defined on a
piecewise smooth graph with each edge colored by
a finite dimensional representation does not correspond to
any classical spacetimes with correct Lorentzian signature.
This may suggest that we should work with the spin network
states i) which is defined on graphs which are not piecewise
analytic and/or ii) each of whose edges are colored by an
infinite dimensional representation.
At present we do not know the correct solution to this \lq\lq issue
of negative squared-area", or even do not know whether
or not we should confront this issue seriously.
However, if we take an optimistic attitude and avoid this issue
by taking only the absolute value of the Casimir,
we obtain the result derived in \cite{RS2}:
\beq
|\hat{A}(D)|\cdot\pi(h_{\alpha}[0,1])=n(D,\alpha)
\sqrt{|\pi(\Delta)|}\pi(h_{\alpha}[0,1]),
\eeq
where $n(D,\alpha)$ is the number of intersections between
the two-dimensional region $D$ and the curve $\alpha$.
The area operators therefore exhibit a discrete spectrum.
It was shown that the volume operators also have
discrete eigenvalues\cite{RS2}. This \lq\lq discreteness
of area and volume" is considered to suggest that the spacetime 
reveals some microstructure in the Planck regime.

Next we consider the volume operators. A naive expression of
the operator $\hat{V}(\cal R)$
which measures the volume of a three-dimensional region
${\cal R}$ is given by
\beq
\hat{V}({\cal R})=\int_{{\cal R}}d^{3}x\left(\frac{1}{3!}
\utep_{abc}\ep^{ijk}\hat{\tpi}^{ai}\hat{\tpi}^{bj}\hat{\tpi}^{ck}
\right)^{\frac{1}{2}}.\label{eq:volume}
\eeq
Because this operator involves the third order functional derivative,
we need some regularization in order to obtain a well-defined result.
Regularized versions of this operator are proposed both
in the continuum case\cite{RS2} and in the discretized case
\cite{loll2}. For detail we refer the reader to the references
and here we briefly explain the essence.

We see from eq.(\ref{eq:volume}) that the nonvanishing
contributions of its action on the spin network states
arise only from vertices. Thus we can concentrate only
on vertices. The results in \cite{RS2}\cite{loll2} tell us
that the action of $\hat{V}$ vanishes on the bi- or trivalent
vertex. In order to have nonvanishing volume, we
have to consider spin network states which are defined
on graphs with at least four-valent vertices.


\subsection{Spin network states in terms of spinor propagators}

In the previous subsections we have developed the framework
of spin network states by using arbitrary finite-dimensional
representations. Because the spin-$\frac{p}{2}$ representation
is expressed as a symmetrized tensor-product of
$p$ copies of spinor representations
\beqy
\pi_{p}(h_{\alpha}[0,1])&\cong& h_{\alpha}[0,1]^{A_{1}}
\LI{(B_{1}}\cdots h_{\alpha}[0,1]^{A_{p}}\LI{B_{p})}\nonumber \\
h_{\alpha}[0,1]^{A}\LI{B}&\equiv&\pi_{1}(h_{\alpha}[0,1])^{A}\LI{B},
\eeqy
we can in principle do our analysis in terms only of spinor
propagators. While it is desirable that the ultimate results
should be described in terms of arbitrary irreducible
representations, it is sometimes more convenient to calculate
purely by means of the spinor representation.
Actually the analysis in the next section is based only on
the spinor representation. Thus it would be proper to develop here
the framework in terms of spinor propagators.

As a result of the defining equation of $SL(2,\BC)$, the
spinor propagator is subject to the following identity:
\beq
\ep^{AD}\ep_{BC}h_{\alpha}[0,1]^{C}\LI{D}=
(h_{\alpha}[0,1]^{-1})^{A}\LI{B}
=h_{\alpha^{-1}}[0,1]^{A}\LI{B}.\label{eq:identity1}
\eeq
This identity is useful because it
enables us to choose relevant orientations
of the propagators according to
a particular problem.

There are three
invariant tensors in the spinor representation, i.e. the
invariant spinors
$\ep_{B}\UI{A}\equiv\delta^{A}_{B}$,
$\ep^{AB}$ and $\ep_{AB}$. Because all the intertwining operators
are constructed from these invariant spinors,
the total rank of an intertwining
operator is even.
Thus the sum of the numbers of spinor propagators
ending and starting at a  vertex is necessarily even.
Using the identity (\ref{eq:identity1}) and the equation
$\ep_{AC}\ep^{BC}=\delta^{B}_{A}$,
we can equalize at each vertex
the number of in-coming propagators
with that of out-going propagators.
As a consequence we can regard any intertwining operator to be a
linear combination of the products of $\delta_{A}^{B}$.

In addition to eq.(\ref{eq:identity1}),
there are three useful identities
in the spinor representation. Two of them
are the two-spinor identity
\beq
\delta_{A}^{B}\delta_{C}^{D}-\delta_{A}^{D}\delta_{C}^{B}
=\ep_{AC}\ep^{BD}
\quad(\mbox{ or }
\phi^{A}\ep^{BC}+\phi^{B}\ep^{CA}+\phi^{C}\ep^{AB}=0)
\label{eq:2spi}
\eeq
and the Fiertz identity
\beq
(J_{i})^{A}\LI{B}(J_{i})^{C}\LI{D}=
-\frac{1}{2}(\delta^{A}_{D}\delta^{C}_{B}
-\frac{1}{2}\delta^{A}_{B}\delta^{C}_{D}).\label{eq:fiertz}
\eeq
The third identity 
\beq
\ep_{ijk}(J_{j})^{A}\LI{B}(J_{k})^{C}\LI{D}=\frac{1}{2}
\{(J_{i})^{A}\LI{D}\delta^{C}_{B}-(J_{i})^{C}\LI{B}
\delta^{A}_{D}\}\label{eq:third}
\eeq
is obtained by combining the
Fiertz identity and the defining equation:
\beq
(J_{i})^{A}\LI{C}(J_{j})^{C}\LI{B}=-\frac{1}{4}\delta_{ij}
\delta^{A}_{B}+\frac{1}{2}\ep_{ijk}(J_{k})^{A}\LI{B}.
\label{eq:defsl2c}
\eeq
The identity (\ref{eq:third}) plays an important role when we
calculate the action of the scalar constraint.

The two-spinor identity tells us that the antisymmetrized
tensor product of the two identical
spinor propagators gives the trivial
representation:
\beq
h_{\alpha}[0,1]^{A}\LI{[B}h_{\alpha}[0,1]^{C}\LI{D]}=
\frac{1}{2}\ep^{AC}\ep_{BD}
\label{eq:2spi2}
\eeq
and that, at an intersection
$\alpha(s_{0})=\beta(t_{0})$ of two curves
$\alpha$ and $\beta$, the following identity holds:
\begin{eqnarray}
h_{\alpha}[0,1]^{A}\LI{B}h_{\beta}[0,1]^{C}\LI{D}-
(h_{\alpha}[0,s_{0}]h_{\beta}[t_{0},1])^{A}\LI{D}
(h_{\beta}[0,t_{0}]h_{\alpha}[s_{0},1])^{C}\LI{B} \nonumber \\
=(h_{\alpha}[0,s_{0}]h_{\beta^{-1}}[1-t_{0},1])^{A}\LI{E}\ep^{EC}
(h_{\beta^{-1}}[0,1-t_{0}]h_{\alpha}[s_{0},1])^{F}\LI{B}\ep_{FD}.
\label{eq:2spi3}
\end{eqnarray}
Owing to eq.(\ref{eq:2spi2}) we do not have to
take account of antisymmetrizing
the identical propagators.

In practical calculation it is frequently
convenient to introduce the graphical representation.
We will denote a spinor propagator
$h_{\alpha}[0,1]^{A}\LI{B}$ by an arrow from $\alpha(0)$
to $\alpha(1)$ with its tail (tip) being
equipped with the spinor index
$A$ ($B$). Identities (\ref{eq:identity1}),(\ref{eq:2spi2}) and
(\ref{eq:2spi3}) are then expressed as follows:

\begin{figure}[ht]
\begin{picture}(150,30)
\put(0,0){\makebox(20,30){$\ep^{AD}\ep_{BC}$}}
\put(20,0){\usebox{\LBRA}}
\put(30,5){\vector(0,1){20}}
{\scriptsize
\put(30,2){$C$}
\put(30,25){$D$}}
\put(40,0){\usebox{\RBRA}}
\put(45,0){\makebox(30,30){$=$}}
\put(75,0){\usebox{\LBRA}}
\put(85,25){\vector(0,-1){20}}
{\scriptsize
\put(85,2){$B$}
\put(85,25){$A$}}
\put(95,0){\usebox{\RBRA}}
\put(100,0){\makebox(5,30){,}}
\put(120,0){\makebox(20,30){(\ref{eq:identity1}$)^{\prime}$}}
\end{picture}
\end{figure}

\begin{figure}[ht]
\begin{picture}(150,65)(-15,0)
\put(-5,35){\usebox{\LBRA}}
\put(8,40){\vector(0,1){20}}
\put(12,40){\vector(0,1){20}}
{\scriptsize
\put(8,37){$A$}
\put(8,60){$B$}
\put(12,37){$C$}
\put(12,60){$D$}}
\put(20,35){\makebox(20,30){$-$}}
\put(48,52){\vector(0,1){8}}
\put(48,40){\line(0,1){8}}
\put(52,52){\vector(0,1){8}}
\put(52,40){\line(0,1){8}}
\put(48,48){\line(1,1){4}}
\put(48,52){\line(1,-1){4}}
{\scriptsize
\put(48,37){$A$}
\put(48,60){$B$}
\put(52,37){$C$}
\put(52,60){$D$}}
\put(60,35){\usebox{\RBRA}}
\put(65,35){\makebox(30,30){$=$ $\ep^{AC}\ep_{BD}$ ,}}
\put(110,35){\makebox(20,30){(\ref{eq:2spi2}$)^{\prime}$}}


\put(-5,0){\usebox{\LBRA}}
\put(10,5){\vector(0,1){20}}
\put(0,15){\vector(1,0){20}}
{\scriptsize
\put(10,2){$A$}
\put(10,25){$B$}
\put(0,15){$C$}
\put(20,15){$D$}}
\put(20,0){\makebox(20,30){$-$}}
\put(49,16){\vector(0,1){9}}
\put(40,16){\line(1,0){9}}
\put(51,14){\vector(1,0){9}}
\put(51,5){\line(0,1){9}}
{\scriptsize
\put(51,2){$A$}
\put(49,25){$B$}
\put(40,16){$C$}
\put(60,14){$D$}}
\put(60,0){\usebox{\RBRA}}
\put(65,0){\makebox(10,30){$=$}}
\put(75,0){\makebox(10,30){$\ep^{EC}\ep_{FD}$}}
\put(85,0){\usebox{\LBRA}}
\put(101,16){\vector(0,1){9}}
\put(101,16){\line(1,0){9}}
\put(99,14){\vector(-1,0){9}}
\put(99,5){\line(0,1){9}}
{\scriptsize
\put(99,2){$A$}
\put(101,25){$B$}
\put(110,16){$F$}
\put(90,14){$E$}}
\put(110,0){\usebox{\RBRA}}
\put(115,0){\makebox(15,30)[l]{. (\ref{eq:2spi3}$)^{\prime}$}}
\end{picture}
\end{figure}


\section{Combinatorial solutions to
the Wheeler-De Witt equation}

In the last section we have prepared mathematical tools
to deal with spin network states. A virtue of using
spin network states is that the action of local operators
can be described purely in terms of local manipulations
on the colored graphs under consideration, while we
have to allow for changes of global properties when we deal with
Wilson loops. In this section we will explicitly see
this in the context of extending the solution found by Jacobson and
Smolin\cite{jacob}, which is a linear combination of spinor Wilson
loops, to arbitrary finite dimensional representations\cite{ezawa3}.

For analytical simplicity we will restrict the type of
vertices $\{v\}$ appearing in the graph $\Gamma$
to those at which two smooth curves intersect.
We further assume that the tangent vectors of the two curves
are linearly independent. We will henceforth call
those vertices \lq\lq regular four-valent vertices".
In this and the next sections we consider the case
where the cosmological constant vanishes, i.e. $\Lambda=0$.


\subsection{Action of the scalar constraint on spin network states}

In order to construct
solutions to the Wheeler-De Witt equation
(\ref{eq:nwd}) from spin network states,
we have to evaluate the action of the scalar constraint operator
\footnote{As is explained in \S\S 2.2,
there is an issue on the choice of
operator ordering which is intimately
related to the problem on the closure of the constraint algebra
\cite{kodama}\cite{bori}\cite{brug2}.
We will not discuss on these issues and
simply choose the ordering in which the momenta
are placed to the right of the $SL(2,\BC)$ connections.}
$$
\hat{\CS}(\tN)=\int_{\M3}d^{3}x\frac{\tN(x)}{2}\ep^{ijk}
F^{i}_{ab}(x)\frac{\delta}{\delta A^{j}_{a}(x)}
\frac{\delta}{\delta A^{k}_{b}(x)}
$$
on the spin network states. Because this operator contains second
order functional derivative, some regularization have to be
prescribed. Here we will use the point-splitting regularization
\cite{RS}\cite{bren}\cite{bori}\cite{brug2}:
\beqy
\hat{\CS}^{\ep}(\tN)&\equiv&\int_{\M3}d^{3}x\int_{\M3}d^{3}x
\frac{\tN(x)}{2}\tilde{f}_{\ep}(x,y)\utep_{abc}
\Tr(-4h_{yx}[0,1]\tB^{c}(x)J_{i}h_{xy}[0,1]J_{j})\nonumber \\
& &\quad\times\frac{\delta}{\delta A^{i}_{a}(x)}
\frac{\delta}{\delta A^{j}_{b}(y)},\label{eq:reghami}
\eeqy
where $h_{xy}[0,1]$ is the parallel propagator along a curve from $x$
to $y$ which shrinks to $x$ in the limit $y\rightarrow x$,
$\tB^{c}\equiv\tB^{ci}J_{i}\equiv\frac{1}{2}
\otep^{abc}F^{i}_{ab}J_{i}$
is the magnetic field of the $SL(2,\BC)$ connection and 
$\Tr$ in this section stands for the trace in the spinor
representation. The regulator
$\tilde{f}_{\ep}(x,y)$ is subject to the condition
$$
\tilde{f}_{\ep}(x,y)\stackrel{\ep\rightarrow0}{\longrightarrow}
\delta^{3}(x,y).
$$
In order to define a particular regulator we must fix
a local coordinate frame. Because a usual choice
of the coordinate frame breaks diffeomorphism covariance,
the resulting regularized scalar constraint are usually
not covariant under diffeomorphisms. Here we will use a somewhat
tricky choice of the local coordinate frame which
is similar to that used in the (2+1)-dimensional
version of Ashtekar's formalism\cite{ezawa3}.

It is obvious that the action of $\hat{\CS}^{\ep}(\tN)$ on the spin
network states is nonvanishing only on
the curves belonging to the graph in question. Eq.(\ref{eq:intpdiff})
tells us that these curves are subject to
the action of the diffeomorphism constraint.
We therefore expect that,
if we introduce a local curvilinear coordinate
frame in which these curves play the role of coordinate curves,
we can define a regularization of the scalar constraint which
preserves diffeomorphism covariance.
In this subsection we will execute such a regularization procedure.

In order to fix a curvilinear coordinate frame completely
we need three linearly independent curves.
In the case under our consideration, we have at most two
curves intersecting at a point. So we have to introduce
one or two \lq\lq virtual curves" as is shown in
figure 3.

\begin{figure}[t]
\begin{center}
\epsfig{file=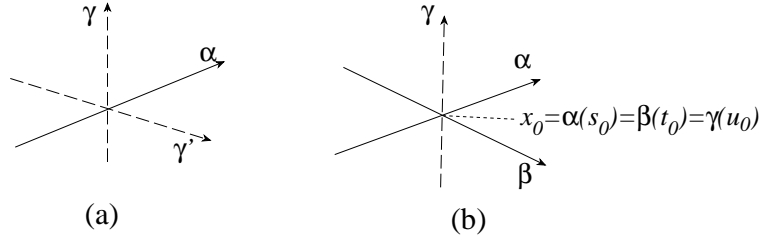,height=4cm}
\end{center}
\caption{Local curvilinear coordinate frames at
an analytic part (a) and at a regular four-valent
vertex (b). The solid and the dashed lines respectively
denote the true curves and the virtual ones in the
graph.}
\end{figure}

Because the scalar constraint involves
two functional derivatives,
it is convenient to separate its action as follows:
\beq
\hat{\CS}(\tN)=\hat{\CS}_{1}(\tN)
+\hat{\CS}_{2}(\tN)\qquad
(\mbox{or}\quad\hat{\CS}^{\ep}(\tN)=\hat{\CS}^{\ep}_{1}(\tN)+
\hat{\CS}^{\ep}_{2}(\tN)),\label{eq:sepa}
\eeq
where $\hat{\CS}_{1}(\tN)$ and $\hat{\CS}_{2}(\tN)$ stand for,
respectively, the action of $\hat{\CS}(\tN)$
on the single spinor propagators and that
on the pairs of spinor propagators. This separation simplifies
considerably the evaluation of the action of
the Hamiltonian constraint on spin-network
states.

Now we demonstrate a few examples of the calculation.

First we consider the action of the regularized Hamiltonian
(\ref{eq:reghami})
on a single spinor propagator $h_{\alpha}[0,1]^{A}\LI{B}$
along a smooth curve $\alpha$. We can easily see that
this vanishes. A functional derivative
on a propagator (in an arbitrary representation)
\beq
\frac{\delta}{\delta A_{a}^{i}(x)}h_{\alpha}[0,1]
=\int_{0}^{1}ds\delta^{3}(x,\alpha(s))\dot{\alpha}^{a}(s)
h_{\alpha}[0,s]J_{i}h_{\alpha}[s,1]
\eeq
necessarily picks a distributional factor $\int_{0}^{1}ds
\delta^{3}(x,\alpha(s))\dot{\alpha}^{a}(s)$
which involves the tangent vector. By carrying out the other
functional derivative and the integrations in eq.(\ref{eq:reghami}),
we find that $\hat{\CS}^{\ep}h_{\alpha}[0,1]=\hat{\CS}^{\ep}_{1}
h_{\alpha}[0,1]$ involves the expression
\beq
\int_{0}^{1}ds\int_{0}^{1}dt\tilde{f}_{\ep}(\alpha(s),\alpha(t))
\utep_{abc}\dot{\alpha}^{a}(s)\dot{\alpha}^{b}(t).
\eeq
In our coordinate choice this expression vanishes
because $\dot{\alpha}^{a}(s)\propto\dot{\alpha}^{a}(t)$
even for $s\neq t$.

By a similar reasoning we see that the action of the regularized
scalar constraint $\hat{\CS}^{\ep}(\tN)$ on a pair of
spinor propagators along a smooth curve $\alpha$ vanishes.
Summarizing these two results we find:
\beq
\hat{\CS}^{\ep}_{1}(\tN)h_{\alpha}[0,1]^{A}\LI{B}=
\hat{\CS}^{\ep}_{2}(\tN)(h_{\alpha}[0,1]^{A}\LI{B}
h_{\alpha}[0,1]^{C}\LI{D})=0.
\eeq
In consequence we have only to pay attention to points on
the graph at which the analyticity of the curve breaks,
namely to vertices and kinks.

Next we calculate the action on a single spinor propagator
$$
h_{\alpha_{1}\cdot\beta_{2}}[0,1]^{A}\LI{B}\equiv(h_{\alpha}
[0,s_{0}]h_{\beta}[t_{0},1])^{A}\LI{B}
$$
along a curve $\alpha_{1}\cdot\beta_{2}$ with a kink\footnote{
In this section we assume that the curves $\alpha$ and
$\beta$ intersect with each other at
$\alpha(s_{0})=\beta(t_{0})=x_{0}$.
$\alpha_{1}\subset\alpha$ is the curve
from $\alpha(0)$ to $\alpha(s_{0})$
and $\beta_{2}\subset\beta$ is the
curve from $\beta(t_{0})$ to $\beta(1)$. We also assume that
the virtual curve $\gamma$ intersects with $\alpha$ and $\beta$
at $\gamma(u_{0})=x_{0}$.}:
\beqy
&&\!\!\!\!\hat{\CS}^{\ep}_{1}(\tN)
h_{\alpha_{1}\cdot\beta_{2}}[0,1]^{A}\LI{B}\nonumber \\
&&\!\!\!\!=\frac{1}{2}\int_{0}^{s_{0}}ds\int_{t_{0}}^{1}dt
\utep_{abc}\dot{\alpha}^{a}(s)\dot{\beta}^{b}(t)
\tilde{f}_{\ep}(\alpha(s),\beta(t))\nonumber \\
&&\!\!\!\!\times\left\{\tN(\alpha(s))
\Tr(-4h_{ts}\tB^{c}(\alpha(s))J_{j}h_{st}J_{k})
h_{\alpha}[0,s]J_{j}h_{\alpha}[s,s_{0}]
h_{\beta}[t_{0},t]J_{k}h_{\beta}[t,1]\right.\nonumber \\
&&\!\!\!\!-\left.
\tN(\beta(t))\Tr(-4h_{st}\tB^{c}(\beta(t))J_{j}h_{ts}J_{k})
h_{\alpha}[0,s]J_{k}h_{\alpha}[s,s_{0}]
h_{\beta}[t_{0},t]J_{j}h_{\beta}[t,1]\right\}^{A}\LI{B},
\label{eq:hamikink}
\eeqy
where we have abbreviated $h_{\alpha(s)\beta(t)}[0,1]$ as $h_{st}$.
In order to go further we have to determine explicitly the regulator
$\tilde{f}_{\ep}$. By introducing the curvilinear coordinate
and by exploiting the fact that $\delta^{3}(x,y)$ transforms as
a scalar density of weight $+1$, we define $\tilde{f}_{\ep}$ as
\beq
\tilde{f}_{\ep}(\alpha(s),\beta(t))\equiv
|\utep_{abc}\dot{\alpha}^{a}(s_{0})\dot{\beta}^{b}(t_{0})
\dot{\gamma}^{c}(u_{0})|^{-1}\frac{1}{(2\ep)^{3}}\theta(\ep-
|s-s_{0}|)\theta(\ep-|t-t_{0}|)\theta(\ep-|u-u_{0}|).
\eeq
By plugging this definition into eq.(\ref{eq:hamikink}) and
by using $(h_{st})^{A}\LI{B}=\delta^{A}_{B}+O(\ep)$ and
similar approximations, we find
\beqy
&&\hat{\CS}^{\ep}_{1}(\tN)h_{\alpha_{1}\cdot\beta_{2}}
[0,1]^{A}\LI{B}\nonumber \\
&&=\frac{1}{16\ep}\{2\UT{n}\cdot\tB^{i}(x_{0})\ep_{ijk}
h_{\alpha}[0,s_{0}]J_{j}J_{k}h_{\beta}[t_{0},1]+O(\ep)\}^{A}\LI{B},
\label{eq:hamikink2}
\eeqy
where we have used the notations
$\UT{n}\cdot\tB^{i}(x_{0})\equiv\UT{n}_{c}(x_{0})\tB^{ci}(x_{0})$
and
$$
\UT{n}_{c}(x_{0})\equiv\tN(x_{0})\utep_{cab}\dot{\alpha}^{a}
(s_{0})\dot{\beta}^{b}(t_{0})|\utep_{abc}\dot{\alpha}^{a}(s_{0})
\dot{\beta}^{b}(t_{0})\dot{\gamma}^{c}(u_{0})|^{-1}.
$$
One might claim that the $O(\ep)$ part between the braces in
eq.(\ref{eq:hamikink2}) is in fact crucial because this part yields
the $O(1)$ contribution if we naively take the limit of
sending $\ep$ to zero. However, this part does not survive
if we consider the \lq\lq flux-tube regularization"\cite{jacob}
in which each curve is replaced by some extended object
such as a ribbon or a tube. This is expected to hold also
when we consider the \lq\lq extended loop" regularization\cite{bart}
in which the distributional expression $\int ds\delta^{3}(x,
\alpha(s))\dot{\alpha}^{a}(s)$ is replaced by a smooth
vector field $X^{a}(x)$ subject to some defining equations.
Taking these into account it seems to be appropriate to
consider the \lq\lq renormalized" scalar constraint
\cite{RS}\cite{bren}\cite{bori}:
\beq
\hat{\CS}^{ren}(\tN)\equiv\lim_{\ep\rightarrow0}
16\ep\hat{\CS}^{\ep}(\tN) \qquad
\left(\hat{\CS}^{ren}_{I}(\tN)\equiv\lim_{\ep\rightarrow0}
16\ep\hat{\CS}^{\ep}_{I}(\tN)\quad(I=1,2)\right).
\label{eq:renohami}
\eeq
Now by using the identity (\ref{eq:third}), we can further simplify
eq.(\ref{eq:hamikink}) and obtain the final result
\beqy
\hat{\CS}^{ren}_{1}(\tN)h_{\alpha_{1}\cdot\beta_{2}}[0,1]^{A}\LI{B}
&=&\{h_{\alpha}[0,s_{0}]\UT{n}\cdot\tB(x_{0})h_{\beta}[t_{0},1]\}
^{A}\LI{B} \nonumber \\
&=&\UT{n}_{c}(x_{0})\frac{1}{2}\otep^{abc}\Delta_{ab}
(\alpha_{1}\cdot\beta_{2},x_{0})
h_{\alpha_{1}\cdot\beta_{2}}[0,1]^{A}\LI{B}.
\eeqy
$\Delta_{ab}$ here stands for the \lq\lq area derivative"
\cite{migdal} acting on the functional of graphs which is defined
by\footnote{We regard $\alpha$ as a part of the graph under
consideration. $\sigma^{ab}\equiv\frac{1}{2}
\int_{\gamma_{x_{0}}}x^{a}dx^{b}$ is the coordinate area element
of the small loop $\gamma_{x_{0}}$.}
\beq
\sigma^{ab}\Delta_{ab}(\alpha,x_{0})\Psi[\alpha,\cdots]\equiv
\Psi[\alpha_{1}\cdot\gamma_{x_{0}}\cdot\alpha_{2},\cdots]-
\Psi[\alpha,\cdots]+O((\sigma^{ab}\sigma^{ab})^{\frac{3}{4}}).
\eeq
for arbitrary small loop $\gamma_{x_{0}}$ based at the point
$x_{0}$.

Action on the other configurations
can be calculated similarly.
We list the action of $\hat{\CS}^{ren}(\tN)$ on all the basic
configurations in Appendix C. There we make
use of the graphical representation.

Since we have at hand the basic action of the scalar constraint,
it is not difficult to calculate its action
on general spin network states. The idea is the following.
Appendix C tells us that the action of $\hat{\CS}^{ren}(\tN)$ has
nonvanishing contributions only at vertices.
Because the action of $\hat{\CS}^{ren}(\tN)$ is local, 
we can separate its action on the individual vertices, i.e.
\beq
\hat{\CS}(\tN)\Psi_{\Gamma}(A)=\sum_{v\in\{v\}\subset\Gamma}
\left(\hat{\CS}^{ren}(\tN)\Psi_{\Gamma}(A)\right)|_{v},
\label{eq:individual}
\eeq
where $\Psi_{\Gamma}(A)$ is the wave function on
$\overline{\CA/\CG}$ which can be expressed by a superposition
of spin network states defined on a graph $\Gamma$.

If we use the separation (\ref{eq:sepa}),
we can easily evaluate the action
on each vertex $v$.
First we add up the contributions from all the kinks
and thus obtain the action of
$\hat{\CS}^{ren}_{1}(\tN)$ on the vertex $v$.
We then compute the sum of the action
of $\hat{\CS}^{ren}_{2}(\tN)$ on all
the pairs of parallel propagators.
The total sum of these contributions yields
the action of $\hat{\CS}^{ren}(\tN)$ at the vertex $v$.
Symbolically we can write:
\begin{eqnarray}
\left(\hat{\CS}^{ren}(\tN)\Psi_{\Gamma}(A)\right)|_{v}&=&
\sum_{k\in K_{v}}\left(\begin{array}{l}
\mbox{the action of $\hat{\CS}^{ren}_{1}(\tN)$}\\
\mbox{ on a kink $k$}\end{array}\right)\nonumber \\
& &\qquad+\sum_{p\in P_{v}}\left(\begin{array}{l}
\mbox{the action of $\hat{\CS}^{ren}_{2}(\tN)$}\\
\mbox{on a pair $p$ of propagators}\end{array}\right),
\label{eq:decomposition}
\end{eqnarray}
where $K_{v}$ ($P_{v}$) is the set of all the kinks (all the pairs
of parallel propagators) at the vertex $v$.
Using eqs.(\ref{eq:individual}),
(\ref{eq:decomposition}) and the equations
in Appendix C, we can therefore
reduce the problem of evaluating the
action of the Hamiltonian constraint
on the spin network states to that of combinatorics.
This applies also to more general operators which
involve the product of a finite number of momenta $\tpi^{i}_{a}$.

We see that the action of $\hat{\CS}^{ren}(\tN)$ defined above
is covariant under spatial diffeomorphisms {\em assuming that
the virtual curves are also subject to the
distortion $\gamma\rightarrow \gamma\circ\phi^{-1}$ under the
action of the integrated diffeomorphism constraint
(\ref{eq:spidiff}).} This regularization is, however,
explicitly depends on the artificial structure,
namely on the choice of the virtual curve and on its
parametrization, through $\UT{n}_{a}$.
In this sense our analysis given here is only heuristic.
In order to obtain more physically solid results, a more
justifiable regularization
method is longed for which does not depends on any artificial
structure and therefore which respects diffeomorphism
covariance of the scalar constraint.


\subsection{Topological solutions}

Now that we have obtained the basic action of the
renormalized scalar constraint $\hat{\CS}^{ren}(\tN)$,
let us construct solutions to the renormalized
Wheeler-De Witt equation (WD equation)
\beq
\hat{\CS}^{ren}(\tN)\Psi[A]=0.\label{eq:renoWD}
\eeq

As we have seen in the last subsection and in Appendix C,
the nonvanishing action of $\hat{\CS}^{ren}(\tN)$ on
spin network states necessarily involves the
area derivative $\Delta_{ab}$ at the vertex $v$.
Hence the action of $\hat{\CS}^{ren}(\tN)$
vanishes if the result of the area
derivative vanishes everywhere,
namely if the $SL(2,\BC)$ connection is flat:
\beq
\tB^{ci}(x)\equiv\frac{1}{2}\otep^{abc}F^{i}_{ab}(x)=0.
\eeq
We therefore find the following \lq distributional' solution
called \lq\lq topological solution"\cite{bren}:
\beq
\Psi_{topo}[A]\equiv
\psi[A]\prod_{x\in\M3}\prod_{a,i}\delta(\tB^{ai}(x)),
\label{eq:7topological}
\eeq
where $\psi[A]$ is an arbitrary gauge-invariant
functional of the $SL(2,\BC)$ connection $A^{i}_{a}$.
Because it has a
support only on flat connections,
the topological solution represented
in terms of spin network states depends only on
homotopy classes of the colored graphs.
As a corollary, it follows that
the topological solution is
invariant under diffeomorphisms and so
it is a solution to all the constraint.
A more detailed analysis on these topological solutions
will be made in \S 6.


\subsection{Analytic loop solutions}

One of the important results obtained in \S\S 4.1 is that the
nonvanishing contributions to the action of $\hat{\CS}^{ren}(\tN)$
are only from vertices, i.e. the points where the analyticity
of the curves breaks down.
From this result we realize that, if we consider
the spin network states consisting only
of Wilson loops evaluated along
smooth loops $\{\alpha_{i}\}$ ($i=1,\cdots,I$)
without any intersection:
\beq
\Psi_{\{(\alpha_{i},\pi_{i})\}}[A]=
\prod_{i=1}^{I}W(\alpha_{i},\pi_{i}),\label{eq:trivial}
\eeq
then these states solve the renormalized WD
equation\cite{jacob}. These states
are related with the solutions to the Hamiltonian constraint which
have been found in the loop representation\cite{RS}.


\subsection{Combinatorial solutions}

If we look at eqs.(C.2), (C.3) and (C.8) carefully,
we find that the following linear combination of spinor
propagators has a vanishing action of $\hat{\CS}^{ren}(\tN)$:
\beq
h_{\alpha}[0,1]^{A}\LI{B}h_{\beta}[0,1]^{C}\LI{D}-2
h_{\alpha_{1}\cdot\beta_{2}}[0,1]^{A}\LI{D}
h_{\beta_{1}\cdot\alpha_{2}}[0,1]^{C}\LI{B}.\label{eq:jacob}
\eeq
This is the solution found by
Jacobson and Smolin\cite{jacob}.
We therefore expect that, even if the graph has
vertices, some appropriate linear
combinations of spin network states
defined on a graph may solve the Hamiltonian constraint.
Here we will find such \lq\lq combinatorial solutions"
which yield a set of extended versions of eq.(\ref{eq:jacob})
to arbitrary finite dimensional representations.
We have seen in eq.(\ref{eq:individual})
that $\hat{\CS}^{ren}(\tN)$ acts on each vertices independently.
In order to be a solution to the renormalized WD equation,
the spin network state must solve this equation
{\em at every vertex.} We can construct a solution
by looking for  intertwining operators
which give the vanishing action of $\hat{\CS}^{ren}(\tN)$ {\em at
each vertex}, and by gluing these solutions at adjacent
(but separate) vertices using an
adequate parallel propagator as a glue.
Thus it is sufficient to concentrate only on one vertex,
say, at $x_{0}$.

Let us now demonstrate explicitly that there exists a set
of solutions to the renormalized WD equation
which are extensions of Jacobson and Smolin's solution
to spin network states defined on
a regular four-valent graph\cite{ezawa3}.

In order to do so we first define the following two sets of
functionals:
\begin{eqnarray}
C^{r}(m,n)^{A_{1}\cdots A_{m};C_{1}\cdots C_{n}}
_{B_{1}\cdots B_{m};D_{1}\cdots D_{n}}&\!\!\!\equiv\!\!\!&
\prod_{i=1}^{m}(h_{\alpha}[0,s_{0}]^{A_{i}}\LI{E_{i}}
h_{\alpha}[s_{0},1]^{F_{i}}\LI{B_{i}})
\prod_{j=1}^{n}(h_{\beta}[0,t_{0}]^{C_{j}}\LI{G_{j}}
h_{\beta}[t_{0},1]^{H_{j}}\LI{D_{j}}) \nonumber \\
\times\sum_{\scriptsize\begin{array}{l}
1\leq k_{1}<\cdots<k_{r}\leq m\\
1\leq l_{1}<\cdots<l_{r}\leq n
\end{array}}
\sum_{\sigma\in P_{r}}&\!\!\!{}\!\!\!&\left(
\prod_{i=1}^{r}
\delta^{E_{k_{i}}}_{H_{l_{\sigma_{i}}}}
\delta^{G_{l_{\sigma_{i}}}}_{F_{k_{i}}}
\prod_{\scriptsize\begin{array}{c}
k^{\prime}=1\\ k^{\prime}\neq k_{1},\cdots,k_{r}
\end{array}}^{m}\delta^{E_{k^{\prime}}}_{F_{k^{\prime}}}
\prod_{\scriptsize\begin{array}{c}
l^{\prime}=1\\ l^{\prime}\neq l_{1},\cdots,l_{r}
\end{array}}^{n}\delta^{G_{l^{\prime}}}_{H_{l^{\prime}}}
\right)\label{eq:configBC} \\
B^{r}(m,n)^{A_{1}\cdots A_{m};C_{1}\cdots C_{n}}
_{B_{1}\cdots B_{m};D_{1}\cdots D_{n}}&\!\!\!\equiv\!\!\!&
\prod_{i=1}^{m}(h_{\alpha}[0,s_{0}]^{A_{i}}\LI{E_{i}}
h_{\alpha}[s_{0},1]^{F_{i}}\LI{B_{i}})
\prod_{j=1}^{n}(h_{\beta}[0,t_{0}]^{C_{j}}\LI{G_{j}}
h_{\beta}[t_{0},1]^{H_{j}}\LI{D_{j}})
\nonumber \\*
&\!\!\!\times\!\!\!&\sum_{\scriptsize\begin{array}{l}
1\leq k_{1}<\cdots<k_{r}\leq m\\
1\leq l_{1}<\cdots<l_{r}\leq n\end{array}}
\sum_{\sigma\in P_{r}}\sum_{i=1}^{r}\left\{
\prod_{\scriptsize\begin{array}{l}
j=1\\j\neq i\end{array}}^{r}
\delta^{E_{k_{i}}}\LI{H_{l_{\sigma_{i}}}}
\delta^{G_{l_{\sigma_{i}}}}\LI{F_{k_{i}}}\right.
\nonumber \\
&\!\!\!{}\!\!\!&\times(\UT{n}\cdot
\tB^{E_{k_{i}}}\LI{H_{l_{\sigma_{i}}}}
\delta^{G_{l_{\sigma_{i}}}}\LI{F_{k_{i}}}
-\delta^{E_{k_{i}}}\LI{H_{l_{\sigma_{i}}}}
\UT{n}\cdot\tB^{G_{l_{\sigma_{i}}}}\LI{F_{k_{i}}})\nonumber \\
&\!\!\!{}\!\!\!&\left.\times\prod_{\scriptsize\begin{array}{c}
k^{\prime}=1\\ \mbox{\tiny $k^{\prime}\neq k_{1},\cdots,k_{r}$}
\end{array}}^{m}\delta^{E_{k^{\prime}}}\LI{F_{k^{\prime}}}
\prod_{\scriptsize\begin{array}{c}
l^{\prime}=1\\ \mbox{\tiny$l^{\prime}\neq l_{1},\cdots,l_{r}$}
\end{array}}^{n}\delta^{G_{l^{\prime}}}\LI{H_{l^{\prime}}}\right\}.
\nonumber
\end{eqnarray}
$P_{r}$ in the above expression is the group of permutations of
$r$ numbers $(l_{1},\cdots,l_{r})$.
We should notice that $B^{0}(m,n)=B^{\min(m,n)+1}(m,n)=0$.
Intuitively, the $C^{r}(n,m)$-functional is obtained as follows:
Prepare $m$ propagators $h_{\alpha}$ along $\alpha$ and $n$
propagators $h_{\beta}$ along $\beta$; then choose $r$ pairs of
$h_{\alpha}$ and $h_{\beta}$; cut and reglue
each pair in the orientation
preserving fashion; and finally sum up the
results obtained from all the
choices of $r$ pairs. The $B^{r}(m,n)$-functional is the result of
the action of $\hat{\CS}^{ren}_{1}(\tN)$
on the $C^{r}(m,n)$-functional.

We will sometimes omit the \lq external'
spinor indices each of which
the tip or the tail of a parallel propagator
$h_{\alpha}[0,1]$ or $h_{\beta}[0,1]$ is equipped with,
because they only serve as the labels of the propagators and  do
not play any essential role in the following calculation.

By using eqs.(C.2-8) and the following useful identity

\vspace{2.4in}

\begin{figure}[ht]
\begin{picture}(150,70)
\put(0,40){\usebox{\LBRA}}
\put(5,55){\line(1,0){8}}
\put(17,55){\vector(1,0){8}}
\put(13,45){\line(0,1){8}}
\put(13,55){\vector(0,1){10}}
\put(17,45){\line(0,1){10}}
\put(17,57){\vector(0,1){8}}
\put(13,53){\line(1,1){4}}
\put(17,55){\circle*{1.5}}
{\scriptsize
\put(5,55){$A$}
\put(25,55){$B$}
\put(13,42){$C$}
\put(13,65){$D$}
\put(17,42){$E$}
\put(17,65){$F$}}
\put(25,40){\makebox(10,30){$-$}}
\put(35,55){\line(1,0){8}}
\put(47,55){\vector(1,0){8}}
\put(43,45){\line(0,1){8}}
\put(43,55){\vector(0,1){10}}
\put(47,45){\line(0,1){10}}
\put(47,57){\vector(0,1){8}}
\put(43,53){\line(1,1){4}}
\put(43,55){\circle*{1.5}}
{\scriptsize
\put(35,55){$A$}
\put(55,55){$B$}
\put(43,42){$C$}
\put(43,65){$D$}
\put(47,42){$E$}
\put(47,65){$F$}}
\put(55,40){\makebox(10,30){$+$}}
\put(65,55){\line(1,0){8}}
\put(77,55){\vector(1,0){8}}
\put(73,45){\line(0,1){6}}
\put(73,57){\vector(0,1){8}}
\put(77,45){\line(0,1){8}}
\put(77,59){\vector(0,1){6}}
\put(73,55){\line(1,1){4}}
\put(73,51){\line(1,1){4}}
\put(73,57){\line(1,-1){4}}
\put(77,55){\circle*{1.5}}
{\scriptsize
\put(65,55){$A$}
\put(85,55){$B$}
\put(73,42){$C$}
\put(73,65){$D$}
\put(77,42){$E$}
\put(77,65){$F$}}
\put(85,40){\makebox(10,30){$-$}}
\put(95,55){\line(1,0){8}}
\put(107,55){\vector(1,0){8}}
\put(103,45){\line(0,1){6}}
\put(103,57){\vector(0,1){8}}
\put(107,45){\line(0,1){8}}
\put(107,59){\vector(0,1){6}}
\put(103,55){\line(1,1){4}}
\put(103,51){\line(1,1){4}}
\put(103,57){\line(1,-1){4}}
\put(103,55){\circle*{1.5}}
{\scriptsize
\put(95,55){$A$}
\put(115,55){$B$}
\put(103,42){$C$}
\put(103,65){$D$}
\put(107,42){$E$}
\put(107,65){$F$}}
\put(115,40){\usebox{\RBRA}}

\put(0,0){\makebox(10,30){$=$}}
\put(10,0){\usebox{\LBRA}}
\put(15,16){\line(1,0){7}}
\put(24,15){\vector(1,0){11}}
\put(24,5){\line(0,1){10}}
\put(27,5){\vector(0,1){20}}
\put(22,16){\vector(0,1){9}}
\put(24,15){\circle*{1.5}}
{\scriptsize
\put(15,16){$A$}
\put(35,15){$B$}
\put(24,2){$C$}
\put(22,25){$D$}
\put(27,2){$E$}
\put(27,25){$F$}}
\put(35,0){\makebox(10,30){$-$}}
\put(45,16){\line(1,0){7}}
\put(54,15){\vector(1,0){11}}
\put(54,5){\line(0,1){10}}
\put(57,5){\vector(0,1){20}}
\put(52,16){\vector(0,1){9}}
\put(52,16){\circle*{1.5}}
{\scriptsize
\put(45,16){$A$}
\put(65,15){$B$}
\put(54,2){$C$}
\put(52,25){$D$}
\put(57,2){$E$}
\put(57,25){$F$}}
\put(65,0){\makebox(10,30){$+$}}
\put(75,15){\line(1,0){11}}
\put(88,14){\vector(1,0){7}}
\put(88,5){\line(0,1){9}}
\put(83,5){\vector(0,1){20}}
\put(86,15){\vector(0,1){10}}
\put(88,14){\circle*{1.5}}
{\scriptsize
\put(75,15){$A$}
\put(95,14){$B$}
\put(83,2){$C$}
\put(83,25){$D$}
\put(88,2){$E$}
\put(86,25){$F$}}

\put(95,0){\makebox(10,30){$-$}}
\put(105,15){\line(1,0){11}}
\put(118,14){\vector(1,0){7}}
\put(118,5){\line(0,1){9}}
\put(113,5){\vector(0,1){20}}
\put(116,15){\vector(0,1){10}}
\put(116,15){\circle*{1.5}}
{\scriptsize
\put(105,15){$A$}
\put(125,14){$B$}
\put(113,2){$C$}
\put(113,25){$D$}
\put(118,2){$E$}
\put(116,25){$F$}}
\put(125,0){\usebox{\RBRA}}
\put(130,0){\makebox(15,30){ \ ,}}
\end{picture}
\end{figure}
\hspace{-0.3in}which is easily proved by means
of the two-spinor identity
(\ref{eq:2spi}), and by performing somewhat lengthy calculations
on permutations and combinations, we find
\beq
\hat{\CS}^{ren}(\tN)C^{r}(m,n)
=(m+n-2r+1)B^{r}(m,n)+2B^{r+1}(m,n).
\label{eq:identity3}
\eeq
Now it is an elementary exercise of the linear algebra
to derive from eq.(\ref{eq:identity3}) the equation
\beq
\hat{\CS}^{ren}(\tN)\left(
\sum_{r=0}^{\min(m,n)}\left(\prod_{i=1}^{r}\frac{-2}{m+n-2i+1}
\right)C^{r}(m,n)\right)=0.
\eeq

Thus we have found a set of solutions to the renormalized
WD equation {\em at one vertex $x_{0}=\alpha(s_{0})=\beta(t_{0})$}:
\beq
\prod_{i=1}^{m}(h_{\alpha}[0,s_{0}]^{A_{i}}\LI{E_{i}}
h_{\alpha}[s_{0},1]^{F_{i}}\LI{B_{i}})
\prod_{j=1}^{n}(h_{\beta}[0,t_{0}]^{C_{j}}\LI{G_{j}}
h_{\beta}[t_{0},1]^{H_{j}}\LI{D_{j}})\times
I^{\circ}(m,n)^{E_{1}\cdots E_{m};G_{1}\cdots G_{n}}
_{F_{1}\cdots F_{m};H_{1}\cdots H_{n}},\label{eq:combisol}
\eeq
where $I^{\circ}(m,n)$ is the relevant intertwining operator:
\begin{eqnarray}
I^{\circ}(m,n)^{E_{1}\cdots E_{m};G_{1}\cdots G_{n}}
_{F_{1}\cdots F_{m};H_{1}\cdots H_{n}}&\equiv&
\sum_{r=0}^{\min(m,n)}
\left(\prod_{i=1}^{r}\frac{-2}{m+n-2i+1}\right)
I^{r}(m,n)^{E_{1}\cdots E_{m};G_{1}\cdots G_{n}}
_{F_{1}\cdots F_{m};H_{1}\cdots H_{n}},
\nonumber \\*
I^{r}(m,n)^{E_{1}\cdots E_{m};G_{1}\cdots G_{n}}
_{F_{1}\cdots F_{m};H_{1}\cdots H_{n}}&\equiv&
\sum_{\scriptsize\begin{array}{l}
1\leq k_{1}<\cdots<k_{r}\leq m\\
1\leq l_{1}<\cdots<l_{r}\leq n
\end{array}}
\sum_{\sigma\in P_{r}}\nonumber \\*
&\times&\left(
\prod_{i=1}^{r}
\delta^{E_{k_{i}}}_{H_{l_{\sigma_{i}}}}
\delta^{G_{l_{\sigma_{i}}}}_{F_{k_{i}}}
\prod_{\scriptsize\begin{array}{c}
k^{\prime}=1\\ k^{\prime}\neq k_{1},\cdots,k_{r}
\end{array}}^{m}\delta^{E_{k^{\prime}}}_{F_{k^{\prime}}}
\prod_{\scriptsize\begin{array}{c}
l^{\prime}=1\\ l^{\prime}\neq l_{1},\cdots,l_{r}
\end{array}}^{n}\delta^{G_{l^{\prime}}}_{H_{l^{\prime}}}
\right).\label{eq:intertwiner}
\end{eqnarray}

We should note that eq.(\ref{eq:combisol})
remains to be the solution even if
we permutate the external indices.
Moreover, from eq.(\ref{eq:2spi2}),
we see that the antisymmetrized tensor product
of two spinor propagators
yields the invariant spinor. As a result
we have only to consider symmetrized tensor products
of the spinor propagators, say
$$
h_{\alpha}[0,s_{0}]^{A_{1}}\LI{(E_{1}}\cdots
h_{\alpha}[0,s_{0}]^{A_{m}}\LI{E_{m})}.
$$
Taking these into account we can provide the procedure
to construct a combinatorial solution defined on a
regular four-valent graph $\Gamma^{reg}$:\\
1) Extract from $\Gamma^{reg}$ all the smooth loops $\alpha_{i}:[0,1]\rightarrow\M3$
$(i=1,2,\cdots,N; \alpha_{i}(0)=\alpha_{i}(1))$.\\
2) Equip each loop $\alpha_{i}$ with a Wilson loop in the
spin-$\frac{p_{i}}{2}$ representation, which is given by the
symmetrized trace of $p_{i}$ spinor propagators:
$$
h_{\alpha_{i}}[0,1]^{A_{1}}\LI{(A_{1}}\cdots
h_{\alpha_{i}}[0,1]^{A_{p_{i}}}\LI{A_{p_{i}})}.
$$
3) At each point where two loops, say,
$\alpha_{i}$ and $\alpha_{j}$
intersect, cut the spinor Wilson loops and
rejoin them by using as a glue
the intertwining operator $I^{\circ}(p_{i},p_{j})$ defined by
eq.(\ref{eq:intertwiner}).

We will denote the combinatorial solution constructed by
this procedure as
$\Psi^{\Gamma^{reg}}_{\{p_{i}\}}[A]$.
If we fix a regular four-valent graph $\Gamma^{reg}$
which consists of loops $\{\alpha_{i}\}$ $(i=1,\cdots,N)$,
we can construct $({\bf Z}_{+})^{N}$ combinatorial solutions
where ${\bf Z}_{+}$ is the set of non-negative integers.
The analytic loop solutions described in the last subsection
are contained in the combinatorial solutions
$\Psi^{\Gamma^{reg}}_{\{p_{i}\}}[A]$
as particular cases
in which all the loops $\alpha_{i}$ in the graph $\Gamma^{reg}$
do not intersect.

Extension of Jacobson and Smolin's solutions to the case
where three or more smooth curves intersect at a point
was explored in ref.\cite{HBP} and several
extended solutions were found in the spinor
representation. While we will not explicitly
examine here, it is in principle possible to generalize these
extended solutions to arbitrary finite dimensional
representations. In order to carry out this, we need the action
of the renormalized scalar constraint on vertices at which
three or more smooth curves intersect.


\subsection{Issues on the combinatorial solutions}

In this section, we have constructed the simplest type of solutions
which are composed of spin network states, namely the combinatorial
solutions $\{\Psi^{\Gamma^{reg}}_{\{p_{i}\}}[A]\}$
each of which are completely determined by fixing
a regular four-valent graph $\Gamma^{reg}$ and a set
of non-negative integers $\{p_{i}\}$ assigned to each smooth
loops in $\Gamma^{reg}$. It turns out, however,
that the classical counterparts of these combinatorial solutions 
are spacetimes with degenerate metric ${\rm det}(q_{ab})=0$.
We can easily see this by the following discussion.

First we are aware that the scalar constraint can be expressed
by the product of the magnetic field $\tB^{ai}$ and the \lq
densitized co-dreibein' $\te^{i}_{a}\equiv\frac{1}{2}\utep_{abc}
\ep^{ijk}\tpi^{bj}\tpi^{ck}$:
\beq
\hat{\CS}=\ep_{ijk}F^{i}_{ab}\hat{\tpi}^{aj}\hat{\tpi}^{bk}=
2\tB^{ai}\hat{\te}^{i}_{a}.\label{dechami}
\eeq
From the way of their construction, we see that the combinatorial
solutions satisfy the following equation
\beq
(\hat{\te}^{i}_{a})^{ren}\Psi^{\Gamma^{reg}}_{\{p_{i}\}}[A]=0,
\label{eq:degenerate}
\eeq
where $(\hat{\te}^{i}_{a})^{ren}$ is the renormalized densitized
co-dreibein
$$
(\hat{\te}^{i}_{a})^{ren}(x)\equiv\lim_{\ep\rightarrow0}
2\ep\int_{\M3}d^{3}y\tilde{f}_{\ep}(x,y)\frac{1}{2}\utep_{abc}
\ep^{ijk}\frac{\delta}{\delta A^{j}_{b}(x)}
\frac{\delta}{\delta A^{k}_{c}(y)}.
$$
On the other hand, the volume operator (\ref{eq:volume})
can be rewritten by using $\hat{\te}^{i}_{a}$
\beq
\hat{V}({\cal R})=\int_{{\cal R}}d^{3}x\left(
\frac{1}{3}\hat{\tpi}^{ai}\hat{\te}^{i}_{a}\right)^{\frac{1}{2}}.
\label{eq:volume2}
\eeq
By comparing eqs.(\ref{eq:degenerate}) and (\ref{eq:volume2})
we conclude that, if there exists a regularization which can deal
with $\hat{\CS}$ and $\hat{V}$ consistently,
the combinatorial solutions correspond to spatial
geometries whose volume element vanishes everywhere in $\M3$.
While at present we do not know such a regularization in the
continuum theory, this situation actually happens for some
lattice version of these operators.

Besides this undesirable feature, there is an argument\cite{mats3}
that the combinatorial solutions are \lq spurious solutions'
which are not genuine solutions to the Wheeler-De Witt equation
$\hat{\CH}\Psi=0$ in quantum general relativity.
The reasoning is the following. Classically the scalar constraint
$\CS$ in Ashtekar's formalism is essentially the volume element
$({\rm det}(\te^{ai}))^{\frac{1}{2}}$ times the Hamiltonian
constraint $\CH$ in the ADM formalism. In the present
operator ordering, $\Psi^{\Gamma^{reg}}_{\{p_{i}\}}[A]$
is eliminated by $\widehat{({\rm det}(\te^{ai}))}^{\frac{1}{2}}$
and not by $\hat{\CH}$. Thus we can consider that the
combinatorial solutions are \lq spurious solutions' which
do not realize correctly the quantum version of
spacetime diffeomorphism invariance.

The above arguments indicate that a correct
solution $\Psi^{corr}[A]$ to the WD equation
must satisfy the conditions
\beq
\hat{\CS}\Psi^{corr}[A]=0\quad\mbox{and}\quad
\hat{\te}^{i}_{a}\Psi^{corr}[A]\neq0.\label{eq:correctsol}
\eeq
In search for these correct solutions in the framework of
spin network states, contributions from the
magnetic field $\tB^{ai}$, or equivalently the area derivative
$\Delta_{ab}$ will play an essential role.
Unfortunately, we do not yet
know how to define the area derivative unambiguously in the continuum
theory. Under this situation, it seems to be useful to investigate
as a heuristic model the lattice discretized formulation,
because it allows us to fix a definition of the area derivative.
So let us explore in the next section a lattice version of
Ashtekar's formalism.


\section{Discretized Ashtekar's formalism}

A lattice discretized version of Ashtekar's formalism
was first proposed by Renteln and Smolin\cite{rente} and
later developed by Loll\cite{loll}\cite{loll2}.
This formulation is based on the idea of
the Hamiltonian lattice gauge theory of Kogut and Susskind\cite{KS}.

We will first explain the setup.
This model is defined on a 3 dimensional
cubic lattice $\Gamma^{N}$ of size N. We will
label lattice sites by $n$ and 
three positive directions of links by $\hat{a}$.
We take the lattice spacing to be $a$.
The $SL(2,\BC)$ connection $A^{i}_{a}(x)$ is replaced by
the link variables $V(n,\hat{a})^{A}\LI{B}\in SL(2,{\bf C})$,
which is regarded as a parallel propagator along a link:
\beqy
V(n,\hat{a})^{A}\LI{B}&=&\CP\exp[a\int_{0}^{1}ds
A^{i}_{\hat{a}}(n+s\hat{a})J_{i}]^{A}\LI{B}\nonumber \\
&=&[{\bf 1}+\frac{a}{2}(A^{i}_{\hat{a}}(n)
+A^{i}_{\hat{a}}(n+\hat{a}))J_{i}+O(a^{2})]^{A}\LI{B}.
\label{eq:latconti}
\eeqy
Next we recall that the conjugate momentum $\hat{\tpi}^{ai}$ acts
on the parallel propagator $h_{\alpha}[0,1]$ as
$$
\hat{\tpi}^{ai}(x)h_{\alpha}[0,1]=-\int_{0}^{1}ds\delta^{3}(x,
\alpha(s))\dot{\alpha}^{a}(s)h_{\alpha}
[0,s]J_{i}h_{\alpha}[s,1].
$$
From this we see that a lattice analog of the conjugate momentum
$\hat{\tpi}^{ia}$ is given by
\beq
\pi^{i}(n,\hat{a})=p_{L}^{i}(n,\hat{a})+p_{R}^{i}(n,\hat{a}),
\label{eq:latmomenta}
\eeq
where $p_{L}^{i}$ and $p_{R}^{i}$ respectively stand for
the left translation and the right translation operators:
\beqy
p_{L}^{i}(n,\hat{a})V(n,\hat{b})^{A}\LI{B}&=&\delta_{\hat{a},
\hat{b}}(J_{i}V(n,\hat{b}))^{A}\LI{B}\nonumber \\
p_{R}^{i}(n,\hat{a})V(n-\hat{b},\hat{b})^{A}\LI{B}&=&
\delta_{\hat{a},\hat{b}}(V(n-\hat{b},\hat{b})J_{i})^{A}\LI{B}.
\eeqy
The relation between momenta in the lattice and continuum theories
is given by
\beq
\pi^{i}(n,\hat{a})=-2a^{2}\hat{\tpi}^{\hat{a}i}(n)+O(a^{3}).
\label{eq:latconti2}
\eeq
While redundant for the completeness of the variables,
it is sometimes convenient to introduce link variables in the
negative direction $V(n,-\hat{a})\equiv V(n-\hat{a},\hat{a})^{-1}$.
Then by using the relation
\beq
p_{R}^{i}(n,\hat{a})=-p_{L}^{i}
(n,-\hat{a}),
\eeq
the lattice momenta are expressed in terms
only of the left translation
\beq
\pi^{i}(n,\hat{a})=p_{L}^{i}(n,\hat{a})-p_{L}^{i}(n,-\hat{a}).
\label{eq:latmomenta2}
\eeq
The lattice counterpart of the curvature $F^{i}_{ab}$ is provided
by plaquette loop variables:
\beqy
V(n,P_{\hat{a}\hat{b}})^{A}\LI{B}
&\equiv& [V(n,\hat{a})V(n+\hat{a},\hat{b})
V(n+\hat{a}+\hat{b},-\hat{a})V(n+\hat{b},-\hat{b})]^{A}\LI{B}
\nonumber \\
&=&[{\bf 1}+a^{2}F^{i}_{\hat{a}\hat{b}}(n)J_{i}+O(a^{3})]
^{A}\LI{B}. \label{eq:plaquette}
\eeqy

In order to formulate a lattice version of Ashtekar's formalism,
we have to translate the constraint operators $\hat{G}^{i}$,
$\hat{\CD}_{a}$ and $\hat{\CS}$ in terms of lattice variables.
Among them, Gauss' law constraint $\hat{G}^{i}$ is solved
by considering only gauge-invariant
functionals of link variables, namely, spin network states
defined on the lattice $\Gamma^{N}$.
While it is difficult in general to impose
the diffeomorphism constraint (\ref{eq:qdiffeo}) in lattice
formulations, we formally solve this constraint by regarding our
lattice to be a purely topological object\cite{loll}.
This is based on the idea of diffeomorphism invariant
states explained in \S\S 3.4. Thus we are left only with the scalar
constraint $\hat{\CS}$.

A plausible candidate for the discretized version of the scalar
constraint which respects the symmetry under
$\frac{\pi}{2}$ rotations around the $\hat{a}$-axis is given by
\cite{ezawa4}
\beqy
\hat{\CS}_{I}(n)&\equiv&\sum_{\hat{a}<\hat{b}}\ep^{ijk}
\Tr(-\widetilde{V}(n,\hat{a}\hat{b})J_{k})\pi^{i}(n,\hat{a})
\pi^{j}(n,\hat{b})\nonumber \\
&=&a^{6}\hat{\CS}(n)+O(a^{7}),\label{eq:scalarI}
\eeqy
where $\widetilde{V}(n,\hat{a}\hat{b})\equiv\frac{1}{4}
(V(n,P_{\hat{a}\hat{b}})+V(n,P_{\hat{b},-\hat{a}})+
V(n,P_{-\hat{a},-\hat{b}})+V(n,P_{-\hat{b},\hat{a}}))$
is the \lq averaged' plaquette loop variable.
The action of $\hat{\CS}_{I}$ on lattice spin network
states can be described in the form which is analogous to
that of $\hat{\CS}^{ren}$ in the last section. The only difference
is that the insertion of $\UT{n}\cdot\tB$ in the latter is replaced
by the insertion of $\frac{1}{4}(\widetilde{V}-\widetilde{V^{-1}})$.
The discretized WD equation using $\hat{\CS}_{I}$ therefore
has as its solutions the lattice version of combinatorial solutions
$\Psi^{\Gamma^{N}}_{\{p_{i}\}}[V]$ in which the smooth loops
are replaced by straight Polyakov loops \cite{loll}.

However we are interested only in finding a clue to
construct \lq correct'
solutions which become solutions only after
the action of the area derivative is taken into account.
For this purpose $\hat{\CS}_{I}$ does not seem to be suitable
because it contains $16\times3$ terms at a vertex.
So we will henceforth use in our analysis
the \lq truncated' scalar constraint
\beq
\hat{\CS}_{II}(n)=\sum_{\hat{a}<\hat{b}}
\sum_{\eta_{1},\eta_{2}=\pm}
\ep^{ijk}\Tr(-V(n,P_{\eta_{1}\hat{a},\eta_{2}\hat{b}})J_{k})
p_{L}^{i}(n,\eta_{1}\hat{a})p_{L}^{j}(n,\eta_{2}\hat{b}).
\label{eq:scalarII}
\eeq
The number of terms appearing in $\hat{\CS}_{II}$ is a quarter
of that in $\hat{\CS}_{I}$.
This enables us to simplify the analysis.

We do not think that our discretized WD equation yield
solutions which become in the continuum limit genuine
solutions to the continuum WD equation. From our analysis, however,
we expect to extract some instructive information on the correct
solutions.

We can define measures in the lattice formulation
in quite an analogous manner to that in which
the measures for spin network states are defined.
For example, the induced Haar measure is defined by
eq.(\ref{eq:HAAR}) with the edges $\{e\}$in the graph $\Gamma$ being
replaced by the links $\{\ell\}$ in the lattice $\Gamma^{N}$.
The induced heat-kernel measure is also defined in a similar way by
using eq.(\ref{eq:HKmeasure}). In this case, however, we need to
fix a length function $l(\ell)$. A convenient and reasonable choice
is to set $l(\ell)=a$. In the following we will adopt this choice.


\subsection{Multi-plaquette solutions}

Let us now search for non-trivial solutions
to the truncated WD equation\cite{ezawa4}
\beq
\hat{\CS}_{II}(n)\Psi[V]=0\quad\mbox{for}\quad^{\forall}
n\in\Gamma^{N}. \label{eq:trunwd}
\eeq
For this aim we have to evaluate the action of $\hat{\CS}_{II}(n)$
on lattice spin network states. This is executed in essentially
the same way as that of calculating the action of
$\hat{\CS}^{ren}(\tN)$. In particular, we can separate its action as
\beq
\hat{\CS}_{II}(n)=\hat{\CS}_{II}(n)_{1}+\hat{\CS}_{II}(n)_{2},
\eeq
where $\hat{\CS}_{II}(n)_{1}$ and $\hat{\CS}_{II}(n)_{2}$ denote
respectively the action on the single link-chains and that on
the pairs of link-chains.

For example, the action on kinks $V(n-\hat{a},\hat{a})V(n,\hat{b})$
($\hat{a}\neq\hat{b}$) is computed as follows
\begin{eqnarray}
\hat{\CS}_{II}(n)_{1}\cdot
\{V(n-\hat{a},\hat{a})V(n,\hat{b})\}^{A}\LI{B}
&\!\!\!\!=&\!\!\!\!
\frac{1}{4}\{V(n-\hat{a},\hat{a})
(V(n,P_{\hat{b},-\hat{a}})-
V(n,P_{-\hat{a},\hat{b}}))V(n,\hat{b})\}^{A}\LI{B},\nonumber \\
\hat{\CS}_{II}(n)_{2}\cdot\left(
\{V(n-\hat{a},\hat{a})V(n,\hat{b})\}\right.
&\!\!\!\!^{A}\LI{B}&\!\!\!\!\left.\{
V(n-\hat{a},\hat{a})V(n,\hat{b})\}^{C}\LI{D}\right)=0.
\label{eq:trukink}
\end{eqnarray}
Topologically, the former action can be interpreted as taking the
difference of the results of inserting plaquettes with the
opposite orientations (figure 4).
The action on the other types of vertices can also be interpreted in
terms of combinatorial topology of the lattice graphs.

\begin{figure}[t]
\begin{center}
\epsfig{file=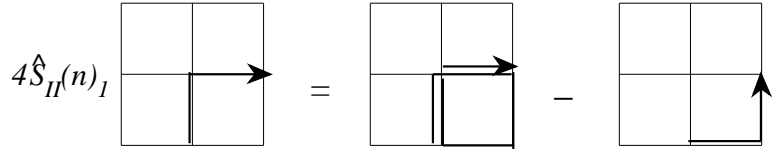,height=3.5cm}
\end{center}
\caption{The action of the truncated scalar constraint on a single
kink.}
\end{figure}

Now we are in a position to construct the simplest \lq\lq nontrivial
solutions" on which the action of the area derivative essentially
contributes to the vanishing action of $\hat{\CS}_{II}$.

We first consider the action on the trace of the $p$-th power of a
plaquette. This is calculated by using eq.(\ref{eq:trukink})
\beq
\hat{\CS}_{II}(n)\cdot\Tr( V(n,P_{\hat{a}\hat{b}})^{p})=
\frac{p}{4}\left(\Tr( V(n,P_{\hat{a}\hat{b}})^{p+1})
-\Tr( V(n,P_{\hat{a}\hat{b}})^{p-1})\right).
\eeq
This equation is reinterpreted as
\beq
\hat{\CS}_{II}(n)\cdot\Tr F(V(n,P_{\hat{a}\hat{b}}))=\Tr
\left[\frac{V(n,P_{\hat{a}\hat{b}})^{2}-1}{4}
\frac{d}{dV}F(V(n,P_{\hat{a}\hat{b}}))\right],
\eeq
where $F$ denotes an arbitrary polynomial.
We can readily extend this equation
to the case where $F$ is a function which can be expressed by
a Laurent series. Thus we find
\beq
\hat{\CS}_{II}(n)\cdot\Tr
\log\left(\frac{1-V(n,P_{\hat{a}\hat{b}})}
{1+V(n,P_{\hat{a}\hat{b}})}\right)=1.\label{eq:log}
\eeq
As for the action of $\hat{\CS}_{II}(m)$
with $m\neq n$, the following can be said. When $m$ coincides with
one of the vertices of the plaquette $P_{\hat{a}\hat{b}}$,
the result is identical to eq.(\ref{eq:log}) owing to the symmetry
of $\hat{\CS}_{II}$. When $m$ does not coincide, on the other hand,
the action necessarily vanishes.
These results are summarized as
\beq
\hat{\CS}_{II}(m)\cdot
\Tr\log\left(\frac{1-V(n,P_{\hat{a}\hat{b}})}
{1+V(n,P_{\hat{a}\hat{b}})}\right)=\left\{
\begin{array}{cc}2&
\mbox{for $m=n,n+\hat{a},n+\hat{b},n+\hat{a}+\hat{b}$,}\\
0&\mbox{for $m\neq n,n+\hat{a},n+\hat{b},n+\hat{a}+\hat{b}$.}
\end{array}\right. \label{eq:master}
\eeq
Now we can provide the prescription for constructing
\lq\lq multi-plaquette solutions" on which
the action of the area derivative is essential\cite{ezawa4}:
i) prepare a connected set of plaquettes $\{P\}$in which each vertex
belongs to at least two plaquettes; ii) assign to
each plaquette $P$ a weight factor $w(P)$ so that the sum of
weight factors of the plaquettes which meet at each vertex vanishes;
iii) the following expression yields a solution
to eq.(\ref{eq:trunwd})
\beq
<V|\{w(P)\}>\equiv
\sum_{P\in\{P\}}w(P)\Tr\log\left(\frac{1-V(P)}{1+V(P)}\right).
\label{eq:mpla}
\eeq

Now we will investigate whether or not the multi-plaquette solutions
are normalizable w.r.t. some measures on the lattice gauge theory.
It is convenient to rewrite the result by means of the Clebsch-Gordan
decomposition. We see that $\log(\frac{1-V}{1+V})$ is
proportional to the \lq\lq one-plaquette state"\cite{ezawa4}
\beq
<V|\log(\frac{1-P}{1+P})>
\equiv\sum_{p=0}^{\infty}\frac{1}{(2p+3)(2p+1)}
\Tr S(V(P)^{2p+1}),\label{eq:spla}
\eeq
where $\Tr S(V^{m})$ is the symmetrized trace
$$
\Tr S(V^{m})\equiv V^{A_{1}}\LI{(A_{1}}\cdots V^{A_{m}}\LI{A_{m})}
=\Tr\pi_{m}(V).
$$

First we examine the normalizability w.r.t. the induced Haar measure
$d\mu_{H}(V)$.
Owing to the consistency of the measure and because
one-plaquette states defined on different plaquettes are orthogonal
with each other, it is sufficient to investigate the norm
of a one-plaquette state
$$
\|<V|\log(\frac{1-P}{1+P})>\|_{d\mu_{H}}=\int
d\mu_{H}(V(P))|<A|\log(\frac{1-P}{1+P})>|^{2}.
$$
Using the formula (\ref{eq:spihaar}) this is easily evaluated as
\beq
\|<V|\log(\frac{1-P}{1+P})>\|_{d\mu_{H}}=\sum_{p=0}^{\infty}
\frac{1}{(2p+3)^{2}(2p+1)^{2}}<\infty,
\eeq
namely the multi-plaquette solutions are normalizable w.r.t.
the induced Haar measure.

Next we consider the induced heat-kernel measure $d\nu_{t}^{l}(V)$.
The above reduction of the problem holds also in this case.
By taking $l(P)=4a$ into account and by using
eq.(\ref{eq:heatkernel}) we find
\beqy
\|<V|\log(\frac{1-P}{1+P})>\|_{d\nu^{l}_{t}}&\equiv&
\int d\nu_{t}^{l}(V(P))|<V|\log(\frac{1-P}{1+P})>|^{2}\nonumber \\
&=&\sum_{p=0}^{\infty}
\frac{e^{at(2p+3)(2p+1)}}{(2p+3)^{2}(2p+1)^{2}}
\rightarrow\infty.
\eeqy
The multi-plaquette solutions therefore turn out to be
non-normalizable w.r.t. the induced heat-kernel measure.


\subsection{Lessons to the continuum theory}

In this section we have investigated a lattice version of
the WD equation and obtained a set of nontrivial solutions
to the truncated WD equation (\ref{eq:trunwd}), namely,
multi-plaquette solutions (\ref{eq:mpla}).

We should note that, while these solutions are nontrivial
in the sense that they have non-vanishing action of the densitized
co-triad operator, they still correspond to three dimensional
manifolds with degenerate metric. This is because the bi- and
trivalent vertices have zero eigenvalue of the volume operator
\cite{loll2}. In this sense we have achieved only a half of our goal.
In order to find solutions with non-degenerate metric
we have to deal with lattice spin network states containing at
least four-valent vertices. This seems to be a quite hard task
and is left to the future investigation.

We expect, however, that these multi-plaquette solutions already
provide us with some lessons which are helpful in searching for
correct solutions to the continuum WD equation.

We have seen in the last subsection that the multi-plaquette
solutions are not normalizable w.r.t. the induced heat-kernel
measure. This can be traced back to the fact that
each multi-plaquette solution is expressed by a sum of
infinitely many spin network states. This seems to be indispensable
when we want to have a non-trivial mechanism of cancelling the
action of the scalar constraint.
Thus we conjecture that
the correct solutions are, if any, not normalizable
w.r.t. the induced heat-kernel measure on $SL(2,\BC)$
while they may be
normalizable w.r.t. the induced Haar measure on $SU(2)$.
But this should not be taken so seriously, because
the induced heat-kernel measure at most serves as a kinematical
measure as is explained in \S\S 3.3.
Or we may rather use this non-normalizability w.r.t. the induced
heat-kernel measure as a criterion for determining whether
an obtained solution is physical or not.

We can expect that a \lq correct' physical state in the
lattice formulation consists of an infinite number of lattice
spin network states each of which have their \lq partners'.
The partners are different from the original term
by addition or removal of two plaquettes. If we try to
reproduce this situation in the continuum theory,
we have to extend our framework of the spin network states
so that we can deal with graphs which are not piecewise analytic
\footnote{A piecewise analytic graph is a graph which can be
composed of a finite number of smooth edges without intersection.
While most of the results on the spin network states are verified
on the piecewise analytic graphs, it is desirable that they
should be extended to the piecewise smooth graphs or more
general contexts. A piecewise smooth graph may contain smooth
curves with infinite number of intersections in a finite interval.}.
This seems to be a difficult but essential subject left to us. 


\section{Topological solutions}

In the last two sections we have seen that any solutions
which have constructed so far by means of spin network states
do not correspond to spacetimes with non-degenerate metric.
From this one might suspect that canonical quantization of
general relativity do not yield any physically meaningful results
and thus it is not worthwhile investigating Ashtekar's formulation
for quantum gravity much further.
It is the existence of topological solutions which sweeps off
this suspicion. In this section we will look into these
topological solutions focusing on their spacetime geometrical
aspects.


\subsection{$SL(2,\BC)$ BF theory and topological solutions}

In order to see the existence of topological solutions,
it is the easiest to show the relation between Ashtekar's formalism
and $SL(2,\BC)$ BF theory\cite{horo}.
As we have seen in \S\S 2.1,
Ashtekar's formalism is obtained from $SL(2,\BC)$ BF theory
\beq
-iI_{BF}=\int_{M}(\Sigma^{i}\wedge F^{i}+\frac{\Lambda}{6}
\Sigma^{i}\wedge \Sigma^{i}) \label{eq:bfac}
\eeq
by imposing the algebraic constraint (\ref{eq:alcon1})
which is equivalent to
\beq
\Sigma^{i}\wedge\Sigma^{j}=\frac{1}{3}\delta^{ij}
\Sigma^{k}\wedge\Sigma^{k}.\label{eq:alcon2}
\eeq
The first class constraints $(G^{i},{\cal V}_{a},
\CS)$ in Ashtekar's formalism can therefore be expressed as linear
combinations of those $(G^{i},\Phi^{ai})$ in $SL(2,{\bf C})$
BF theory:
\beqy
G^{i}&=&G^{i}\nonumber \\
{\cal V}_{a}&=&{\cal C}_{ab}^{i}\Phi^{bi}\nonumber \\
\CS&=&\widetilde{{\cal C}}_{ai}\Phi^{ai},
\eeqy
where ${\cal C}_{ab}^{i}$ and $\widetilde{{\cal C}}_{ai}$ are
functionals only of the momenta $\tpi^{ai}$.

In consequence, if we take the operator ordering with momenta
to the left, all the solutions to the quantum $SL(2,\BC)$ BF
constraints
\beqy
\hat{G}^{i}\Psi[A]&=&
-D_{a}\left(\frac{\delta}{\delta A^{i}_{a}}\Psi[A]\right)=0
\nonumber \\
\hat{\Phi}^{ai}\Psi[A]&=&\left(\frac{1}{2}\otep^{abc}F^{i}_{bc}
-\frac{\Lambda}{3}\frac{\delta}{\delta A^{i}_{a}}\right)\Psi[A]
=0 \label{eq:BFEQ}
\eeqy
are contained in the solution space of quantum Ashtekar's formalism.
They are the topological solutions, which take different forms
according to whether the cosmological constant $\Lambda$
vanishes or not.

For $\Lambda\neq0$ we have \lq\lq Chern-Simons solutions"
\cite{kodama}\cite{CS}
\beq
\Psi^{CS}_{I}[A]=I[A]\exp\{\frac{3}{2\Lambda}S_{CS}[A]\},
\label{eq:cssol}
\eeq
where $I[A]$ is some topological invariant of the $SL(2,\BC)$
connection and $S_{CS}[A]$ is the Chern-Simons invariant
\beq
S_{CS}[A]\equiv\int_{\M3}\left(A^{i}dA^{i}+\frac{1}{3}\ep^{ijk}
A^{i}\wedge A^{j}\wedge A^{k}\right).\label{eq:csinv}
\eeq
We see from eq.(\ref{eq:cssol}) that the number of linearly
independent Chern-Simons
solutions is equal to that of principal $SL(2,\BC)$ bundles
over $\M3$. This is consistent with the classical result
because the reduced phase space for $SL(2,\BC)$ BF theory
with $\Lambda\neq0$ is a set of discrete points each of which
represents a principal $SL(2,\BC)$ bundle over $\M3$\cite{horo}.

For $\Lambda=0$, all the gauge-invariant
wavefunctions with their support only on
flat connections become solutions to eq.(\ref{eq:BFEQ}).
This was first pointed out by Brencowe\cite{bren}.
Each of these solutions is formally described as
\beq
\Psi_{topo}[A]=\psi[A]\prod_{x\in\M3}\left(\prod_{i,a}\delta
(\otep^{abc}F^{i}_{bc}(x))\right),
\eeq
where $\psi[A]$ is a gauge-invariant functional of the $SL(2,\BC)$
connection. Because the defining region of $\psi[A]$ is reduced to
the space of flat connections, we can regard $\psi[A]$ as a function
on the moduli space $\CN$ of flat $SL(2,\BC)$ connections on $\M3$
modulo small gauge transformations.
This allows us to give an alternative expression for $\Psi_{topo}[A]$
\beqy
\Psi_{topo}[A]&=&\int_{\CN}dn\psi(n)\Psi_{n}[A]\label{eq:Toposol}\\
\Psi_{n}[A]&=&\int[dg(x)]\prod_{x}\prod_{a,i}\delta(A_{a}^{i}(x)=
g[A_{0n}]^{i}_{a}(x)),\label{eq:toposol}
\eeqy
where $A_{0n}$ is a flat $SL(2,\BC)$ connection which represent
a point $n$ on the moduli space $\CN$ and
$$
g[A]^{i}_{a}(x)J_{i}\equiv g(x)A^{i}_{a}(x)J_{i}g^{-1}(x)+g(x)
\partial_{a}g^{-1}(x)
$$
denotes the gauge-transformed $SL(2,\BC)$ connection.
This expression of the topological solutions is essentially
a kind of Fourier transform of the expression given in
ref.\cite{mats}. It was discussed in \cite{mats} that
it is sufficient to take the integration region in
eq.(\ref{eq:toposol}) to be the space
of small $SU(2)$ gauge transformations.

Because the topological solutions are solutions to $SL(2,\BC)$
BF theory, most of their properties can be clarified by investigating
the quantization of this theory. Some of the works of this kind
can be seen, for example, in
\cite{CS}\cite{chang}\cite{lee}\cite{cotta}\cite{baez} and
several interesting results have been found.
However, if we investigate the pure BF theory we cannot extract
any information on the spacetime geometry in which the
general relativitist are the most interested.
In order to investigate spacetime
geometrical feature of the solutions
to quantum gravity, however, we need a physical inner product
and physical observables which measure quantities necessary to
reconstruct spacetimes. This is left to the future investigation
because we do not know any
physical inner product and any such physical
observables. We can nevertheless
provide particular types of solutions
with spacetime geometrical interpretations {\em semiclassically}, and
this is the case with the topological solutions.

In the following we will give semiclassical interpretations for
these topological solutions. We first review the $\Lambda\neq0$ case
and then look into the $\Lambda=0$ case.


\subsection{WKB orbits of the Chern-Simons solutions}

It is a well known fact that the Chern-Simons solution takes
the form of a WKB wavefunction
\beq
\Psi^{CS}_{I}[A]=I[A]e^{iW[A]},\label{eq:WKB}
\eeq
where $W[A]=-i\frac{3}{2\Lambda}S_{CS}[A]$
is regarded as a Hamilton principal
functional subject to the Hamilton-Jacobi equation:
\beq
\frac{\partial}{\partial t}W[A]+H(A_{a}^{i},\tpi^{ai}=-i
\frac{\delta}{\delta A^{i}_{a}}W[A])=0.\label{eq:HJ}
\eeq
Hamiltonian in eq.(\ref{eq:HJ}) is given by
\beqy
H&=&-G(A_{t}-N^{a}A_{a})+\CD(\vec{N})+S(\tN)\nonumber \\
&=&i\int_{\M3}d^{3}x(D_{a}A_{t}^{i}\tpi^{ai}-\Sigma^{i}_{ta}
(\tpi^{bj})
\Phi^{ai}),\label{eq:hamiltonian}
\eeqy
where $\Sigma^{i}_{ta}(\tpi^{bj})$ is the solution to
eq.(\ref{eq:alcon1}) whose explicit form is given by
eq.(\ref{eq:lagmul}). Now we can semiclassically interpret the WKB
wavefunction (\ref{eq:WKB}) by investigating a family of
corresponding classical solutions (WKB orbits). This was first
investigated by Kodama\cite{kodama} in the minisuperspace model.
Here we will review his result in terms of generic spacetimes.
WKB orbits are determined by the following equations
\beqy
\tpi^{ai}&=&-i\frac{\delta}{\delta A^{i}_{a}}W=
-\frac{3}{\Lambda}\tB^{ai}\label{eq:AR}\\
\frac{\partial}{\partial t}A^{i}_{a}&=&\{A^{i}_{a},H\}_{PB}
\nonumber \\
&=&D_{a}A_{t}^{i}-\frac{\Lambda}{3}\Sigma^{i}_{ta}(\tpi^{bj})
-i\int_{\M3}d^{3}x\{A^{i}_{a},\Sigma^{i}_{ta}(\tpi^{bj})\}_{PB}
\Phi^{ai}. \label{eq:tempoASD}
\eeqy
Eq.(\ref{eq:AR}) is called the Ashtekar-Renteln ansatz\cite{AR}.
Because the last term in eq.(\ref{eq:tempoASD}) vanishes
on account of eq.(\ref{eq:AR}), these equations can be summarized
into a covariant form
\beq
F^{i}_{\mu\nu}=-\frac{\Lambda}{3}\Sigma^{i}_{\mu\nu}.\label{eq:ASD}
\eeq
If we stick to the region of Lorentzian signature, the WKB orbits are
subject to further restrictions. Namely, by imposing the (covariant)
reality conditions (\ref{eq:creal1})(\ref{eq:alcon1}), we find
\beq
R^{\alpha\beta}=\frac{\Lambda}{3}e^{\alpha}\wedge e^{\beta}.
\label{eq:desitter}
\eeq
This tells us that the Chern-Simons solutions correspond to
a family of spacetimes which are locally de Sitter\footnote{
For simplicity we have assumed that the cosmological constant
$\Lambda$ is positive.}:
\beq
ds_{DS}^{2}=\frac{3}{\Lambda}[-d\xi^{2}+\cosh^{2}\xi
(d\Omega_{(3)})^{2}],
\eeq
where $(d\Omega_{(3)})^{2}$ denotes the standard line element on
$S^{3}$.

However, this is not the whole story. From the Ashtekar-Renteln
ansatz (\ref{eq:AR}), we can read off the 3-metric $q_{ab}$
which is given by
\beq
(q^{-1})^{ab}=-\frac{\Lambda}{3}(\det(\tB^{ck}))^{-1}
\tB^{ai}\tB^{bi}.
\eeq
Sticking to the Lorentzian signature thus amounts to restricting
ourselves to the region $\det(\tB^{ck})<0$. However, because
$\Psi_{I}^{CS}[A]$ is well-defined on the whole space of
$SL(2,\BC)$ connections, there seems to be no natural reason to
rule out the region with $\det(\tB^{ck})>0$ (i.e. with $\det(q_{ab})
<0$)\cite{kodama}.
Spacetimes in such a region corresponds to Euclidean manifolds
with signature $(-,-,-,-)$ which represent spacetimes fluctuating
about $H^{4}$ (the 4 dimensional hyperbolic hypersurface):
\beq
ds_{H^{4}}^{2}=-\frac{3}{\Lambda}[d\xi^{2}+\sinh^{2}
\xi(d\Omega_{(3)})^{2}].
\eeq
Moreover, if we consider  the pure imaginary time
$\xi^{\prime}=i\xi$ ($\xi$: real) also,
we obtain the Euclidean spacetimes
with signature $(+,+,+,+)$ fluctuating about the four-sphere $S^{4}$:
\beq
ds_{S^{4}}^{2}=\frac{3}{\Lambda}[d\xi^{2}+\sin^{2}\xi
(d\Omega_{(3)})^{2}].
\eeq

By appropriately assembling these results,
the Chern-Simons solution $\Psi_{I}^{CS}[A]$ is considered to
correspond with spacetimes fluctuating about the sequence
$$
H^{4}\rightarrow S^{4}\rightarrow dS^{4}.
$$
This gives
the picture of \lq\lq creation of the Lorentzian spacetimes
($dS^{4}$) from a mother Euclidean space ($H^{4}$) by way of a
temporary bridge ($S^{4}$)"\cite{kodama}.

Because the deSitter space is a vacuum solution to the Einstein
equation in the case of a positive cosmological constant,
we expect that the Chern-Simons solutions are vacuum states
of quantum gravity with $\Lambda>0$.
This expectation was confirmed by investigating stability of
the classical background (\ref{eq:AR}) w.r.t. small perturbations
\cite{smol2}.

Here we will make a remark.
In this section the reality conditions are
imposed {\em classically} in order
to extract information on the real quantities $(e^{\alpha},\omega
^{\alpha\beta})$. In the actual quantum theory, however,
reality conditions are imposed {\em quantum mechanically} through
the inner product. As a result the self-dual part
$C^{(+)\rho}_{\qquad\sigma\mu\nu}$ of the
conformal curvature may fluctuate\cite{smol2}, while its anti-self
dual part $C^{(-)\rho}_{\qquad\sigma\mu\nu}$
and the Ricci tensor $R_{\mu\nu}$ are fixed almost completely by
eq.(\ref{eq:ASD}) to be
$$
C^{(-)\rho}_{\qquad\sigma\mu\nu}=0,
\quad R_{\mu\nu}=\Lambda g_{\mu\nu}.
$$


\subsection{Semiclassical interpretation
of $\Psi_{topo}[A]$}

Let us now look into the case of a vanishing cosmological constant.
Because $\Psi_{topo}[A]$ does not possess a WKB structure,
one may consider that we cannot interpret $\Psi_{topo}[A]$
semiclassically. We can nevertheless give $\Psi_{topo}[A]$
a semiclassical interpretation.

To see this, let us first consider the simplest example, namely
the harmonic oscillator with the hamiltonian $h=\frac{1}{2}p^{2}
+\frac{1}{2}q^{2}$. If we introduce the complex variable $z\equiv
q+ip$, the ground state $<q|0>=\psi_{0}(q)$ of this system satisfies
\beq
\hat{z}\cdot\psi_{0}(q)=(q+\hbar \frac{d}{dq})\psi_{0}(q)=0.
\eeq
The normalized ground state is therefore given by
\beq
\psi_{0}(q)=(\pi\hbar)^{-1/4}e^{-\frac{1}{2\hbar}q^{2}}.
\eeq
Thus the probability density that the coordinate
takes the value $q$ and the one that the momentum takes the value
$p$ are, respectively, given by
\beqy
|<q|0>|^{2}&=&\frac{1}{\sqrt{\pi\hbar}}e^{\frac{1}{\hbar}q^{2}}
\stackrel{\hbar\rightarrow 0}{\longrightarrow}\delta(q),\nonumber \\
|<p|0>|^{2}&=&\frac{1}{\sqrt{\pi\hbar}}e^{\frac{1}{\hbar}p^{2}}
\stackrel{\hbar\rightarrow 0}{\longrightarrow}\delta(p).
\eeqy
We can conclude that the ground state $|0>$ semiclassically
corresponds to the origin $(q,p)=(0,0)$ of the phase space.
Of course this is not the case in the quantum sense
because $\hat{\overline{z}}\cdot\psi_{0}(q)$ does not vanish.
The essential thing is that the quantum commutator
$[\hat{z},\hat{\overline{z}}]=2\hbar$ is negligible in the
semiclassical region and thus we can \lq simultaneously
diaginalize' $\hat{z}$ and $\hat{\overline{z}}$ (or equivalently
$\hat{q}$ and $\hat{p}$) under the semiclassical approximation.

This \lq semiclassical interpretation' is expected to apply
also to the topological solutions for Ashtekar's formalism.
From its definition, it is obvious that the topological solutions
(\ref{eq:Toposol}) correspond to classical solutions with
\beq
F^{i}_{ab}=0.
\eeq
Because the temporal components $A^{i}_{t}$ only play the role of
gauge parameters, they can be arbitrary chosen to yield
\beq
F^{i}_{\mu\nu}=0.\label{eq:ASD0}
\eeq
Now, by classically imposing the reality conditions
(\ref{eq:creal1}), we see that $\Psi_{topo}[A]$ correspond to
spacetimes which are flat\footnote{
As in the case of $\Psi_{I}^{CS}[A]$, self-dual part
of the conformal curvature still survives slightly in the full
quantum theory because the quantum reality conditions cannot
completely fix this part.}
\beq
R^{\alpha\beta}=0.\label{eq:flatsp}
\eeq

This suggests that the topological solutions represent
vacuum states of quantum gravity in the $\Lambda=0$ case.
Some affirmative evidence for this suggestion was provided
in ref.\cite{mats}, which appeals to the analogy of
$\Psi_{topo}[A]$ to the typical vacuum states of quantum
field theories. For example, that $\Psi_{topo}[A]$ carry the
topological degrees of freedom (the moduli $n\in\CN$ of flat
$SL(2,\BC)$ connections) is essentially the same as that
the vacuum states in a quantum field theory carry the similar
ones (e.g. the soliton numbers).

Unlike in the case with $\Lambda\neq0$ we can
exploit from the topological solutions for $\Lambda=0$
more detailed information on the spacetime.
As is immediately seen in eq.(\ref{eq:toposol}), we can
at least in principle distinguish different moduli by means of the
\lq characteristic' topological solutions $\Psi_{n}[A]$.
Giving semiclassical interpretation to $\Psi_{n}[A]$
thus amounts to finding a family of flat spacetimes
corresponding to the moduli $n$.

This is the situation similar
to that which we have encountered in (2+1)-dimensional
gravity\cite{witt}. There the reduced phase space is
the moduli space of flat connections in a noncompact gauge
theory and a point on the moduli space is related to a
(2+1)-dimensional spacetime through a geometric structure
\cite{carl}.\footnote{For a detailed explanation of the
geometric structure, see for example, refs.\cite{gold}\cite{mess}.}
This relation was explicitly established in the cases
of simple topologies with $\Lambda=0$\cite{carl2}\cite{louko},
$\Lambda>0$\cite{ezawa} and with $\Lambda<0$\cite{ezawa2}.

Difference between the problem at hand and that in (2+1)-dimensions
is that, in the former,
we need only to work out the relation of the space $\CN$ of canonical
coordinates to the spacetimes,
while in the latter we have established the relation
between the phase space and the set of flat
(or maximally symmetric) spacetimes.
We can therefore expect easily that the relation between $\CN$ and
the space of flat spacetimes is one to infinity
(not countable).\footnote{
The relation between the reduced phase space of $SL(2,\BC)$
BF theory and the space of flat spacetimes are much more complicated
\cite{ezawa6}. In this paper we will not go into its detail.}

Let us investigate in practice what spacetimes correspond to a
point $n$ on the moduli space $\CN$\cite{ezawa6}.
For this purpose we will first provide a parametrization of $\CN$.

We should note that the moduli space $\CN$ in general consists
of disconnected sectors which are related with one another
by large gauge transformations\cite{horo},
namely by gauge transformations
which cannot be continuously connected to the identity.
Because we are now interested in a family of spacetime metrics
which corresponds to a point $n\in\CN$, it is sufficient to
investigate only one of these sectors. For simplicity, we will choose
the sector $\CN_{0}$ which is connected to the trivial connection
$A^{i}_{a}=0$.

In order to parametrize $\CN_{0}$, it is convenient to use holonomies
along non-contractible loops:
\beq
H(\alpha)\equiv h_{\alpha}[0,1]\equiv
\CP\exp(\int_{0}^{1}ds\dot{\alpha}^{a}(s)A_{a}),
\eeq
where $\alpha:[0,1]\rightarrow\M3$ ($\alpha(0)=\alpha(1)=x_{0}$)
is a loop on $\M3$.
Because the connection $A_{a}$ is flat the holonomy depends
only on the homotopy class $[\alpha]$ of the loop $\alpha$:
\beq
H(\alpha)=H(\alpha^{\prime})\equiv H[\alpha]\quad\mbox{if}\quad
[\alpha]=[\alpha^{\prime}].
\eeq
The (large or small) gauge transformation
$A_{a}(x)\rightarrow g(x)A_{a}(x)
g^{-1}(x)+g(x)\partial_{a}g^{-1}(x)$ on the connection $A_{a}(x)$
is cast into the conjugation by $g(x_{0})$ on the holonomy
\beq
H[\alpha]\rightarrow g(x_{0})H[\alpha]g^{-1}(x_{0}).
\eeq
The moduli space $\CN_{0}$ is therefore identical to the space
of equivalence classes of homomorphisms from the fundamental group
$\pi_{1}(\M3)$ to the group $SL(2,{\bf C})$ modulo conjugations:
\beq
\CN_{0}={\rm Hom}(\pi_{1}(\M3),SL(2,{\bf C}))/\sim.
\eeq

Once we have parametrized the moduli space $\CN_{0}$ by using
holonomies, we can work out the explicit relation between
the moduli $n$ and flat spacetimes.
Because we know from eq.(\ref{eq:ASD0}) that the $SL(2,\BC)$
connection $A^{i}_{a}$ is flat,
it can be written in the form of a pure gauge
on the universal covering $\TM\approx{\bf R}\times
\widetilde{\M3}$ \footnote{The universal covering
$\widetilde{\M3}$ of $\M3$ is
the space of all the homotopy classes of curves
in $\M3$ starting from, say, the origin $x=0$. We will
decompose the point $\tx$ on $\widetilde{\M3}$ as
$\tx=\gamma+x$, which means the curve which first passes
along the loop $\gamma$ beginning at the origin $x=0$
and then goes from
the origin to the point $x$ on $\M3$ by way of the shortest
path measured by some positive-definite background metric
on $\M3$. We will denote by $\tx^{\mu}=(t,\tx)$ a point on $\TM$.}
\beq
A^{i}(\tx^{\mu})J_{i}=\Lambda^{-1}(\tx^{\mu})d\Lambda(\tx^{\mu}),
\label{eq:flatsl2c}
\eeq
where $\Lambda(\tx^{\mu})\in SL(2,\BC)$ is the integrated connection
which is subject to the periodicity condition
\beq
\Lambda(\gamma+\tx^{\mu})=H[\gamma]\Lambda(\tx^{\mu}).
\eeq
Next by imposing the reality conditions
(\ref{eq:creal1}) classically,
we can (up to gauge degrees of freedom)
completely fix a spin connection $\omega^{\alpha}\LI{\beta}
(\tx^{\mu})$ on $\TM$ as
\beq
\omega^{\alpha}\LI{\beta}(\tx^{\mu})=
(\Lambda^{-1}(\tx^{\mu}))^{\alpha}\LI{\gamma}
d\Lambda(\tx^{\mu})^{\gamma}\LI{\beta},\label{eq:flatlorentz}
\eeq
where $\Lambda(\tx^{\mu})^{\alpha}\LI{\beta}\equiv
(\CP\exp\int_{0}^{\tx^{\mu}}\omega(\tilde{y}^{\nu}))^{\alpha}
\LI{\beta}$ is the integrated spin connection which is related to
$\Lambda(\tx^{\mu})$ in eq.(\ref{eq:flatsl2c}) by
the last two equations in Appendix A.

In order to construct spacetime metrics which are related to
this spin connection, we have to search for the vierbein $e^{\alpha}$
which are single-valued on $M$ and
which satisfy the torsion-free condition
$$
de^{\alpha}(\tx^{\mu})+\omega^{\alpha}\LI{\beta}(\tx^{\mu})
\wedge e^{\beta}(\tx^{\mu})=0.
$$
Using eq.(\ref{eq:flatlorentz}), the torsion-free condition
is cast into the closedness condition
$$
d[\Lambda(\tx^{\mu})^{\alpha}
\LI{\beta}e^{\beta}(\tx^{\mu})]=0.
$$
Because $\TM$ is simply-connected
by definition, this equation is completely solved by
\beq
e^{\alpha}(\tx^{\mu})=(\Lambda^{-1}(\tx^{\mu}))^{\alpha}\LI{\beta}
dX^{\beta}(\tx^{\mu}),\label{eq:master1}
\eeq
where the set $\{X^{\alpha}(\tx^{\mu})\}$ is considered to be the
embedding of $\TM$ into (the universal covering of) an adequate
subspace of the (3+1)-dimensional Minkowski space $M^{3+1}$.
In order for the vierbein $e^{\alpha}(\tx^{\mu})$ to be single-valued
on $M$, it must satisfy the \lq periodicity condition'
\beq
e^{\alpha}(\gamma+\tx^{\mu})=e^{\alpha}(\tx^{\mu})\quad\mbox{for}
\quad^{\forall}[\gamma]\in\pi_{1}(M),\label{eq:peri1}
\eeq
(plus some conditions necessary when $\TM$ is not contractible
to a point).
By substituting eq.(\ref{eq:master1}) into eq.(\ref{eq:peri1})
we find
\beq
dX^{\alpha}(\gamma+\tx^{\mu})=d(H[\gamma]^{\alpha}\LI{\beta}
X^{\beta}(\tx^{\mu})),\label{eq:exact1form}
\eeq
where $H[\gamma]^{\alpha}\LI{\beta}$ stands for the holonomy
of the spin connection $\omega^{\alpha}\LI{\beta}$ evaluated
along the loop $\gamma$. The relation between
$H[\gamma]^{\alpha}\LI{\beta}$ and $H[\gamma]$ is also given by
the last two equations in Appendix A.
By integrating eq.(\ref{eq:exact1form})
we obtain the most important result:
\beq
X^{\alpha}(\gamma+\tx^{\mu})=H[\gamma]^{\alpha}\LI{\beta}
X^{\beta}(\tx^{\mu})+V[\gamma]^{\alpha},\label{eq:Lorentzian}
\eeq
which states that the periodicity condition of the embedding
functions $\{X^{\alpha}(x^{\mu})\}$ is given by Poincar\'{e}
transformations which are isometries of the Minkowski space.
Consistency among the periodicity conditions imposed on
all the loops in $\pi_{1}(M)\cong\pi_{1}(\M3)$ requires that
the set of Poincar\'{e} transformations
$\{(H[\gamma]^{\alpha}\LI{\beta},V[\gamma]^{\alpha})|
\quad\gamma\in\pi_{1}(\M3)\}$
should give a homomorphism from $\pi_{1}(\M3)$ to the
(3+1)-dimensional Poincar\'{e} group. This is precisely the
Lorentzian structure on the manifold $M\approx{\bf R}\times\M3$
\cite{gold}.
Thus we can say that the moduli of flat $SL(2,{\bf C})$
connections on a spacetime manifold $M\approx{\bf R}\times\M3$
specifies a
Lorentz transformation part of Lorentzian structures
each of which belongs to
$$
{\rm Hom}(\pi_{1}(\M3),(\CP^{3+1})^{\uparrow}_{+})/\sim,
$$
where $(\CP^{3+1})^{\uparrow}_{+}$ denotes the
proper orthochronous Poincar\'{e}
group in (3+1)-dimensions and $\sim$ stands for the equivalence under
the conjugation by (proper orthochronous)
Poincar\'{e} transformations.

Here we should note that a physically permissible Lorentzian
structure must be of rank 3 \footnote{The rank here is meant to be
the maximal number of generators in the subgroup.}
in order to avoid untamable spacetime
singularities, and that its action must be spacelike
in order to render the spacetime subject to the strong
causality condition\cite{hawk}.
We will see in the next subsection that these conditions
restrict severely the semiclassically permissible region of
the moduli space $\CN$, by examining the simple case 
in which the spatial 3-manifold is homeomorphic to a 3-torus $T^{3}$.


\subsection{The case of $\M3\approx T^{3}$}

Let us now consider the $\M3\approx T^{3}$ case\cite{ezawa6}.
We first construct the moduli space $\CN_{0}$.

Because $\pi_{1}(T^{3})\cong{\bf Z}\oplus{\bf Z}\oplus
{\bf Z}$ is an abelian group, the holonomy group on $T^{3}$
is generated by three commuting elements $(H[\alpha],H[\beta],
H[\gamma])$ of $SL(2,\BC)$.
By taking appropriate conjugations we find that the moduli
space $\CN_{0}$ consists of several \lq disconnected' sectors
\footnote{As in (2+1)-dimensional cases \cite{louko}\cite{ezawa}
\cite{ezawa2}, these sectors are in fact connected in a
non-Hausdorff manner. However, we will not go into this problem.}:
\beq
\CN_{0}=\CN_{S}\oplus\left(\bigoplus_{n_{1},n_{2},n_{3}\in\{0,1\}}
\CN_{F}^{n_{1},n_{2},n_{3}}\right).
\eeq
The standard sector $\CN_{S}$ is characterized by
\beqy
H[\alpha]&=&\exp((u+ia)J_{1})\nonumber \\
H[\beta]&=&\exp((v+ib)J_{1})\nonumber \\
H[\gamma]&=&\exp((w+ic)J_{1}),\label{eq:nst}
\eeqy
where $u,v,w$ are real numbers defined modulo $4\pi$ and
$a,b,c\in{\bf R}$. Gauge equivalence under
$g(x_{0})=\exp(\pi J_{2})$ further imposes the following
equivalence condition
\beq
(a,b,c;u,v,w)\sim -(a,b,c;u,v,w).\label{eq:Z2}
\eeq
As a consequence the standard sector $\CN_{S}$ has the topology
$(T^{3}\times{\bf R}^{3})/{\bf Z}_{2}$. A simple connection
which represent the point (\ref{eq:nst}) on $\CN_{S}$ is obtained
by making an appropriate gauge choice and an adequate choice of
periodic coordinates $(x,y,z)$ with period $1$.
We find
\beq
A_{a}dx^{a}=[(u+ia)dx+(v+ib)dy+(w+ic)dz]J_{1}\equiv
(dW_{1}+idW_{2})J_{1}=dWJ_{1}.\label{eq:NST}
\eeq

The flat sectors $\CN_{F}^{n_{1},n_{2},n_{3}}$ are parametrized
by the following holonomies:
\beqy
H[\alpha]&=&(-1)^{n_{1}}\exp(\xi(J_{2}+iJ_{1}))\nonumber \\
H[\beta]&=&(-1)^{n_{2}}\exp(\eta(J_{2}+iJ_{1}))\nonumber \\
H[\gamma]&=&(-1)^{n_{3}}\exp(\zeta(J_{2}+iJ_{1})),\label{eq:nf}
\eeqy
where $\xi,\eta,\zeta$ are complex numbers which do not vanish
simultaneously. Gauge equivalence under
$g(x_{0})=\exp(-i\kappa J_{3})$ with $\kappa\in{\bf C}$ tells
us that $(\xi,\eta,\zeta)$ provide the homogeneous coordinates
on ${\bf CP}^{2}$:
\beq
(\xi,\eta,\zeta)\sim e^{\kappa}(\xi,\eta,\zeta).\label{eq:CP2}
\eeq
Thus we find $\CN_{F}^{n_{1},n_{2},n_{3}}\approx{\bf CP}^{2}$.
Because two flat sectors are related with each other by
a large local Lorentz
transformation like $g(x)=\exp(2\pi(\delta n_{1}x+
\delta n_{2}y+\delta n_{3}z)J_{3})$, they are considered as
corresponding
to the same set of spacetimes. Thus in the following
we will consider only one flat sector $\CN_{F}\equiv\CN_{F}^{0,0,0}$.
As in $\CN_{S}$ we can find a connection which represent a
point (\ref{eq:nf}) in $\CN_{F}$. The result is
\beq
A_{a}dx^{a}=(\xi dx+\eta dy+\zeta dz)(J_{2}+iJ_{1})
\equiv d\Xi(J_{2}+iJ_{1})=(d\Xi_{1}+id\Xi_{2})(J_{2}+iJ_{1}).
\label{eq:NF}
\eeq

Now we can discuss what Lorentzian structures correspond to each
point on $\CN_{0}^{\prime}\equiv\CN_{S}\oplus\CN_{F}$.
We should bear in mind that, because
$\pi_{1}(T^{3})\cong{\bf Z}\oplus{\bf Z}\oplus{\bf Z}$,
the Lorentzian structure in question is generated by
three Poincar\'{e} transformations which mutually commute.

First we consider the standard sector $\CN_{S}$.
Because the parameters $(u,v,w)$ and $(a,b,c)$ respectively
give rise to rotations in the $(X^{2},X^{3})$-plane
and Lorentz boosts in the $(X^{0},X^{1})$-plane, we have to
separately consider the four cases.

I) For $(u,v,w)\neq\vec{0}\neq(a,b,c)$, the Lorentzian structures
are necessarily embedded into the rank 2 subgroup
$\{R_{23},L_{1}\}$ of
the Poincar\'{e} group\footnote{$R_{ij}$ and $L_{k}$ respectively
stand for the group of rotations on the $(X^{i},X^{j})$-plane
and that of Lorentz boosts in the $(X^{0},X^{k})$-plane}.
We therefore expect that
the spacetime given by one of these
Lorentzian structures suffers from singularities.
This is indeed the case. Such spacetime
inevitably has singularities at which the metric degenerates
and through which the orientation of the local Lorentz frame
defined by $\{e^{\alpha}\}$ is inverted\cite{ezawa6}.
Thus these spacetimes
cannot be considered to be physical, at least classically.

II) For $(a,b,c)=\vec{0}\neq(u,v,w)$, we can embed the Lorentzian
structure into a rank 3 subgroup of the Poincar\'{e} group
$\{R_{23},T^{0},T^{1}\}$, where $T^{\alpha}$
denotes the group of translations
in the $X^{\alpha}$-direction. This subgroup,
however, includes time-translations. The corresponding spacetimes
thus involve timelike-tori and so
they are not considered to be physical.
Of course we could use the Lorentzian structure which is embedded in
the rank 2 subgroup $\{R_{23},T^{1}\}$. But in this case
singularities similar to those appeared in case I) always exist in
the resulting spacetime. This case therefore corresponds to
a set of spacetimes not allowed in general relativity.

III) For $(u,v,w)=\vec{0}\neq(a,b,c)$, the situation drastically
changes. The Lorentzian structure in this case can be embedded
into the rank 3 subgroup $\{L_{1},T^{2},T^{3}\}$.
Fortunately the action of this subgroup on the region $\{(X^{0})^{2}-
(X^{1})^{2}>0\}\in M^{3+1}$ is spacelike. This indicates that
this case corresponds to a set of well-behaved spacetimes.

More precisely, by plugging into eq.(\ref{eq:master1}) the
integrated spin connection constructed from $A=idW_{2}J_{1}$,
we find the equation
\beqy
\cosh W_{2}e^{0}-\sinh W_{2}e^{1}&=&dX^{0}\nonumber \\
-\sinh W_{2}e^{0}+\cosh W_{2}e^{1}&=&dX^{1}\nonumber \\
e^{2}\qquad&=&dX^{2}\nonumber \\
e^{3}\qquad&=&dX^{3}.\label{eq:23}
\eeqy
A choice of embedding functions
which yields well-behaved spacetimes is:
$$
(X^{\alpha})=(\tau\cosh(W_{2}+\alpha),-\tau\sinh(W_{2}+\alpha),
\vec{\beta}\cdot\vec{x}+\psi_{2},\vec{\gamma}\cdot\vec{x}+\psi_{3}),
$$
where $\vec{\beta}$ and $\vec{\gamma}$ are constant vectors in
${\bf R}^{3}$. $\tau$, $\alpha$,
$\psi_{2}$ and $\psi_{3}$ are single-valued functions
on $M$. Substituting this into eq.(\ref{eq:23}),
we find
\beqy
e^{0}&=&d\tau\cosh\alpha+\tau\sinh\alpha d(W_{2}+\alpha)\nonumber \\
e^{1}&=&-d\tau\sinh\alpha-\tau\cosh\alpha
d(W_{2}+\alpha) \nonumber \\
e^{2}&=&dX^{2}=\vec{\beta}\cdot d\vec{x}+d\psi_{2} \nonumber \\
e^{3}&=&dX^{3}=\vec{\gamma}\cdot d\vec{x}+d\psi_{3}.
\eeqy
This vierbein gives a physically permissible spacetime whose only
pathology is the initial singularity at $\tau=0$:
\beq
ds^{2}=-d\tau^{2}+\tau^{2}d(W_{2}+\alpha)^{2}+(dX^{2})^{2}+
(dX^{3})^{2}.
\eeq

IV) For $(a,b,c)=(u,v,w)=\vec{0}$. In this case also we can embed
the geometric structure into the rank 3 subgroup
$\{T^{1},T^{2},T^{3}\}$, whose action on the whole Minkowski space
is spacelike. We expect this case to correspond to spacetimes
without any singularity. Indeed we see that the spacetimes which
correspond to this case take the following form
\beq
ds^{2}=-dT^{2}+dX^{i}dX^{i},\label{eq:zerometric}
\eeq
where $X^{i}\equiv\vec{\alpha}^{i}\cdot\vec{x}$ and
each $\vec{\alpha}^{i}$ is a constant vector in ${\bf R}^{3}$.
These spacetimes in general have no singularity.

Next we consider the flat sector $\CN_{F}$.
We deal with two cases separately.

I)' For ${\rm Re}(\vec{\xi})$ and ${\rm Im}(\vec{\xi})$
being linear independent,
with $\vec{\xi}\equiv(\xi,\eta,\gamma)$, the Lorentzian structure
in question is embedded into the subgroup $\{N^{1+},N^{2+},T^{+}\}$,
where $T^{+}$ is the translation in the $X^{+}(\equiv X^{0}+X^{3})$-
direction and $N^{\hat{i}+}$ ($\hat{i}=1,2$) is the null-rotation
which stabilizes $X^{+}$ and $X^{\hat{i}}$. Because this subgroup
contains translation in the null-direction, there is a possibility
that closed null curves appear. An inspection \cite{ezawa6}
shows that, while the corresponding spacetimes\footnote{
$Z_{+}$ is defined as $Z_{+}\equiv\vec{\alpha}_{+}\cdot\vec{x}$
with $\vec{\alpha}_{+}$ being a constant vector in ${\bf R}^{3}$.
$\overline{\Xi}$ denotes the complex conjugate of $\Xi$.}
\beq
ds^{2}=-dt^{2}-e^{t}dtdZ_{+}+e^{2t}d\Xi d\overline{\Xi}
\label{eq:nullmetric}
\eeq
may or may not have closed causal curves\footnote{
In fact this metric has at worst closed null curves.},
they certainly violate the strong causality condition\cite{hawk}.
These spacetimes are therefore not so desirable in
general relativity.

II)' For ${\rm Im}(\vec{\xi})=0\neq{\rm Re}(\vec{\xi})$
(or equivalently for $\vec{\xi}=\frac{1}{\cos\phi}{e^{i\phi}}{\rm Im}
(\vec{\xi})$, $\phi\in(-\frac{\pi}{2},\frac{\pi}{2})$),
the Lorentzian structure is embedded into
the rank 3 subgroup $\{N^{2+},T^{+},T^{2}\}$. Also in this case
the resulting spacetimes spoil the strong causality
condition in the same way that the spacetime (\ref{eq:nullmetric})
does.

To summarize, each point on $\CN_{0}^{\prime}=\CN_{S}\oplus
\CN_{F}$ yields a family of Lorentzian structures whose
projection onto the Lorentz group is given by
the corresponding $SL(2,\BC)$ holonomy group.
While there is a subspace $\{(u,v,w)=0\}\in\CN_{S}$
each point on which yield a family of physically permissible
spacetimes,
most of the points on $\CN_{0}^{\prime}$
give rise to a family of spacetimes
which suffer from singularities or which ruin the strong causality
condition. 
Most of the characteristic topological solutions $\Psi_{n}[A]$
therefore correspond to a family of spacetimes which are
not permissible from the viewpoint of general relativity.

How to resolve this issue? Two methods are conceivable.
One is to impose a selection rule which extract only the subspace
$\{(u,v,w)=0\}\in\CN_{S}$. However, this seems to be
artificial and so not desirable
compared to the situation of the Chern-Simons solutions.
The other is to consider the singularities to be
annihilated owing to the uncertainty principle in
quantum gravity. This is possible because any topological solutions
$\Psi_{topo}[A]$ cannot fix a spacetime (or a Lorentzian structure)
and, in particular, $\Psi_{n}[A]$ corresponds to a family of
spacetimes. As for the flat sector $\CN_{F}$ which violates
the strong causality condition, taking account of the fact that
this sector \lq almost degenerates'
with the origin of $\CN_{S}$ in a
non-Hausdorff manner, corresponding spacetimes
are in fact some \lq superposition" of (\ref{eq:zerometric})
and (\ref{eq:nullmetric}) which may restore the strong causality.

In the actual quantum gravity, we have to take account also of the
\lq\lq right-handed graviton mode"
$C^{(+)\rho}_{\qquad\sigma\mu\nu}$. To elucidate what actually
happens to the topological solutions is thus postponed
until its full treatment in quantum gravity will be made possible.


\subsection{The Euclidean case}

So far we have investigated what family of spacetimes specify
the topological solution $\Psi^{CS}_{I}[A]$ or $\Psi_{n}[A]$
putting emphasis on the case of Lorentzian signature.
For completeness it would be proper to mention the
Euclidean signature case. The case with signature $(-,-,-,-)$
has not yet been studied so much. The case with signature
$(+,+,+,+)$ has been investigated in ref.\cite{ANU}. We will briefly
review the results.

We first note that, in the Euclidean case, we cannot exploit reality
conditions to extract the full information on the spin connection
$\omega^{\alpha\beta}$ from that on the anti-self-dual connection
$$
A_{E}^{i}=-\frac{1}{2}\ep^{ijk}\omega^{jk}+\omega^{0i}.
$$
Some of the degrees of freedom concerning the self-dual connection
thus remains intact even if we impose the torsion-free
condition $de^{\alpha}+\omega^{\alpha}\LI{\beta}\wedge e^{\beta}=0$.

For $\Lambda\neq0$, the Chern-Simons solutions
\beq
\Psi_{I}^{CS}[A_{E}]=I[A_{E}]e^{-\frac{3}{2\Lambda}S_{CS}[A_{E}]}
\label{eq:ECSsol}
\eeq
are the topological solutions also for the Euclidean case and
the equation for the WKB orbits
\beq
F_{E}^{i}=\frac{\Lambda}{3}\Sigma_{E}^{i}=
\frac{\Lambda}{3}(-\frac{1}{2}\ep^{ijk}
e^{j}\wedge e^{k}+e^{0}\wedge e^{i})
\eeq
yields the conformally self-dual Euclidean manifolds with
a fixed Ricci curvature $R_{\mu\nu}=\Lambda g_{\mu\nu}$.

Thus the family of WKB orbits corresponding to the Chern-Simons
solutions are equal to the moduli space of conformally
self-dual Riemannian manifolds, which was shown by
Hitchin\cite{besse} to yield either $S^{4}$ with the standard
metric or ${\bf CP}^{2}$ with the Fubini-Study metric
in the case where the four dimensional manifold $M$ is compact and
$\Lambda$ is positive.

For $\Lambda=0$, the topological solutions $\Psi_{topo}[A_{E}]$
have their support only on flat anti-self-dual connections
$$
F^{i}_{E}=0,
$$
namely on self-dual four dimensional manifolds $M$.
If $M$ is compact, then by the theorem of Hitchin\cite{besse}
the topology of $M$ is limited to that of $T^{4}$, $K3$ or
their quotients by some discrete groups.
His theorem further informs us that $\Psi_{topo}[A_{E}]$
is characterized by the moduli of Einstein-K\"{a}hlerian manifolds
with vanishing real first Chern class.

As a result of the flatness of the connection $A^{i}_{E}$,
the equations which determine the two-form $\Sigma^{i}_{E}$ can
be written as
\beqy
\Sigma^{i}_{E}\wedge\Sigma^{j}_{E}&=&
\frac{1}{3}\delta^{ij}\Sigma^{k}\wedge\Sigma^{k} \nonumber\\
d\Sigma^{i}_{E}&=&0.
\eeqy
In general the second equation holds only locally.
If the canonical bundle over $M$ is trivial, however,
these equations hold also globally. In this case the corresponding
manifolds are called hyper-K\"{a}hlerian. On the other hand,
manifolds whose canonical bundles are non-trivial are
called locally hyper-K\"{a}hlerian.


\subsection{Topological solutions for $N=1$ supergravity}

Up to here in this paper we have explored several aspects
of Ashtekar's formulation for pure gravity.
As a matter of fact Ashtekar's formulation can be applied
also to supergravities with $N=1$\cite{jacob2}\cite{capo}
and with $N=2$\cite{kuni}. These Ashtekar's formulation for
supergravities can also be cast into
(graded) BF theories with the two-forms subject to the algebraic
constraints like eq.(\ref{eq:alcon2})\cite{ezawa5}.
This relation elegantly explains the existence
of topological solutions also for $N=1$ supergravity
\cite{sano}\cite{mats} and for $N=2$ supergravity
\cite{sano}\cite{ezawa5}.
Here we will briefly see this in the $N=1$ case only. The similar
arguments hold also in the $N=2$ case. For a detail,
we refer the reader to ref.\cite{ezawa5}.

Ashtekar's formulation for $N=1$ supergravity was first derived by
Jacobson\cite{jacob2}. Later a more convenient derivation by way of
the chiral action was discovered in ref.\cite{capo}. We will follow
the latter case.\footnote{
We will follow the convention used in refs.\cite{sano}\cite{sano2}.}

Our starting point is the following chiral action
with a cosmological constant $\Lambda=g^{2}$ (possibly with $g=0
$)\footnote{To avoid the confusion we will denote in this subsection
the $SL(2,\BC)$ generator in the spinor representation as
$(\frac{\sigma_{i}}{2i})^{A}\LI{B}$, with $\sigma_{i}$ being
the Pauli matrix.}
\beq
-iI^{N=1}_{BF}=\int\left(\begin{array}{r}
\Sigma^{i}\wedge(F^{i}+\lambda g(\frac{\sigma^{i}}{2i})_{AB}
\psi^{A}\wedge\psi^{B})+2\chi^{A}\wedge D\psi_{A}\\
-\frac{g^{2}}{6}\Sigma^{i}\wedge\Sigma^{i}
-\frac{g}{3\lambda}\chi_{A}\wedge\chi^{A}
\end{array}\right),\label{eq:N1ASH1}
\eeq
where $A^{i}$ and $\Sigma^{i}$ denote as in the bosonic case
the self-dual part of the spin connection and the self-dual
two-form respectively, and 
$\psi^{A}$ and $\chi^{A}$ respectively stand for
the left-handed gravitino and the left-handed two-form which
corresponds to the right-handed gravitino\footnote{
Note that $\chi^{A}$ and $\psi^{A}$ are
Grassmann odd fields. Whether an object is Grassmann even or odd
can be determined by whether the number of its Lorentz
spinor indices is even or odd.}. In consequence
$\Sigma^{AB}\equiv\Sigma^{i}(\frac{\sigma_{i}}{2i})^{AB}$
and $\chi^{A}$ are subject to the algebraic constraints
\beqy
\Sigma^{ABCD}&\equiv&\Sigma^{(AB}\wedge\Sigma^{CD)}=0,\nonumber \\
\Xi^{ABC}&\equiv&\Sigma^{(AB}\wedge\chi^{C)}=0.\label{eq:N1alcon}
\eeqy

We can easily show that eq.(\ref{eq:N1ASH1}) actually coincides
with the BF action with the gauge group being the graded group
$GSU(2)$:
\beq
-iI^{N=1}_{BF}=\int\Str(\CB\wedge\CF-\frac{g^{2}}{6}\CB\wedge\CB),
\label{eq:BFac}
\eeq
where $\CB=\Sigma^{i}J_{i}-\frac{1}{\lambda g}\chi^{A}J_{A}$
is a $GSU(2)$-valued two-form and $\CF=d\CA+\CA\wedge\CA$ is
the curvature two form of the $GSU(2)$ connection $\CA=A^{i}J_{i}+
\psi^{A}J_{A}$. $(J_{i},J_{A})$ are the generators
of the graded Lie algebra $GSU(2)$\cite{pais}:
\beq
[J_{i},J_{j}]=\ep_{ijk}J_{k},\quad
[J_{i},J_{A}]=(\frac{\sigma^{i}}{2i})_{A}\UI{B}J_{B},\quad
\{J_{A},J_{B}\}=-2\lambda g(\frac{\sigma^{i}}{2i})_{AB}J_{i},
\label{eq:GSU2}
\eeq
where $\{\mbox{ , } \}$ denotes the anti-commutation relation.
$\Str$ stands for the $GSU(2)$ invariant bilinear form
which is unique up to an overall constant factor
\beqy
&&\Str(J_{i}J_{j})=\delta_{ij},\quad
\Str(J_{A}J_{B})=-2\lambda g\ep_{AB},\nonumber \\
&&\Str(J_{A}J_{i})=\Str(J_{i}J_{A})=0.\label{eq:bin}
\eeqy

Similarly to the bosonic case Ashtekar's formalism is attainable
by performing (3+1)-decomposition of the action (\ref{eq:N1ASH1}).
We first obtain
\beqy
-iI^{N=1}_{BF}&\!\!\!=\!\!\!&
\int dt\int_{\M3}\Str(\tPi^{a}\dot{\CA}_{a}+\CA_{t}
{\bf G}+\CB_{ta}{\bf \Phi}^{a})\nonumber \\
&\!\!\!=\!\!\!&\int dt\int_{\M3}
(\tpi^{ai}\dot{A}^{i}_{a}+2\tpi^{A}\dot{\psi}_{aA}+A_{t}^{i}G^{i}
-2\psi_{tA}L^{A}+\Sigma_{ta}^{i}\Phi^{ai}-2\chi_{taA}\Phi^{aA}),
\label{eq:N1caac1}
\eeqy
where we have set $\tPi^{a}=\frac{1}{2}\otep^{abc}\CB_{bc}=
\tpi^{ai}J_{i}-\frac{1}{\lambda g}\tpi^{aA}J_{A}$.
From this we can read two types of constraints in $GSU(2)$
BF theory, one is Gauss' law constraint which generates
the $GSU(2)$ gauge transformations
\beqy
{\bf G}&=&\CD_{a}\tPi^{a}
\equiv\partial_{a}\tPi^{a}+[\CA_{a},\tPi^{a}]
=G^{i}J_{i}-\frac{1}{\lambda g}L^{A}J_{A},
\nonumber \\
G^{i}&=&D_{a}\tpi^{ai}-2(\frac{\sigma^{i}}{2i})_{AB}
\psi^{A}_{a}\tpi^{aB}\nonumber \\
L^{A}&=&D_{a}\tpi^{aA}+\lambda g\tpi^{ai}
(\frac{\sigma^{i}}{2i})^{A}\LI{B}\psi^{B}_{a}.\label{eq:gauss1}
\eeqy
And the other is the constraint which
generates the extended Kalb-Ramond symmetry\cite{KR}
\beqy
{\bf \Phi}^{a}&=&\frac{1}{2}\otep^{abc}\CF_{bc}
-\frac{g^{2}}{3}\tPi^{a}\nonumber \\
&=&\Phi^{ai}J_{i}+\Phi^{aA}J_{A},\nonumber \\
\Phi^{ai}&=&\frac{1}{2}\otep^{abc}(F^{i}_{bc}+2\lambda g
(\frac{\sigma^{i}}{2i})_{AB}\psi^{A}_{b}\psi^{B}_{c})
-\frac{g^{2}}{3}\tpi^{ai}\nonumber \\
\Phi^{aA}&=&\otep^{abc}D_{b}\psi_{c}^{A}+
\frac{g}{3\lambda}\tpi^{aA}.\label{eq:ckr1}
\eeqy
 
Next we solve the algebraic constraints (\ref{eq:N1alcon})
for the lagrange multiplier $(\Sigma^{i}_{ta},\chi^{A}_{ta})$.
By substituting the results into eq.(\ref{eq:N1caac1}) we find
that Gauss' law constraint remains intact while the
other constraint survives only partially, namely, what survive are
only the following linear combinations
\beqy
R^{A}&=&\frac{1}{2}\utep_{abc}\ep^{ijk}\tpi^{bj}\tpi^{ck}
(\frac{\sigma^{i}}{2i})^{AB}\Phi^{a}_{B},\nonumber \\
\CS&=&\frac{1}{4}\utep_{abc}\ep^{ijk}\tpi^{bj}\tpi^{ck}\Phi^{ai}
-2\utep_{abc}\tpi^{bi}(\frac{\sigma^{i}}{2i})_{AB}\tpi^{cB}\Phi^{aA},
\nonumber \\
\CV_{a}&=&\frac{1}{2}\utep_{abc}\tpi^{bi}\Phi^{ci}+
\utep_{abc}\tpi^{bA}\Phi^{c}_{A}.
\eeqy
Physically, $R^{A}$ generates right-supersymmetry
transformations, $\CS$ generates many-fingered time evolutions, and
$\CV_{a}$ generates spatial diffeomorphisms. In passing,
among Gauss' law constraint, $G^{i}$ generates local Lorentz
transformations for left-handed fields and $L^{A}$ generates
left-supersymmetry transformations.

Quantization also proceeds as in the bosonic case
by replacing $i$ times the Poisson brackets with
the quantum commutators. If we use the representation in which
$A^{i}_{a}$ and $\psi^{A}_{a}$ are diagonalized, the action
of the operators $\hat{\tpi}^{ai}$ and $\hat{\tpi}^{aA}$
on the wavefunction $\Psi[\CA_{a}]$ is represented by
functional derivatives:
\beq
\hat{\tpi}^{ai}(x)\cdot\Psi[\CA_{a}]=
-\frac{\delta}{\delta A_{a}^{i}(x)}\Psi[\CA],\quad
\hat{\tpi}^{aA}(x)\cdot\Psi[\CA_{a}]=
\frac{1}{2}\frac{\delta}{\delta\psi_{aA}(x)}\Psi[\CA_{a}].
\eeq

The physical wave functions in quantum $N=1$
supergravity have to satisfy
the first class constraint equations:
\beqy
\hat{G}^{i}\Psi[\CA_{a}]&=&\hat{L}^{A}\Psi[\CA_{a}]=0
\label{eq:N1gauss}\\
\hat{R}^{A}\Psi[\CA_{a}]&=&\hat{\CS}\Psi[\CA_{a}]
=\hat{\CV}_{a}\Psi[\CA_{a}]=0.\label{eq:diffrh}
\eeqy
Gauss' law constraint (\ref{eq:N1gauss}) requires the wavefunctions
to be invariant under the $GSU(2)$ gauge transformations. When we
solve the constraints $(R^{A},\CS,\CV_{a})$, we should note that
these constraints are linear combinations of the constraints
$(\Phi^{ai},\Phi^{aA})$ in BF theory, with the coefficients
being polynomials in the momenta $(\tpi^{ai},\tpi^{aA})$.
Thus, as in the bosonic case, if we take the ordering with the
momenta to the left, solutions to $GSU(2)$
BF theory are involved in the solution space for quantum
$N=1$ supergravity in the Ashtekar form. They are
the topological solutions for $N=1$ supergravity.

For $g\neq0$, we have the $N=1$ Chern-Simons solutions
\cite{sano}\cite{sano2}\cite{fulo}
\beq
\Psi^{CS}_{I}[\CA_{a}]=
I[\CA_{a}]e^{-\frac{3}{2g^{2}}S_{CS}^{N=1}[\CA_{a}]},
\label{eq:N1CS1}
\eeq
where $I[\CA_{a}]$ is some topological invariant
and $S_{CS}^{N=1}[\CA_{a}]$ is the Chern-Simons functional for the
$GSU(2)$ connection $\CA$
\beqy
S_{CS}^{N=1}&=&\int_{\M3}\Str(\CA d\CA+
\frac{2}{3}\CA\wedge\CA\wedge\CA)\nonumber \\
&=&\int_{\M3}(A^{i}dA^{i}+\frac{1}{3}\ep^{ijk}A^{i}\wedge A^{j}
\wedge A^{k}-2\lambda g\psi^{A}\wedge D\psi_{A}).
\eeqy

For $g=0$, we have the following formal expression for the
topological solutions
\beq
\Psi_{topo}[\CA_{a}]=F[\CA_{a}]\prod_{x\in\M3}
\left(\prod_{a,i}\delta(\otep^{abc}F^{i}_{bc}(x))
\prod_{a,A}\delta(\otep^{abc}D_{b}\psi^{A}_{c}(x))\right),
\label{eq:N1topo1}
\eeq
where $F[\CA_{a}]$ is an arbitrary $GSU(2)$-invariant functional
of the connection $\CA_{a}$. Owing to the delta functions
$F[\CA_{a}]$ reduces to the function on the moduli space
${\cal GN}$ of flat $GSU(2)$ connections.
Alternatively, we can write $\Psi_{topo}[\CA_{a}]$ formally as
\beqy
\Psi_{topo}[\CA_{a}]&=&\int_{gn\in{\cal GN}}d(gn)F(gn)\Psi_{gn}
[\CA_{a}]\nonumber \\
\Psi_{gn}[\CA_{a}]&\equiv&\int[d{\cal G}(x)]
\prod_{x\in\M3}\prod_{a,(i,A)}\delta(\CA_{a}(x)=
{\cal G}[\CA_{0,gn}]_{a}(x)),\label{eq:N1toposol}
\eeqy
where $\CA_{0,gn}$ is a flat $GSU(2)$ connection which represent
a point $gn$ on the moduli space ${\cal GN}$ and
$$
{\cal G}[\CA]_{a}(x)\equiv {\cal G}(x)\CA_{a}(x){\cal G}^{-1}(x)
+{\cal G}(x)\partial_{a}{\cal G}^{-1}(x)
$$
denotes the gauge-transformed $GSU(2)$ connection.
The integration region in
eq.(\ref{eq:N1toposol}) is taken to be the space
of small $GSU(2)$ gauge transformations.
This expression of topological solutions is essentially
a kind of Fourier transform of the $N=1$ topological solutions
given in ref.\cite{mats}.

The solutions (\ref{eq:N1CS1}) and (\ref{eq:N1topo1}) are
considered to be vacuum states of $N=1$ supergravity
for the same reason as in the bosinic case\cite{mats}.
It is expected, however, that the spacetimes corresponding
to these topological solutions are considerably different
from those in the bosonic case\cite{fulo},
because in supergravity
the existence of the gravitino influences the nonvanishing
value of the torsion through the equation:
\beq
de^{\alpha}+\omega^{\alpha}\LI{\beta}\wedge e^{\beta}=
-\frac{i}{2}\sigma^{\alpha}_{A\bar{A}}\overline{\psi}^{\bar{A}}
\wedge\psi^{A},
\eeq
where $\overline{\psi}^{\bar{A}}$ is the right-handed
gravitino and $\sigma^{\alpha}_{A\bar{A}}$ stands for the
\lq\lq soldering form" in the Minkowski space $M^{3+1}$
\cite{penrose2}.
This was explicitly established in the minisuperspace model
for $g\neq0$\cite{sano2}. There it was demonstrated that,
due to the effect of the gravitino,
behavior of the WKB orbits in general deviates from
that of the deSitter space.


\subsection{Summary and discussion on \S 6}

In this section we have looked into the topological solutions.
Semiclassically these solutions correspond to a family of
vacuum solutions to general relativity i.e.
spacetimes with no graviton excitation,
and thus they are expected to be
vacuum states of quantum gravity. This expectation is
confirmed by the fact that the topological solutions are
also physical states in a topological field theory, $SL(2,\BC)$
BF theory, and that they naturally carry the topological degrees of
freedom such as the moduli of flat $SL(2,\BC)$ connections
which yield a family of Lorentzian structures on the flat spacetime.
These topological solutions as vacuum states of
quantum gravity naturally extends to $N=1,2$ supergravities.
In supergravities, however, corresponding spacetimes
are nontrivial because of the existence of the gravitino.

The existence of the topological solutions is intimately related
with the fact that Ashtekar's formulation for canonical gravity
can be cast into the form of $SL(2,\BC)$ BF theory
with the two-form field $\Sigma^{i}$ subject to the algebraic
constraint. In association with this
there are several attempts to interpret
Einstein gravity as a kind of
\lq\lq unbroken phase" of some topological field theories
\cite{medi}\cite{suga}. This idea is interesting
because it may provide us with some information on a more
mathematically tractable formulation on the quantum gravity
as in the case of (2+1)-dimensional gravity\cite{witt}.
Now from the result in \S\S 6.6 we expect that
these idea can be  extended to
supergravities. Probably this deserves studying because
the existence of the supersymmetry is believed by
most of the high energy phenomenologists and from such a
viewpoint supergravities seem to be more realistic than
pure gravity.


\section{Other solutions and remaining issues}

In this paper we have given an outline of two types of solutions,
namely the Wilson loop solutions and the topological solutions,
for quantum Ashtekar's formalism.
As far as the author knows there are further two types of solutions
for canonical quantum gravity. Here we will make a brief introduction
of these solutions. After that we will discuss remaining issues
on Ashtekar's formulation for canonical quantum gravity.


\subsection{Loop representation and the Jones polynomial}

In this paper we have worked solely
with the connection representation
in which the wavefunctions are represented by holomorphic
functionals $\Psi[A^{i}_{a}]$ of the $SL(2,\BC)$ connection.
As a matter of fact, there is
another representation which is extensively explored in Ashtekar's
formulation for quantum gravity. It is the loop representation
\cite{RS} in which the loop
functionals $\Psi[\gamma]$ play the role of wavefunctions.
Relation between the connection representation and
the loop representation is given by the formal
\lq\lq loop transform"\cite{RS}
\beq
\Psi[\gamma]=\int[dA^{i}_{a}(x)]\overline{W[\gamma]}
\Psi[A^{i}_{a}],\label{eq:looptfm}
\eeq
where $W[\gamma]\equiv W[\gamma,\pi_{1}]$ is the Wilson loop
in the spinor representation and $[dA^{i}_{a}(x)]$ is some
gauge- and diffeomorphism invariant measure on the space of
connections. If we use spin network states in stead of
Wilson loops, this formal transformation can be made
rigorous\cite{RS3} at least in the Euclidean case if we use as an
invariant measure $[dA^{i}_{a}(x)]$ the induced Haar measure
$d\mu_{H}(A)$ explained in \S\S 3.2:
\beq
\Psi[\Gamma,\{\pi_{e}\},\{i_{v}\}]=\int_{\overline{\CA/\CG}}
d\mu_{H}\overline{<A^{i}_{a}|
\Gamma,\{\pi_{e}\},\{i_{v}\}>}\Psi[A^{i}_{a}].
\eeq
To find out a mathematically rigorous transformation in the
Lorentzian case is left to the future investigation
because we do not know a well-defined invariant measure
in this case.

By means of the transformation (\ref{eq:looptfm}),
the operator $\hat{O}_{C}$ in the connection representation
is inherited to the loop representation as an operator $\hat{O}_{L}$
which is subject to the relation:
\beq
\hat{O}_{L}\overline{W[\gamma]}=\overline{\hat{O}_{C}^{\dagger}
W[\gamma]},\label{eq:connloop}
\eeq
where $\dagger$ denotes the adjoint operator w.r.t. the measure
$[dA^{i}_{a}(x)]$. We can exploit this relation when we define the
constraint operators in the loop representation.
Owing to the existence of $\dagger$, operator ordering in the
loop representation is the opposite to
that in the connection representation.
A merit of the loop representation is that we can easily implement
the diffeomorphism constraint (\ref{eq:qdiffeo}) by
dealing only with the functionals of diffeomorphism equivalence
classes of loops (or of colored graphs). So we have only to
concentrate on the scalar constraint. For simplicity
we will consider the Euclidean case in which the
reality conditions for $(\hat{A}_{E})^{i}_{a}$ and
$(\hat{\tpi}_{E})^{ai}$ are given by
the usual self-adjointness condition.

Owing to the relation (\ref{eq:connloop}) loop transform of
the solutions in the connection representation all yields the
solutions in the loop representation. For example,
the Wilson loop solutions tranforms to the characteristic
functions of particular colored graphs\cite{RS}.

Let us now consider the loop transform of the Chern-Simons solution
(\ref{eq:ECSsol}) with $I[A_{E}]=1$\cite{GP}\cite{brug}.
According to ref.\cite{witt2}, the loop transform of
$\Psi_{1}^{CS}[A_{E}]$ coincides with the Kauffman bracket
${\rm KB}_{\Lambda}[\gamma]$\footnote{
In quantum gravity we consider the knot invariants to be extended
to the knots with kinks and/or intersections.}
\beqy
\Psi_{1}^{CS}[\gamma]&=&\int[dA_{E}]\overline{W[\gamma]}
e^{-\frac{3}{2\Lambda}S_{CS}[A_{E}]} \nonumber \\
&=&{\rm KB}_{\Lambda}[\gamma]\nonumber \\
&=&e^{\Lambda{\rm Gauss}[\gamma]}
{\rm JP}_{\Lambda}[\gamma],
\eeqy
where ${\rm Gauss}[\gamma]$ and ${\rm JP}_{\Lambda}[\gamma]$ denote
respectively the Gauss self-linking number and
the Jones polynomial\footnote{In deriving the second equation
we have used the property of $SU(2)$ Wilson loops:
$\overline{W[\gamma]}=W[\gamma^{-1}]=W[\gamma]$.}.
The Kauffman bracket
${\rm KB}_{\Lambda}[\gamma]$ is thus considered to be
a solution to the WD equation in the case where the cosmological
constant $\Lambda$ is nonvanishing.
It was also shown \cite{GP}
that $\exp(\Lambda{\rm Gauss}[\gamma])$ is also
annihilated by the scalar constraint $\hat{\CS}$ with
$\Lambda\neq0$. From this
we can easily show that the second coefficient of
the Jones Polynomial is a solution to the WD equation with
$\Lambda=0$\cite{BGP}\cite{GP2}.
From this fact a conjecture
arose that the coefficients of
the Jones polynomial are solutions to the WD equation
with vanishing cosmological constant\cite{GP}.
It was shown, however, that the third
coefficient does not solve the WD equation\cite{griego}.

This approach of \lq\lq finding solutions to
the WD equation out of knot invariants" is expected to
develop as a method which complements the approach of
constructing solutions from the spin network states.
The extended loop representation\cite{bart}
is considered to provide us with a systematic mathematical
tool for pursuing this avenue.

In this approach it is indispensable to find an invariant measure
on the space of $SL(2,\BC)$ connections which makes the transform
(\ref{eq:looptfm}) well-defined. Otherwise the loop representation
cannot be qualified as a representation of Lorentzian
quantum gravity.


\subsection{The wormhole solution in $N=1$ supergravity}

As we have seen so far, finding solutions to the constraint
equations in Ashtekar's formalism has made a steady progress.
What about other formalisms of canonical quantum gravity?
In supergravity the Hamiltonian constraint and the momentum
constraint are expressed as appropriate linear combinations of
quantum commutators between the left-SUSY generator $\hat{L}_{A}$
and the right-SUSY generator $\hat{\overline{L}}_{\bar{A}}$.
It is therefore possible to easily solve the constraints
without recourse to Ashtekar's new variables.
In fact, starting from the ansatz
\beq
\Psi[e_{a}^{i}]=\prod_{x\in\M3}\hat{L}^{A}(x)\hat{L}_{A}(x)
g(e^{\alpha}_{a}),
\eeq
where $g(e^{\alpha}_{a})$ is
a local Lorentz invariant functional of the vierbein
$e^{\alpha}_{a}$,
we can get a solution\cite{csordas} which is somewhat
similar to the topological solutions discussed in \S\S 6.6:
\beq
\Psi_{WH}[e_{a}^{i}]=\int[d\ep^{A}(x)]\int[d\theta^{i}(x)]
\exp(\int_{\M3}d^{3}x\ep^{A}\hat{L}_{A})\cdot
\exp(-\frac{1}{2}\int_{\M3}d^{3}x \otep^{abc}(e^{\theta})^{\alpha}
_{a}\partial_{b}(e^{\theta})_{c\alpha}),
\eeq
where $[d\ep^{A}(x)]$ and $[d\theta^{i}(x)]$
are $SU(2)$ invariant measures
and $(e^{\theta})^{\alpha}_{a}$ denotes the vierbein which is
obtained from $e^{\alpha}_{a}$ by the left-handed local Lorentz
transformation generated by $\theta^{i}(x)$.
By investigating its reduction to the spatially homogeneous model,
it turns out that this solution should be interpreted as a
worm-whole state\cite{csordas}.


\subsection{Remaining issues}

In this paper we have looked into the developments in
the program of solving
the Wheeler-De Witt equation. We have seen that, up to now, we have
found several types of solutions to the WD equation.
In order to complete the program of canonical quantum gravity
in terms of Ashtekar's new variables, however,
there still remain several problems aside from finding
complete set of solutions. We will conclude this paper
by listing a few of these problems:

1) Regularization and operator ordering. The solutions obtained
so far are formal in the sense that we have worked with
constraint equations which are not, or which are
at most incompletely, regularized.
In order to obtain mathematically rigorous solutions, the
constraint operators must be regularized correctly.
The problem of operator ordering considered in \S\S 2.2 is
in fact meaningful only when we regularize the constraints.
It is desirable that the regularized constraint operators
should form a closed commutator algebra and, in particular,
that they should respect the covariance under spatial
diffeomorphisms.\footnote{In Dirac's quantization
we usually require the quantum commutator algebra of
constraints to close in order to ensure the solvability
of constraint equations. One might consider this closedness
condition to be redundant in quantum gravity because we have
already found solutions at least heuristically. The author thinks,
however, that the existence of the Wilson-loop solutions
is not sufficient to guarantee the existence of {\em phisically
intriguing solutions}. While there may have not been
any signals of anomalous terms in the analysis made so far on
the spin network states, this is possibly due to
the simple structure of the space of spin network states
defined on {\em piecewise-analytic graphs}.}
It is likely that the regularization which satisfies
this condition, if any, should be independent of artificial
background structures. At present, the extended loop
representation\cite{bart} seems to be the only candidate
which enjoys this property. However, it is possible that
anomalies inevitably appear in the quantum commutator algebra.
Because Dirac's quantization cannot apply in this case,
a modified treatment of the constraint is required\cite{mats}
as that of the Virasoro constraints in string theory.
Anyway, a remarkable statement was made
in ref.\cite{mats} that the topological
solutions still remains physical states even in the presence
of anomalies, if an appropriate modification of
the quantization procedure is made.

2) How to deal with the diffeomorphism constraint?
In \S 4 and \S 5 we have solved only the integrated
diffeomorphism constraint (\ref{eq:spidiff}). In order to
assert that we have solved the diffeomorphism constraint
(\ref{eq:qdiffeo}), however, this is not sufficient.
The action of $\hat{\cal V}_{a}$ on the parallel propagator
is calculated as
\beq
\int d^{3}xN^{a}(x)\hat{\cal V}_{a}(x)h_{\alpha}[0,1]
=\int_{0}^{1}dsN^{a}(\alpha(s))\dot{\alpha}^{b}(s)\Delta_{ab}
(\alpha,\alpha(s))h_{\alpha}[0,1].\label{eq:infinidiff}
\eeq
This is visualized for example in figure 5 as the difference
between the results of inserting
infinitesimal loops which are sitting
in the direction of $\dot{\alpha}^{a}(s)$ and which are
stretching toward the opposite directions.
Thus in a microscopic viewpoint this may change the
differential topology of curves. When we deal with 
piecewise analytic graphs, no problem arises because
this operation is not well-defined (appendix C of
\cite{ALMMT2}). However, because it is almost
definite that in quantum gravity we should work with graphs which
are not piecewise analytic, it is possible
that different treatments yield different results.
Roughly speaking there are two options: i) consider
only the integrated version (\ref{eq:spidiff}) as relevant;
ii) solve the entire equation (\ref{eq:qdiffeo}) by appropriately
regularizing the expression like (\ref{eq:infinidiff}).
It is then possible that only combinatorial topology of the graph
is relevant and that some solutions are given by approriate
extensions of the combinatorial solutions to non-piecewise analytic
graphs.
While in \S 4 and \S 5 we have followed the attitude i),
the discussion in ref.\cite{GP2} seems to based on ii).
Anyway, what is the correct treatment of $\hat{\cal V}_{a}$ will
be finally determined by some physical requirement.

\begin{figure}[t]
\begin{center}

\begin{picture}(150,40)(-5,0)
\put(0,0){\makebox(20,30){$\int d^{3}xN^{a}
\hat{{\cal V}}_{a}$}}
\put(20,0){\usebox{\LBRA}}
\put(30,5){\vector(0,1){20}}

\put(35,0){\usebox{\RBRA}}
\put(40,0){\makebox(5,30){$=$}}
\put(45,13){$\int dsN^{a}(\alpha(s))$}
\put(70,13){$\lim$}
\put(70,10){\scriptsize$\delta\rightarrow0$}
\put(78,13){\Large$\frac{1}{\pi\delta^{2}}$}
\put(85,0){\usebox{\LBRA}}

\put(100,5){\line(0,1){8}}
\put(100,17){\vector(0,1){8}}
\put(100,15){\oval(4,4)[r]}
\put(105,0){\makebox(15,30){$-$}}
\put(125,5){\line(0,1){8}}
\put(125,17){\vector(0,1){8}}
\put(125,15){\oval(4,4)[l]}

\put(135,0){\usebox{\RBRA}}
\end{picture}

\end{center}
\caption{Action of the vector constraint on
a parallel propagator.}
\end{figure}

3) The physical inner product. This problem is a central
problem in Ashtekar's formalism because it is closely related
with that of implementing the reality conditions.
As is briefly explained in \S\S 3.4 a plausible candidate
for the physical inner product is the Wick rotated measure
\cite{thie}. The problem is whether it is appropriately
regularized because this measure is induced by the Wick rotator
which involves a non-polynomial functional of conjugate momenta.

4) Finding physical observables. Not only the physical
inner product but also physical observables are
necessary to extract quantum gravitational information
from solutions to the constraints. We know however little about
them because the diffeomorphism invariance severely restricts
the possible form of observables. In the matter-coupled gravity,
area and volume operators are expected to provide a large set of
physical observables. In the topological sector, namely the sector
which is described purely by $SL(2,\BC)$ BF theory, a set
of nonlocal observables can be constructed by using the BRST
descent sequences\cite{chang}. However, what we are interested in
is a set of physical observables which measure e.g. graviton
excitations efficiently. Unfortunately, as far as the author knows,
such a set has not yet been discovered.

It is not likely that these problems are resolved separately,
because they are intimately related with each other.
While it seems to take great efforts to solve these formidable
issues, we have a flash of hope.
The existence of the topological solutions guarantees that
our labors toward this direction will not be wasted.
We expect that a true theory of quantum gravity
will appear as an extension of this program of nonperturbative
canonical quantum gravity.


\vspace{0.2in}

\noindent {\large\bf Acknowledgments}

I would like to thank Professors K. Kikkawa,
K. Higashijima, A. Sato, H. Itoyama, T. Kubota and H. Kunitomo 
for useful discussions and careful readings of the manuscript.
I am also grateful to Prof. L. Smolin for his
complements on this thesis and informing me of
some literature, and Prof. R. Loll
for her helpful comments on my previous work (ref.\cite{ezawa4}).
This work is supported by the Japan Society for the
Promotion of Science.

\newpage
\setcounter{section}{0}
\setcounter{section}{0}


\appendix{The projector $P^{(-)i}_{\alpha\beta}$}

Here we provide the definition and some properties of the
projector $P^{(-)i}_{\alpha\beta}$. First we define the
projection operator $P^{(-)\alpha\beta}_{\quad\gamma\delta}$
into the space of anti-self-dual Lorentz bi-vectors:
\beq
P^{(-)\alpha\beta}_{\quad\gamma\delta}=
\frac{1}{4}(\delta^{\alpha}_{\gamma}\delta^{\beta}_{\delta}-
\delta^{\alpha}_{\delta}\delta^{\beta}_{\gamma}
-i\ep^{\alpha\beta}_{\quad\gamma\delta}),
\eeq
where $\ep^{\alpha\beta\gamma\delta}$ is the totally anti-symmetric
pseudo tensor with $\ep^{0123}=\ep^{123}=1$. We use the
metric $(\eta_{\alpha\beta})=(\eta^{\alpha\beta})=
{\rm diag}(-1,1,1,1)$ to raise or lower the Lorentz indices.
This projection operator possesses the following properties
\beq
P^{(-)\alpha\beta}_{\quad\gamma\delta}=
-\frac{i}{2}\ep^{\alpha\beta}_{\quad\alpha^{\prime}\beta^{\prime}}
P^{(-)\alpha^{\prime}\beta^{\prime}}_{\quad\gamma\delta}=-
\frac{i}{2}P^{(-)\alpha\beta}_{\quad\gamma^{\prime}\delta^{\prime}}
\ep^{\gamma^{\prime}\delta^{\prime}}_{\quad\gamma\delta}=
P^{(-)\alpha\beta}_{\quad\alpha^{\prime}\beta^{\prime}}
P^{(-)\alpha^{\prime}\beta^{\prime}}_{\quad\gamma\delta}.
\eeq
The projector $P^{(-)i}_{\quad\alpha\beta}$ is defined as
\beqy
P^{(-)i}_{\quad\alpha\beta}&\equiv&
\frac{1}{2}(\delta^{0}_{\alpha}\delta^{i}_{\beta}-
\delta^{0}_{\beta}\delta^{i}_{\alpha}-i
\ep^{0i}_{\quad\alpha\beta})\nonumber \\
&=&2P^{(-)0i}_{\quad\alpha\beta}=-i\ep^{ijk}
P^{(-)jk}_{\quad\alpha\beta}.
\eeqy
This projector satisfies the following identities
\beqy
P^{(-)i}_{\quad\gamma\delta}P^{(-)i\alpha\beta}&=&
-P^{(-)\alpha\beta}_{\quad\gamma\delta}\label{eq:projection} \\
\eta^{\beta\delta}P^{(-)i}_{\quad\alpha\beta}
P^{(-)j}_{\quad\delta\gamma}
&=&\frac{i}{2}\ep^{ijk}P^{(-)k}_{\quad\alpha\gamma}+
\frac{1}{4}\delta^{ij}\eta_{\alpha\gamma}.
\eeqy
Using this projector we can give the relation between $SO(3,1)$
representation $\Lambda^{\alpha}\LI{\beta}$ and $SO(3,{\bf C})$
representation $\Lambda^{ij}$ of the (proper orthochronous)
Lorentz group:
\beq
\Lambda^{ij}=-P^{(-)i}_{\quad\alpha\beta}\Lambda^{\alpha}
\LI{\gamma}\Lambda^{\beta}\LI{\delta}P^{(-)j\gamma\delta}.
\eeq
This $SO(3,{\bf C})$ representation is obtained as the adjoint
representation of $SL(2,{\bf C})$:
\beq
(e^{\theta^{k}J_{k}})^{ij}\Phi^{j}J_{i}=e^{\theta^{k}J_{k}}
\Phi^{j}J_{j}e^{-\theta^{k}J_{k}},
\eeq
where $(J_{k})^{ij}=\ep^{ikj}$ is the $SL(2,{\bf C})$ generator
in the adjoint representation.


\appendix{Invariant measures on $SU(2)$ and $SL(2,{\bf C})$}

Here we give useful formulae for the invariant measures on
$SU(2)$ and $SL(2,{\bf C})$, namely, the Haar measure and
the (averaged) heat-kernel measure. As for detailed discussion
on measures on general
compact Lie groups and on their complexifications, we refer
the reader to refs.\cite{creu} and \cite{hall}.

For simplicity we will consider only finite dimensional
representations, namely unitary representations in $SU(2)$.
Because any unitary representation is equivalent to the
direct sum of a finite number of irreducible representations,
it is sufficient to deal only with irreducible representations,
i.e. spin-$\frac{p}{2}$ representations $\pi_{p}$ ($p\in{\bf Z}$).

\subsection{The Haar measure on $SU(2)$}

The Haar measure $d\mu(g)$ on $SU(2)$  ($g\in SU(2)$)is completely
determined by the following two conditions\cite{creu}:
\beqy
\mbox{normalization}&:&\int d\mu(g)=1 \\
\mbox{bi-$SU(2)$ invariance}&:&d\mu(h_{1}gh_{2})=d\mu(g),
\eeqy
where $h_{1}$ and $h_{2}$ are some fixed elements of $SU(2)$.
By exploiting the unitarity $\overline{\pi_{p}(g)_{I}\UI{J}}
=\pi_{p}(g^{-1})_{J}\UI{I}$ ($I,J=1,\ldots,p+1$),
we immediately find
\beq
\int d\mu(g)\overline{\pi_{p}(g)_{I}\UI{J}}
\pi_{q}(g)_{I^{\prime}}\UI{J^{\prime}}=\frac{1}{p+1}\delta_{p,q}
\delta_{I^{\prime}}^{I}\delta^{J^{\prime}}_{J}.\label{eq:haar}
\eeq
This formula plays a fundamental role when we consider the
measure on $SU(2)$ gauge theories. 

While eq.(\ref{eq:haar}) suffices for any computation of
inner products or expectation values in the framework of
spin network states, it is sometimes convenient if we know
formulae in terms only of the spinor representation
$V^{A}\LI{B}\equiv\pi_{1}(V)^{A}\LI{B}$. Their computations
are straightforward if we use the property $\ep_{AC}V^{A}\LI{B}
V^{C}\LI{D}=\ep_{BD}$. We provide only two main results. One is
\beq
\int d\mu(V)V^{A_{1}}\LI{B_{1}}V^{A_{2}}\LI{V_{2}}\cdots
V^{A_{2m}}\LI{B_{2m}}=
\frac{1}{(m+1)!m!2^{m}}\sum_{\sigma\in P_{2m}}\prod_{k=1}^{m}
\ep^{A_{\sigma_{2k-1}}A_{\sigma_{2k}}}
\ep_{B_{\sigma_{2k-1}}B_{\sigma_{2k}}}.
\eeq
where $m$ is a non-negative integer and $P_{n}$ denotes the
group of permutations of $n$ entries. The integration vanishes
if the integrand is the product of
an odd number of copies of spinor representation.
The other is
\beq
\int d\mu(V)\overline{V^{A_{1}}\LI{(B_{1}}\cdots
V^{A_{n}}\LI{B_{n})}}V^{C_{1}}\LI{(D_{1}}\cdots
V^{C_{m}}\LI{D_{m})}=\frac{1}{n+1}\delta_{n,m}
\delta^{C_{1}}_{(A_{1}}\cdots\delta^{C_{n}}_{A_{n})}
\delta^{(B_{1}}_{D_{1}}\cdots\delta^{B_{n})}_{D_{n}}.
\label{eq:spihaar}
\eeq
In deriving the latter equation we have used the unitarity
$$
\overline{V^{B}\LI{A}}=(V^{-1})^{A}\LI{B}=
\ep_{BC}\ep^{AD}V^{C}\LI{D}.
$$
We should notice that eq.(\ref{eq:haar}) and eq.(\ref{eq:spihaar})
are two different realizations of the same formula.


\subsection{The heat-kernel measure on $SL(2,{\bf C})$}

While the Haar measure provides a useful basis for the inner product
on compact gauge theories, its analogue does not exist for
complexified gauge groups such as $SL(2,{\bf C})$.
However, there is a candidate for the well-defined measure on
complexified gauge groups. This was discovered by Hall\cite{hall}
in the context of extending the coherent state transform
to compact Lie groups.

The key ingredient is the coherent state transform
$$
C_{t}:L^{2}(SU(2),d\mu)\rightarrow L^{2}(SL(2,{\bf C}),d\nu_{t})
\cap\CH(SL(2,{\bf C}))
$$
($t>0$) from the space of functions on $SU(2)$ which are
square-integrable w.r.t. the Haar measure $d\mu$ to the space of
holomorphic functions on $SL(2,{\bf C})$ which are square-integrable
w.r.t the \lq\lq averaged" heat-kernel measure $d\nu_{t}$.
Explicitly $C_{t}$ is given by
\beq
C_{t}[F](g)=\int_{SU(2)}d\mu(x)F(x)\rho_{t}(x^{-1}g),
\eeq
where $F(x)\in L^{2}(SU(2),d\mu)$ and $g\in SL(2,{\bf C})$.
$\rho_{t}(g)$ is the analytic continuation to $SL(2,{\bf C})$
of the heat kernel for the Casimir operator $\Delta=X_{i}X_{i}$
on $SU(2)$:\footnote{$X_{i}$ stands for the left-invariant
vector field corresponding to the $SU(2)$ generator $J_{i}$.
$\Delta$ can be regarded as the Laplace-Beltrami operator on
$SU(2)$}
\beqy
\frac{d}{dt}\rho_{t}(x)&=&\frac{1}{2}\Delta\rho_{t}(x)\nonumber \\
\lim_{t\rightarrow0}\rho_{t}(x)&=&\delta(x,{\bf 1})
\eeqy

For our purpose the following theorem is the most important:

{\em {\bf Theorem2 (Hall \cite{hall})}: The coherent
state transform $C_{t}$ is an isometric isomorphism
of $L^{2}(SU(2),d\mu)$ onto $L^{2}(SL(2,{\bf C}),d\nu_{t})
\cap\CH(SL(2,{\bf C}))$}.

We can express this theorem by
\beq
\int_{SL(2,{\bf C})/SU(2)}d\nu_{t}(g)\overline{C_{t}[F](g)}
C_{t}[F^{\prime}](g)=\int_{SU(2)}d\mu(x)\overline{F(x)}
F^{\prime}(x).
\eeq

In order to obtain the explicit formula
we further need eq.(30) of ref.\cite{hall}
\beq
C_{t}[\pi(x)](g)=e^{\frac{t}{2}\pi(\Delta)}\pi(g).\label{eq:coherent}
\eeq
Using these results and eq.(\ref{eq:haar})
we find the desired formula:
\beq
\int d\nu_{t}(g)\overline{\pi_{p}(g)_{I}\UI{J}}\pi_{q}
(g)_{I^{\prime}}\UI{J^{\prime}}=
\frac{\delta_{p,q}}{p+1}e^{t\frac{p(p+2)}{4}}
\delta^{I}_{I^{\prime}}\delta^{J^{\prime}}_{J}.\label{eq:heatkernel}
\eeq
For the explicit expression of $d\nu_{t}$, we refer the reader
to \cite{hall}. Here we only say that $d\nu_{t}$ is
essentially the heat-kernel measure on $SL(2,{\bf C})/SU(2)$
which is bi-$SU(2)$ invariant and left-$SL(2,{\bf C})$ invariant,
and that they are defined by using the Laplace-Beltrami operator
on $SL(2,{\bf C})$.


\appendix{Basic action of the renormalised scalar constraint}

Here we list the action of the
renormalized scalar constraint $\hat{\CS}^{ren}(\tN)$
on the basic configurations in terms of the graphical notation.
We will denote the parallel propagators along
the curves $\alpha$ and $\beta$ by
an upward vertical arrow and a
horizontal arrow oriented to the right respectively.
For example, we write the tensor product
of the propagators along $\alpha$ and $\beta$ as follows
\footnote{We will
assume that $\alpha$ and $\beta$ intersects once at $x_{0}=
\alpha(s_{0})=\beta(t_{0})$.}:

\begin{figure}[h]
\begin{picture}(150,30)(-10,0)
\put(0,0){\makebox(40,30){%
$h_{\alpha}[0,1]_{A}\UI{B}h_{\beta}[0,1]_{C}\UI{D}$}}
\put(40,0){\makebox(20,30){$=$}}
\put(65,0){\usebox{\LBRA}}
\put(90,5){\vector(0,1){20}}
\put(80,15){\vector(1,0){20}}
{\scriptsize
\put(90,2){$A$}
\put(90,25){$B$}
\put(80,15){$C$}
\put(100,15){$D$}}
\put(110,0){\usebox{\RBRA}}
\put(115,0){\makebox(5,30){.}}
\put(120,0){(C.1)}
\end{picture}
\end{figure}

As is seen in \S\S 4.1,
we have only to concentrate on the action at vertices.
In the present case the only vertex
is at $x_{0}$.
In the following we provide some of the
basic actions on the regular
four-valent vertex $x_{0}$
in the graphical notation. The dot in the diagram denotes the
the location in which
the magnetic field $\UT{n}\cdot\tB$ is inserted.
Using equations (C.2-8) and
identities(\ref{eq:identity1})(\ref{eq:2spi2})
(\ref{eq:2spi3}), we can write out all the basic action,
namely the action of the
renormalized scalar constraint $\hat{\CS}^{ren}(\tN)$ on
any single propagator and any pair of propagators,
in the regular four-valent graph.

\begin{figure}[h]
\begin{picture}(150,70)(-10,0)
\put(0,40){\makebox(15,30){$\hat{\CS}^{ren}_{1}(\tN)$}}
\put(15,40){\usebox{\LBRA}}
\put(30,45){\line(0,1){10}}
\put(30,55){\vector(1,0){10}}
{\scriptsize
\put(30,42){$A$}
\put(40,55){$B$}}

\put(45,40){\usebox{\RBRA}}
\put(50,40){\makebox(10,30){$=$}}
\put(60,40){\usebox{\LBRA}}

\put(80,45){\line(0,1){10}}
\put(80,55){\vector(1,0){10}}
\put(80,55){\circle*{1.5}}
{\scriptsize
\put(80,42){$A$}
\put(90,55){$B$}}

\put(110,40){\usebox{\RBRA}}
\put(115,40){\makebox(5,30){,}}
\put(120,40){(C.2)}


\put(0,0){\makebox(15,30){$\hat{\CS}^{ren}_{2}(\tN)$}}
\put(15,0){\usebox{\LBRA}}
\put(30,5){\vector(0,1){20}}
\put(20,15){\vector(1,0){20}}
{\scriptsize
\put(30,2){$A$}
\put(30,25){$B$}
\put(18,15){$C$}
\put(40,15){$D$}}

\put(40,0){\usebox{\RBRA}}
\put(45,0){\makebox(10,30){$=$ $2$}}
\put(55,0){\usebox{\LBRA}}

\put(60,15){\line(1,0){10}}
\put(71,14){\vector(1,0){9}}
\put(70,15){\vector(0,1){10}}
\put(71,5){\line(0,1){9}}
\put(71,14){\circle*{1.5}}
{\scriptsize
\put(70,2){$A$}
\put(70,25){$B$}
\put(58,15){$C$}
\put(80,14){$D$}}
\put(80,0){\makebox(15,30){$-$}}
\put(95,16){\line(1,0){9}}
\put(105,15){\vector(1,0){10}}
\put(104,16){\vector(0,1){9}}
\put(105,5){\line(0,1){10}}
\put(104,16){\circle*{1.5}}
{\scriptsize
\put(105,2){$A$}
\put(105,25){$B$}
\put(93,16){$C$}
\put(115,15){$D$}}

\put(115,0){\usebox{\RBRA}}
\put(120,0){\makebox(5,30){,}}
\put(125,0){(C.3)}
\end{picture}
\end{figure}

\begin{figure}[p]
\begin{picture}(150,190)(-10,-40)
\put(0,120){\makebox(15,30){$\hat{\CS}^{ren}_{2}(\tN)$}}
\put(15,120){\usebox{\LBRA}}
\put(27,125){\vector(0,1){20}}
\put(30,125){\line(0,1){10}}
\put(30,135){\vector(1,0){10}}
{\scriptsize
\put(27,122){$A$}
\put(27,145){$B$}
\put(30,122){$C$}
\put(40,135){$D$}}

\put(40,120){\usebox{\RBRA}}
\put(45,120){\makebox(10,30){$=$}}
\put(55,120){\usebox{\LBRA}}

\put(66,135){\vector(0,1){10}}
\put(70,135){\vector(1,0){10}}
\put(70,125){\line(0,1){6}}
\put(66,125){\line(0,1){6}}
\put(66,131){\line(1,1){4}}
\put(66,135){\line(1,-1){4}}
\put(70,135){\circle*{1.5}}
{\scriptsize
\put(66,122){$A$}
\put(66,145){$B$}
\put(70,122){$C$}
\put(80,135){$D$}}
\put(80,120){\makebox(15,30){$-$}}
\put(101,135){\vector(0,1){10}}
\put(105,135){\vector(1,0){10}}
\put(105,125){\line(0,1){6}}
\put(101,125){\line(0,1){6}}
\put(101,131){\line(1,1){4}}
\put(101,135){\line(1,-1){4}}
\put(101,135){\circle*{1.5}}
{\scriptsize
\put(101,122){$A$}
\put(101,145){$B$}
\put(105,122){$C$}
\put(115,135){$D$}}

\put(115,120){\usebox{\RBRA}}
\put(120,120){\makebox(5,30){,}}
\put(125,120){(C.4)}


\put(0,80){\makebox(15,30){$\hat{\CS}^{ren}_{2}(\tN)$}}
\put(15,80){\usebox{\LBRA}}
\put(30,85){\vector(0,1){20}}
\put(27,95){\vector(0,1){10}}
\put(20,95){\line(1,0){7}}
{\scriptsize
\put(20,95){$A$}
\put(27,105){$B$}
\put(30,82){$C$}
\put(30,105){$D$}}

\put(40,80){\usebox{\RBRA}}
\put(45,80){\makebox(10,30){$=$}}
\put(55,80){\usebox{\LBRA}}

\put(66,99){\vector(0,1){6}}
\put(60,95){\line(1,0){6}}
\put(70,85){\line(0,1){10}}
\put(70,99){\vector(0,1){6}}
\put(66,95){\line(1,1){4}}
\put(66,99){\line(1,-1){4}}
\put(66,99){\circle*{1.5}}
{\scriptsize
\put(60,95){$A$}
\put(66,105){$B$}
\put(70,82){$C$}
\put(70,105){$D$}}
\put(80,80){\makebox(15,30){$-$}}
\put(101,99){\vector(0,1){6}}
\put(95,95){\line(1,0){6}}
\put(105,85){\line(0,1){10}}
\put(105,99){\vector(0,1){6}}
\put(101,95){\line(1,1){4}}
\put(101,99){\line(1,-1){4}}
\put(105,99){\circle*{1.5}}
{\scriptsize
\put(95,95){$A$}
\put(101,105){$B$}
\put(105,82){$C$}
\put(105,105){$D$}}

\put(115,80){\usebox{\RBRA}}
\put(120,80){\makebox(5,30){,}}
\put(125,80){(C.5)}


\put(0,40){\makebox(15,30){$\hat{\CS}^{ren}_{2}(\tN)$}}
\put(15,40){\usebox{\LBRA}}
\put(32,55){\vector(1,0){8}}
\put(28,55){\vector(-1,0){8}}
\put(28,45){\line(0,1){10}}
\put(32,45){\line(0,1){10}}
{\scriptsize
\put(28,42){$A$}
\put(20,55){$B$}
\put(32,42){$C$}
\put(40,55){$D$}}

\put(40,40){\usebox{\RBRA}}
\put(45,40){\makebox(10,30){$=$}}
\put(55,40){\usebox{\LBRA}}

\put(72,55){\vector(1,0){8}}
\put(68,55){\vector(-1,0){8}}
\put(68,45){\line(0,1){6}}
\put(72,45){\line(0,1){6}}
\put(68,51){\line(1,1){4}}
\put(68,55){\line(1,-1){4}}
\put(72,55){\circle*{1.5}}
{\scriptsize
\put(68,42){$A$}
\put(60,55){$B$}
\put(72,42){$C$}
\put(80,55){$D$}}
\put(80,40){\makebox(15,30){$-$}}
\put(107,55){\vector(1,0){8}}
\put(103,55){\vector(-1,0){8}}
\put(103,45){\line(0,1){6}}
\put(107,45){\line(0,1){6}}
\put(103,51){\line(1,1){4}}
\put(103,55){\line(1,-1){4}}
\put(103,55){\circle*{1.5}}
{\scriptsize
\put(103,42){$A$}
\put(95,55){$B$}
\put(107,42){$C$}
\put(115,55){$D$}}

\put(115,40){\usebox{\RBRA}}
\put(120,40){\makebox(5,30){,}}
\put(125,40){(C.6)}


\put(0,0){\makebox(15,30){$\hat{\CS}^{ren}_{2}(\tN)$}}
\put(15,0){\usebox{\LBRA}}
\put(32,15){\line(1,0){8}}
\put(20,15){\line(1,0){8}}
\put(28,15){\vector(0,1){10}}
\put(32,15){\vector(0,1){10}}
{\scriptsize
\put(28,25){$B$}
\put(20,15){$A$}
\put(32,25){$D$}
\put(40,15){$C$}}

\put(40,0){\usebox{\RBRA}}
\put(45,0){\makebox(10,30){$=$}}
\put(55,0){\usebox{\LBRA}}

\put(72,15){\line(1,0){8}}
\put(60,15){\line(1,0){8}}
\put(68,19){\vector(0,1){6}}
\put(72,19){\vector(0,1){6}}
\put(68,15){\line(1,1){4}}
\put(68,19){\line(1,-1){4}}
\put(72,15){\circle*{1.5}}
{\scriptsize
\put(68,25){$B$}
\put(60,15){$A$}
\put(72,25){$D$}
\put(80,15){$C$}}
\put(80,0){\makebox(15,30){$-$}}
\put(107,15){\line(1,0){8}}
\put(95,15){\line(1,0){8}}
\put(103,19){\vector(0,1){6}}
\put(107,19){\vector(0,1){6}}
\put(103,15){\line(1,1){4}}
\put(103,19){\line(1,-1){4}}
\put(103,15){\circle*{1.5}}
{\scriptsize
\put(103,25){$B$}
\put(95,15){$A$}
\put(107,25){$D$}
\put(115,15){$C$}}

\put(115,0){\usebox{\RBRA}}
\put(120,0){\makebox(5,30){,}}
\put(125,0){(C.7)}


\put(0,-40){\makebox(15,30){$\hat{\CS}^{ren}_{2}(\tN)$}}
\put(15,-40){\usebox{\LBRA}}
\put(31,-26){\vector(1,0){9}}
\put(20,-24){\line(1,0){9}}
\put(29,-24){\vector(0,1){9}}
\put(31,-35){\line(0,1){9}}
{\scriptsize
\put(29,-15){$B$}
\put(20,-24){$C$}
\put(31,-38){$A$}
\put(40,-26){$D$}}

\put(40,-40){\usebox{\RBRA}}
\put(45,-40){\makebox(10,30){$=$}}
\put(55,-40){\makebox(15,30){$\hat{\CS}^{ren}_{2}(\tN)$}}
\put(70,-40){\usebox{\LBRA}}

\put(78,-23){\vector(1,0){12}}
\put(81,-26){\vector(1,0){9}}
\put(78,-35){\line(0,1){12}}
\put(81,-35){\line(0,1){9}}
{\scriptsize
\put(90,-23){$B$}
\put(78,-38){$A$}
\put(90,-26){$D$}
\put(81,-38){$C$}}
\put(95,-40){\usebox{\RBRA}}
\put(100,-40){\makebox(20,30){$=0$ .}}
\put(125,-40){(C.8)}
\end{picture}
\end{figure}


\end{document}